\documentclass[aps,prb,twocolumn,longbibliography,preprintnumbers,amsmath,amssymb,superscriptaddress]{revtex4-1}

\usepackage{graphicx}
\usepackage[T1]{fontenc}
\usepackage[colorlinks,bookmarks=false,citecolor=blue,linkcolor=red,urlcolor=blue]{hyperref}
\usepackage[dvipsnames]{xcolor}
\usepackage{times}

\usepackage{bbm}

\newcommand{\be}{\begin{eqnarray}}
\newcommand{\ee}{\end{eqnarray}}

\newcommand{\nn}{\nonumber } 
\newcommand{\Eqref}[1]{Eq.~\eqref{#1}}

\begin{document}

\author{Daniel D. Scherer and Brian M. Andersen}
\affiliation{Niels Bohr Institute, University of Copenhagen, Lyngbyvej 2, DK-2100 Copenhagen, Denmark}

\title{Effects of spin-orbit coupling on spin-fluctuation induced pairing in iron-based superconductors}

\begin{abstract}

We perform a theoretical study of the leading pairing instabilities and the associated superconducting gap functions within the spin-fluctuation mediated pairing scenario in the presence of spin-orbit coupling (SOC). Focussing on iron-based superconductors (FeSCs), our model Hamiltonian consists of a realistic density functional theory (DFT)-derived ten-band hopping term, spin-orbit coupling, and electron-electron interactions included via the multi-orbital Hubbard-Hund Hamiltonian. We perform an extensive parameter sweep and investigate different doping regimes including cases with only hole- or only electron Fermi pockets. In addition, we explore two different bandstructures: a rather generic band derived for LaFeAsO but known to represent standard DFT-obtained bands for iron-based superconductors, and a band specifically tailored for FeSe which exhibits a notably different Fermi surface compared to the generic case. It is found that for the generic FeSCs band, even rather large SOC has negligible effect on the resulting gap structure; the $s_{+-}$ (pseudo-)spin singlet pairing remains strongly favored and SOC does not lead to any SOC-characteristic gap oscillations along the various Fermi surfaces. By contrast in the strongly hole-doped case featuring only hole-pockets around the $\Gamma$-point, the leading solution is $d$-wave pseudo-spin singlet, but with a notable SOC-driven tendency towards helical pseudo-spin triplet pairing, which may even become the leading instability. In the heavily electron doped situation, featuring only electron pockets centered around the $M$-point, the leading superconducting instabilities are pseudo-spin singlets with SOC favoring the $s$-wave case as compared to $d$-wave pairing, which is the favored gap symmetry for vanishing SOC. We end with a discussion pertaining to the role of SOC on the gap structure relevant for FeSe from a weak-coupling point of view. As our formalism, described at length in the supplementary material, is rather general, our study is of relevance to other multiband superconductors as well.
 
\end{abstract}

\maketitle

\section{Introduction}
\label{sec:intro}

The strive for a detailed understanding of the electronic properties of a growing class of materials, broadly dubbed `unconventional superconductors', currently represents one of the most important research challenges in condensed matter physics. Here, the word unconventional refers to superconductivity generated not by the standard BCS phonon mechanism, and often features superconducting pairing gaps with sign-changing order parameters in momentum space. Important material families considered to belong to this fascinating class of materials include, for example, heavy fermion compounds, cuprates, organic Bechgaard materials, strontium ruthenate oxide compounds, and iron-based superconductors including both the iron-pnictides and iron-chalcogenides~\cite{scalapinoreview,Taillefer2019}. In addition, it is currently under intense discussion whether twisted bilayer graphene may also belong to the class of unconventional superconductors~\cite{TBG}.

For unconventional superconductivity the pairing mechanism is of central importance, i.e., what is the microscopic origin of superconductivity in the above-mentioned materials. Due to the existence of significant electron-electron interactions, the presence of attractive `glue' for Cooper pairing has often been pursued from purely repulsive (bare) models such as the Hubbard model and its close cousins. These models have been studied theoretically both from a strong-coupling and a weak-coupling perspective, and in both cases the role of over-screening and magnetic fluctuations have been highlighted. Indeed, the ability of Coulomb repulsions to dress the bare onsite Coulomb interaction, and generate a Cooper instability in higher angular momentum channels is well-known, and derives from seminal work in the mid 1960s by Kohn and Luttinger,~\cite{kohn} and by Berk and Schrieffer~\cite{berk66}, along with other theoretical extensions found in Refs.~\onlinecite{scalapino86,beal-monod86,miyake86}, and reviewed e.g. in Ref.~\onlinecite{maiti}.

The superconducting pairing gap is the most direct measurable quantity directly derived from the pairing kernel, and has therefore always exhibited a prominent position in the study of unconventional superconductivity. This is because of the fact that the irreducible representation to which the leading superconducting instability belongs, has close ties to the physical properties of the fluctuations generating the superconducting state. Importantly, however, the gap structure contains much more information than just the basic transformation properties of the gap in momentum- (or real-) space. These include e.g. the detailed form of the gap along the Fermi surface, the possibility of accidental gap nodes, or the existence of sign-changes between different Fermi surface sheets in multiband materials~\cite{hirschfeld2011,chubukov2012}. The study of such gap details is largely driven by improved spectroscopic experimental probes with high (sub meV) resolution in both energy-  and momentum-space. Important examples where peculiar gap structures have been experimentally detected, and sparked numerous subsequent theoretical investigations, are found within the family of iron-based superconductors (FeSCs). In this class of superconductors the observed gap structures can be very rich and may vary substantially between the different specific compounds. While it is currently established that several of the FeSCs host sign-changing gap functions~\cite{hirschfeld2015,martiny,sprau,HHWen}, a plethora of other gap details have been mapped out by various spectroscopic probes~\cite{hirschfeld2011,chubukov2012,hirschfeld2016}.

For example, the material LiFeAs has been shown to exhibit a puzzling superconducting gap structure~\cite{hirschfeld2016}. The Fermi surface consists of two small and one large hole pocket centered at the $\Gamma$ point and two electron pockets centered at the $M$ point of the 2-Fe Brillouin zone (BZ)~\cite{borisenkolifeas1,borisenkolifeas2}. Both ARPES and STM measurements reported a significant gap anisotropy as a function of the Fermi surface angle along the different pockets~\cite{borisenkolifeas1,borisenkolifeas2,Umezawa,Allan}. For example, the two large hole pockets were found by STM quasi-particle interference (QPI) measurements to exhibit pronounced oscillations of the gap amplitude with four distinct minima (maxima) along the Fe-Fe bond direction for the largest (smallest) hole pocket. Likewise the two electron pockets also exhibited a notable oscillatory gap behavior along the rim of their respective Fermi pockets~\cite{Allan}. Another interesting material with unusual gap structure is KFe$_2$As$_2$, which has a Fermi surface made up from three hole pockets centered at $\Gamma$, and smaller hole barrels near the $M$ point of the BZ~\cite{Evtushinsky,Terashima}. The pairing symmetry of KFe$_2$As$_2$ has been a topic of significant controversy with some experimental probes finding $d$-wave gap symmetry while others advocated for a fully gapped $s$-wave state~\cite{Dong,Reid,Okazaki,Bohm,Hardy}. Finally we point out the gap structure of FeSe, recently mapped out in detail by QPI measurements~\cite{sprau}. FeSe is an electronic nematic system featuring only one small elliptical-shaped hole pocket at $\Gamma$ and two tiny electron peanut-shaped pockets at the $M$ point~\cite{Coldeareview,Bohmerreview,Kostin}. The gap structure is extremely anisotropic on both the hole- and the electron pockets despite the small C$_2$-breaking of the underlying lattice~\cite{sprau}. Such strong variation of the gap amplitude is highly unusual and surprising given the tiny extent of the Fermi pockets in momentum space, and has given rise to several recent theoretical investigations~\cite{onari,kreisel2017,she,kang2018,benfatto,rhodes,kreisel2018}.

The above examples of some of the peculiar superconducting gap structures known to exist in FeSCs highlight a significant theoretical challenge to explain these gap features. Naively, any successful theoretical model capturing the overall pairing symmetry and the gap details needs to incorporate the correct physical mechanism driving the superconductivity. However, an important concern relates to the fact that it is unclear at present to what extent gap details are universal and directly related to the underlying mechanism, or whether they originate from other physical effects not typically included in theoretical descriptions. For example, one might ask how the superconducting gap depends on e.g. inclusion of longer-range Coulomb repulsion or spin-orbit coupling (SOC). For example, incorporating the non-local effect of longer-range Coulomb repulsion has been shown to rather straightforwardly produce the small Fermi surface pockets of FeSe~\cite{Jiang,scherer2017}. Surprisingly, for the FeSCs the role of SOC on the superconducting gap structure has not been systematically investigated theoretically from the perspective of spin-fluctuation mediated superconductivity, despite a sizable SOC in these materials~\cite{borisenko,day}. There have been, however, microscopic studies of SOC-effects on the gap structure within the so-called orbital-spin fluctuation theory~\cite{saito}. Such studies concluded that SOC is important for explaining e.g. the observed $k_z$-dependence of the superconducting gap structure~\cite{saito,liu_2017}. In addition, there exists a number of theoretical works applying effective models starting from projected band structures that point out interesting possible consequences of SOC on the superconducting pairing of FeSCs~\cite{khodas,cvetkovic,vafek,eugenio,Boker}. For example, in the case of strong hole-doping, Vafek and Chubukov found a leading triplet $s$-wave pairing gap in the presence of SOC, and in a parameter regime where the Hund's coupling exceeds the intraorbital Hubbard repulsion $U'$~\cite{vafek}. This study included only some of the hole pockets and relies on interorbital triplet pairing, and found a nodal gap structure reminiscent to that found by laser ARPES in KFe$_2$As$_2$~\cite{Okazaki}. More recently, the preferred pairing relevant to the highly electron-doped case was also investigated theoretically. This study is relevant to e.g. monolayers of FeSe on STO and the lithium hydroxide intercalated FeSe compounds such as Li$_{1-x}$Fe$_x$OHFeSe. Starting again from an orbitally projected band model based on symmetry constraints from the crystalline symmetry of FeSe, Eugenio and Vafek classified the allowed pairing symmetries in the presence of SOC for the case of two $M$-centered electron pockets~\cite{eugenio}. In this case, the resulting Cooper-pairing symmetries can choose between $s$-wave, $d$-wave, and helical $p$-wave states, but pinpointing which is the favored state in the actual materials is still unsettled. Due to SOC, all three pair states are mixtures of physical spin-triplet and spin-singlet pairs. B\"{o}ker {\it et al.}, recently calculated the QPI patterns relevant to these three distinct states, and concluded from a comparison to experimental data, that only the triplet-dominated even parity A$_{1g}$ $s$-wave or the $B_{2g}$ $d$-wave states are consistent with currently available experiments on Li$_{1-x}$Fe$_x$(OHFe)$_{1-y}$Zn$_y$Se~\cite{Boker}.

Here we perform a detailed theoretical study of the pairing problem within spin-fluctuation mediated superconductivity relevant for FeSCs in the presence of SOC. We are particularly interested in the doping dependence of the pairing problem, including cases with only electron or only hole pockets at the Fermi surface. We apply the itinerant weak-coupling RPA framework and use a realistic ten-band model including SOC relevant to iron-based materials, and electron interactions included via the multi-orbital Hubbard-Hund Hamiltonian. The SOC is incorporated both in the bandstructure and in the two-particle pairing vertex providing the pairing kernel of the problem. We are motivated partly by the existence of significant gap anisotropy in some of the FeSCs, as discussed above, and the question to which extent SOC may (or may not) play a role in generating such unusual gap details. Another motivation comes from our own earlier theoretical investigations of the effects of SOC on the magnetic fluctuations in FeSCs~\cite{scherer2018,scherer2019}. In Ref.~\onlinecite{scherer2018} we focussed on the paramagnetic phase and found overall agreement between the theoretical results and the experimental data in terms of material-variability, doping-, temperature-, and energy-dependence of the polarization of the low-energy magnetic fluctuations~\cite{christenson2008,lumsden,dai,Inosov2016}. More recently, this was extended to the superconducting phase where the spin-space anisotropy of the neutron resonance was investigated, and again overall consistency was found to the experimental data~\cite{scherer2019}. These works give credence to the itinerant approach and the inclusion of SOC within our theoretical framework. Since the SOC breaks spin-rotational invariance, it picks out a preferred fluctuation direction of the low-energy spin excitations which could be of importance in determining the superconducting state. For an example where the inclusion of SOC and the associated spin anisotropy can change the leading superconducting instability from a spin-singlet state to a helical spin-triplet phase, we point to a recent theoretical study of Sr$_2$RuO$_4$~\cite{astrid19}. For Sr$_2$RuO$_4$ the SOC is substantially larger than in FeSCs, but compared to the Fermi energy $E_{\mathrm{F}}$ of some of the small Fermi pockets found in the latter class of materials, SOC can be comparable to $E_{\mathrm{F}}$, and it seems therefore potentially important to include SOC in order to properly describe pairing on such pockets. This highlights the importance of understanding the SOC-induced spin-space differentiation of the magnetic fluctuations and their role in determining the properties of the superconducting state. We note that understanding the effects of SOC in FeSCs has been also recently emphasized in terms of possible topological phases existing in these materials~\cite{hongding1,hongding2,jiang2019,pan}.

The paper is structured as follows. In Sec.~\ref{sec:model} we 
introduce the model Hamiltonian for the electronic system. The linearized
gap equation with a spin-fluctuation mediated pairing vertex, which forms 
the basis of our analysis, is briefly described in Sec.~\ref{sec:LGE}. Further
technial details on the gap equation and the pairing vertex can be found in the 
appendices Secs.~\ref{sec:BdG}-\ref{sec:vertex}. Having described model
and method, we eventually turn to presenting our numerical results for the
leading superconducting instabilities in the presence of SOC for two different bandstructures
in Sec.~\ref{subsec:LaFeAsO} (generic FeSCs band) and Sec.~\ref{subsec:FeSe} (FeSe). Finally, we conclude with a summary of our findings in Sec.~\ref{sec:summary}. 
Additional material on the momentum structure of spin fluctuations and the spin singlet and triplet character of the gap solutions can be found in Secs.~\ref{sec:orbital} and~\ref{sec:orbital_FeSe}.

\section{Model}
\label{sec:model}

In the following, we will define the Hamiltonian describing the electronic degrees of freedom of the $3d$ shell of iron-pnictide and iron-chalcogenide systems. We take a multiorbital Hubbard Hamiltonian $ H = H_{0} - \mu_{0} N + H_{\mathrm{SOC}}  + H_{\mathrm{int}} $, where $H_{0}$ is the hopping Hamiltonian encoding both the electronic bandstructure in the absence of SOC and the orbital character of single-particle states. Defining the fermionic operators $ c_{l i \mu \sigma}^{\dagger}$, $c_{l i \mu \sigma}$ to create and destroy, respectively, an electron on sublattice $ l $ at site $ i $ in orbital $ \mu $ with spin polarization $\sigma$, $H_{0}$ can be written as
\be
\label{eq:hopping}
H_{0} = \sum_{\sigma}\sum_{l,l^{\prime},i,j}\sum_{\mu,\nu} 
c_{li \mu \sigma}^{\dagger} 
t_{li;l^{\prime}j}^{\mu\nu}
c_{l^{\prime}j \nu \sigma},
\ee
where the hopping matrix elements $t_{li;l^{\prime}j}^{\mu\nu}$ are material specific. Working in the grand canonical ensemble, the electronic filling is fixed by the chemical potential $\mu_{0}$ with $N$ from above denoting the electronic number operator. The indices $l,l^{\prime} \in \{A,B\}$ denote the 2-Fe sublattices, corresponding to the two inequivalent Fe-sites in the 2-Fe unit cell due to the pnictogen(Pn)/chalcogen(Ch) staggering about the FePn/FeCh plane, and the indices $i,j$ run over the unit cells of the square lattice. The indices $\mu,\nu$ specify the $3d$-Fe orbitals with $ d_{xz}, d_{yz}, d_{x^2-y^2}, d_{xy} $ and $ d_{3z^{2}-r^{2}} $ symmetry. In the following, we will work in a so-called phase staggered basis, in which the $xz$ and $yz$ orbitals on the $B$ sublattice acquire a phase shift of $\pi$ relative to their counterparts on the $A$ sublattice. In this orbital basis, the angular momentum operator assumes different representations on the two sublattices and consequently, the site-local SOC Hamiltonian reads as
\be 
\label{eq:SOC}
H_{\mathrm{SOC}} = \frac{\lambda_{\mathrm{SOC}}}{2}\sum_{l,i} \sum_{\mu,\nu} \sum_{\sigma,\sigma^{\prime}}
c_{l i \mu \sigma}^{\dagger} [{\bf L}_{l}]_{\mu\nu}\cdot{\boldsymbol \sigma}_{\sigma\sigma^{\prime}}c_{l i \nu \sigma^{\prime}},
\ee
with $ \boldsymbol \sigma $ the vector of Pauli matrices acting in spin space and $ {\bf L}_{l} $ the matrix representations of the angular momentum operator on the $A$ and $B$ sublattice.

The goal of the present manuscript is to present a comprehensive analysis of the influence of SOC on the Cooper instabilities within a spin-fluctuation mediated pairing scenario for two electronic bandstructures, obtained for LaFeAsO and FeSe, respectively. While the bandstructure of LaFeAsO features many of the rather generic properties of iron pnictides (also outside of the 1111 family), the bandstructure of the chalcogenide FeSe is known to display marked differences relative to its pnictide cousins. The hopping parameters $t_{li;l^{\prime}j}^{\mu\nu}$ for the LaFeAsO bandstructure were taken from Ref.~\onlinecite{ikeda2010}, while the corresponding hopping parameters for FeSe were obtained in Ref.~\onlinecite{scherer2017}. Details of the resulting Fermi surfaces will be discussed in Sec.~\ref{subsec:LaFeAsO} and Sec.~\ref{subsec:FeSe}, respectively. The parameter $\lambda_{\mathrm{SOC}}$ entering the SOC Hamiltonian and parameterizing the SOC strength is a free parameter in our model. We have previously shown~\cite{scherer2018}, that the inclusion of onsite SOC can indeed reproduce the magnetic anisotropy of FeSCs as observed with e.g. inelastic neutron scattering. To study the effect of SOC on the spin-fluctuation mediated pairing, we allowed $\lambda_{\mathrm{SOC}}$ to vary in a rather large interval ranging from $ \lambda_{\mathrm{SOC}} = 0\,$meV to $ \lambda_{\mathrm{SOC}} = 100\,$meV.

Finally, the bare interaction Hamiltonian $H_{\mathrm{int}}$ of the $3d$ electrons is given by a local Hubbard-Hund interaction term, parameterized by the interaction parameters $ U $ (Hubbard-$U$) and $ J $ (Hund's coupling). Details on the form of the Hamiltonian can be found in Sec.~\ref{sec:vertex}.

\section{Linearized Gap Equation with SOC}
\label{sec:LGE}

The Fermi-surface projected linearized gap equation (LGE) provides the main
tool for our analysis of the impact of SOC on the leading Cooper instabilities within
the spin-fluctuation mediated pairing formalism. The LGE itself is nothing but the 
Bethe-Salpeter equation in the pairing channel, and for a given 2PI pairing vertex and
corresponding irreducible particle-particle bubble, the LGE allows for the detection of a superconducting instability and determination of the momentum, orbital and spin properties of the superconducting state for $ T \lesssim T_{\mathrm{c}} $. Details on our conventions for the Bogoliubov-de-Gennes Hamiltonian, the corresponding Nambu-Gorkov Greens function and the gap equation as well as our approximation to the 2PI pairing vertex can be found in Secs.~\ref{sec:BdG}-\ref{sec:vertex}. 

In our conventions, the Fermi surface projected LGE, stated as an eigenvalue problem with eigenvalue $ \lambda $ for the so-called pairing kernel appearing on its right-hand side, reads as
\be 
\label{eq:LGE_main}
\lambda \hat{\Delta}_{\varsigma}^{b}({\bf k}_{\mathrm{F}}) \! = \!
\sum_{b^{\prime},\varsigma^{\prime}} \int_{\mathrm{FS}^{\prime}}
\hat{\mathcal{M}}_{b,b^{\prime}}^{\varsigma,\varsigma^{\prime}}({\bf k}_{\mathrm{F}}, {\bf k}_{\mathrm{F}}^{\prime}) \,
\hat{\Delta}_{\varsigma^{\prime}}^{b^{\prime}}({\bf k}_{\mathrm{F}}^{\prime}),
\ee
with the pairing kernel
\be 
\label{eq:kernel}
\hat{\mathcal{M}}_{b,b^{\prime}}^{\varsigma,\varsigma^{\prime}}({\bf k}_{\mathrm{F}}, {\bf k}_{\mathrm{F}}^{\prime}) = - \frac{1}{V_{\mathrm{BZ}}} \frac{dk^{\prime}}{v_{\mathrm{F}}({\bf k}_{\mathrm{F}}^{\prime})}
[\hat{\Gamma}^{\varsigma,\varsigma^{\prime}}({\bf k}_{\mathrm{F}}, {\bf k}_{\mathrm{F}}^{\prime})]_{b^{\prime}b^{\prime}}^{b \,\, b},
\ee
where $ \hat{\Delta}_{\varsigma}^{b}({\bf k}_{\mathrm{F}}) $ denotes the intraband gap function ($ b $ is the band index corresponding to energy band $\epsilon_{b}({\bf k})$ obtained by bringing $ H_{0} + H_{\mathrm{SOC}} $ into a non-hybridizing representation, see Sec.~\ref{sec:BdG}) at the Fermi surface, specified by the Fermi momentum $ {\bf k}_{\mathrm{F}} $. The symbol $dk^{\prime}$ in \Eqref{eq:kernel} is to be understood as a length element of a Fermi surface segment. The label $ \varsigma = 0, x, y, z $ distinguishes (pseudo-)spin singlet ($0$) and triplet ($x,y,z$) gap solutions. Correspondingly, the Fermi surface projected pairing vertex in the static intraband approximation, denoted by $ [\hat{\Gamma}^{\varsigma,\varsigma^{\prime}}({\bf k}_{\mathrm{F}}, {\bf k}_{\mathrm{F}}^{\prime})]_{b^{\prime}b^{\prime}}^{b \,\, b} $, has a matrix structure in (pseudo-)spin singlet and triplet space. In \Eqref{eq:LGE_main} the symbol $ v_{\mathrm{F}}({\bf k}_{\mathrm{F}}) $ denotes the Fermi velocity at Fermi momentum $ {\bf k}_{\mathrm{F}} $. Solving the eigenvalue problem posed by \Eqref{eq:LGE_main}, we obtain a hierarchy of gap-solutions $ \hat{\Delta}_{\varsigma}^{b}({\bf k}_{\mathrm{F}}) $, ordered with respect to the eigenvalue $ \lambda  < 1$. The largest positive $\lambda$ corresponds to the `most attractive' solution. As $ \lambda $ grows (typically upon lowering temperature $ T $ or enhanced spin fluctuations) the system eventually displays a pairing instability, unless an instability in the particle-hole channel cuts off the fluctuations driving the pairing.

The inclusion of SOC leads to two qualitative modifications of the LGE compared to the situation with full SU(2) spin-rotation invariance, where a classification of superconducting instabilities in terms of spin singlet and spin triplet pairing is in general possible. In this work, we employ the random-phase approximation (RPA) to spin-fluctuation mediated pairing. Thus, the 2PI vertex $ \hat{\Gamma}^{\varsigma,\varsigma^{\prime}}({\bf k}_{\mathrm{F}}, {\bf k}_{\mathrm{F}}^{\prime})$ in the particle-particle channel is obtained by summing up diagrams composed of chains of irreducible particle-hole bubbles connected by the bare particle-hole vertex as defined by $ H_{\mathrm{int}} $ (see Sec.~\ref{sec:vertex}). While $ H_{\mathrm{int}} $ itself is spin-conserving and does not generate any anisotropy on its own, SOC enters the electronic Greens function and thereby affects the spin and orbital structure of irreducible bubbles in both particle-hole and particle-particle channels. The qualitative modifications of the LGE can now be attributed to each one of the two irreducible bubbles, where the particle-particle bubble is already contained in the definition of the pairing kernel, cf. Sec.~\ref{sec:gap}.

Due to SOC, the particle-hole bubble contains spin-non-conserving terms, which eventually lead to a differentiation of spin-fluctuations with different spin-polarization. In other words, SOC induces magnetic anisotropy by generating a polarization dependent gap hierarchy (and more generally polarization dependent dispersion) for the paramagnon modes of the metallic normal state system, which in our approximation are contained in the RPA propagator for magnetic fluctuations, and which are responsible for generating an attractive pairing interaction between the electrons. In the case without SOC, all paramagnon modes in different polarization channels have identical energy dispersion and thereby contribute equally to the mediation of the pairing interaction. 

We further note, that while SOC in principle allows for a mixing of charge and spin fluctuations, the degree of mixing we observe at the level of the RPA approximation is indeed small. Thus, the collective particle-hole fluctuations can effectively still be well separated into charge and spin modes. At the RPA level, it is the spin fluctuations that provide the attractive pairing interaction.

The second major qualitative modification of the gap equation enters through the irreducible particle-particle bubble. In \Eqref{eq:LGE_main}, the particle-particle bubble was effectively absorbed by carrying out the Fermi surface projection and formulating the LGE in band space (see Sec.~\ref{sec:gap}). As mentioned above, the absence of SOC allows for the classification of Cooper pairs in terms of their spin structure into spin singlet and spin triplet. With finite SOC, however, due to the locking of angular momentum (which is not free to rotate due to crystal field and hopping terms) and spin, spin-rotation invariance is in general broken and the spin quantum number can no longer be used to label the Fermi surface states participating in the pairing. The spin degree of freedom is replaced by pseudospin, which, carefully constructed, inherits the transformation properties of spin-1/2 under symmetry operations (see Sec.~\ref{sec:pseudospin} for details). The construction further ensures that as $ \lambda_{\mathrm{SOC}} \to 0 $, pseudospin continuously evolves into physical spin. For finite SOC, it is thus possible to classify the pairing solutions according to pseudospin singlet and triplet solutions. In contrast to the fully SU(2) symmetric case, where the vertex $ \hat{\Gamma}^{\varsigma,\varsigma^{\prime}}({\bf k}_{\mathrm{F}}, {\bf k}_{\mathrm{F}}^{\prime}) $ is diagonal in singlet-triplet space and reduces to a singlet and a single triplet component (the components for $ \varsigma = x,y,z $ are identical due to spin-rotation invariance), the pairing vertex will in general have a more involved structure in pseudospin singlet-triplet space.

\section{Influence of SOC on the Leading Cooper Instabilities}
\label{sec:cooper}

In the following, we will present our results for the impact of SOC on 
the leading Cooper instabilities for a generic FeSCs band and FeSe in Sec.~\ref{subsec:LaFeAsO}
and Sec.~\ref{subsec:FeSe}, respectively. Our primary focus of interest
is to uncover the influence of both the SOC-generated magnetic anisotropy
and SOC-driven changes in electronic bandstructure as well as the orbital and spin structure of Fermi surface states on the gap solutions. Therefore, we did not attempt to construct a phase diagram based on an instability analysis by \Eqref{eq:LGE_main}, but rather we identified the main trends in the changes caused by SOC of the solutions of \Eqref{eq:LGE_main}. For a given bandstructure and fixed
chemical potential and temperature $ T = 10 \, $meV, we construct the band-space
pairing kernel $ \hat{\mathcal{M}}_{b,b^{\prime}}^{\varsigma,\varsigma^{\prime}}({\bf k}_{\mathrm{F}}, {\bf k}_{\mathrm{F}}^{\prime}) $ in the RPA approximation and analyze
the changes occurring in the hierarchy of the solutions, and the changes in the momentum-space
structure of the gap function $ \hat{\Delta}_{\varsigma}^{b}({\bf k}_{\mathrm{F}}) $ upon variation of the SOC strength $ \lambda_{\mathrm{SOC}} $ and the interaction strength $ U $, while keeping fixed the ratio $ J = U/4 $.

\subsection{Generic FeSCs band structure (LaFeAsO)}
\label{subsec:LaFeAsO}

The hopping parameters which we apply to the `generic FeSCs band' are derived for LaFeAsO and are specified in Ref.~\onlinecite{ikeda2010}.
We will start the discussion of the results for the undoped system in 
Sec.~\ref{subsubsec:undoped}. The cases of hole- and electron-doped systems will be discussed in Secs.~\ref{subsubsec:hdoped} and \ref{subsubsec:edoped}, respectively.

\subsubsection{Undoped System}
\label{subsubsec:undoped}

\begin{figure}[t!]
\centering
\begin{minipage}{1\columnwidth}
\centering
\includegraphics[width=1\columnwidth]{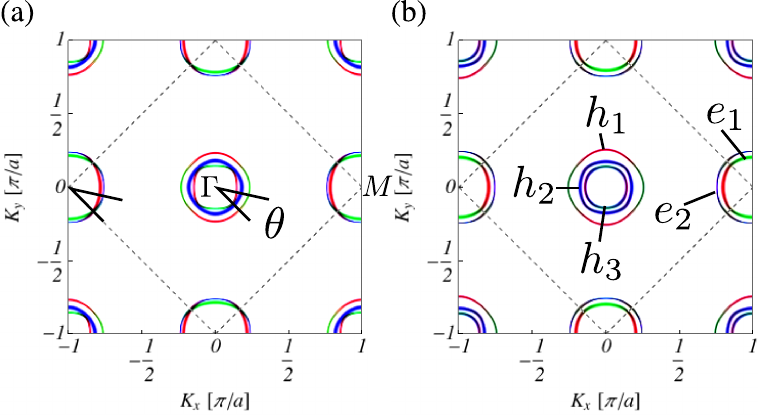}
\end{minipage}
\caption{Normal state Fermi surfaces of the generic FeSC band in the 1-Fe BZ ($(K_{x},K_{y})$ denotes momenta in the 1-Fe BZ coordinate system) extracted from the orbitally resolved contributions to the electronic spectral function with $ \mu_{0} = 0 $\,eV for (a) $\lambda = 0\,$meV and (b) $\lambda = 75\,$meV. The 2-Fe BZ is indicated by the dashed square. The colors refer to $d_{xz}$ (red), $d_{yz}$ (green) and $d_{xy}$ (blue) orbital contributions. SOC leads to a splitting of the states at the 2-Fe BZ boundary. As shown in (b), the inner and outer electron pockets (as seen from the 2-Fe M point) are labeled $e_{1}$ and $e_{2}$, while $h_{1}$, $h_{2}$ and $h_{3}$ refer to the outer, middle and inner hole-pocket (as seen from the $\Gamma$ point). The definition of the pocket angle $\theta$ (with respect to the axes of the 2-Fe coordinate system) is illustrated in panel (a).}
\label{fig:FS_undoped}
\end{figure}

The Fermi surface of the undoped system without and with SOC, as well as the orbital composition of the corresponding Fermi surface states is shown in Fig.~\ref{fig:FS_undoped}(a) and \ref{fig:FS_undoped}(b). In both cases, the Fermi surfaces feature three hole pockets around the $\Gamma$ point and two electron pockets around the $M$ point (in 2-Fe notation). We label the hole and electron pockets as shown in Fig.~\ref{fig:FS_undoped}.

The inner ($h_{3}$) and outer ($h_{1}$) hole pockets are mostly composed of $d_{xz}$ and $d_{yz}$ orbitals, while the middle ($h_{2}$) hole pocket is dominated by the $d_{xy}$ orbital.
The inner ($  e_{1} $) and outer ($ e_{2} $) electron pockets (where `inner' and `outer' are defined with respect to the $M$ point) overlap on the 2-Fe BZ boundary. The outer pockets is $d_{xy}$ dominated, while the inner pocket has dominant $d_{xz}$ and $d_{yz}$ orbital character. 

The Fermi surface of the undoped system features approximate nesting of hole and electron pockets with nesting vectors $ {\bf Q}_{1}=(\pi,0) $ and $ {\bf Q}_{2}=(0,\pi) $. Within a weak coupling approach, it is this nesting property that eventually leads to strong spin fluctuations at these wavevectors. For sufficiently large interactions, the strong spin fluctuations give rise to a spin-density wave (SDW) instability and the associated condensation of SDW order. Due to the repulsive nature of spin-fluctuation mediated interactions at the wavevectors $ {\bf Q}_{1} $ and $ {\bf Q}_{2} $,  these spin fluctuations favor the formation of the $s_{+-}$ pairing state, where the gap features a $\pi$-phase between hole and electron pockets~\cite{hirschfeld2011,chubukov2012}. 

While for pnictides, the superconducting instability typically emerges upon either hole or electron doping, we find that in a moderate doping regime, where the Fermi surface topology does not change and the approximate nesting features persist, SOC influences the LGE solutions obtained at different chemical potentials in basically the same way. Therefore, in the following, we will present results for the undoped case, and comment on changes for lightly doped systems in Secs.~\ref{subsubsec:hdoped} and~\ref{subsubsec:edoped}.

\bigskip

\textit{Vanishing SOC}, $\lambda_{\mathrm{SOC}} = 0$. We start by first focussing on the case of vanishing SOC, i.e., $ \lambda = 0\,$eV, in order to later emphasize the changes induced by finite SOC. To this end, we will present a selection of our numerical results highlighting the impact of the interaction parameter $ U $ on the leading gap solutions. To some extent, we will also discuss changes in subleading solutions as well. We further note, that while we do not observe a leading triplet solution for the undoped and lightly doped cases, we will nevertheless include the triplet components of the effective interaction in our presentation, as their inclusion completes the picture of SOC-induced changes in the framework of spin-fluctuation mediated pairing.

\begin{figure}[t!]
\centering
\begin{minipage}{0.89\columnwidth}
\centering
\includegraphics[width=1\columnwidth]{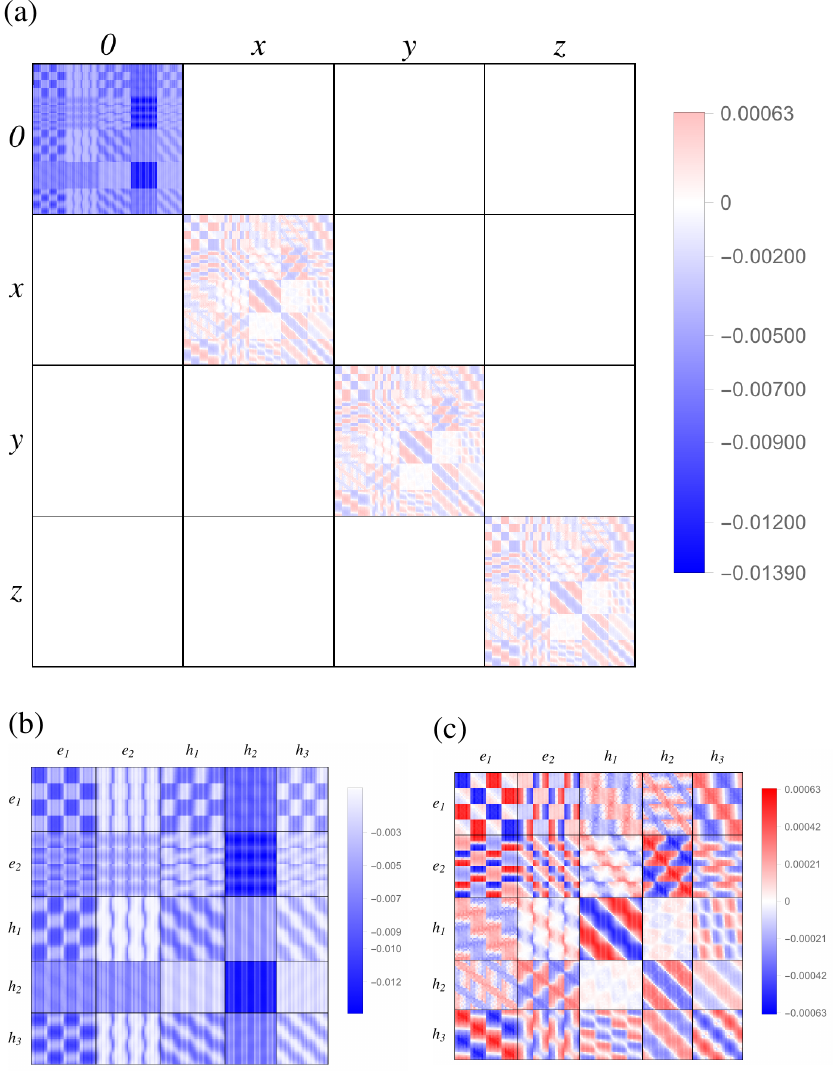}
\end{minipage}
\caption{Visualization of the pairing kernel entering the LGE. The spin-fluctuation mediated pairing interaction was determined for $\mu = 0\,$eV, $ T = 0.01\,$eV, $U = 0.50\,$eV and $ J = U/4 $ for vanishing SOC. (a) Full pairing kernel in singlet-triplet space, where $\varsigma = 0$ refers to the singlet component and $\varsigma = x,y,z$ to the triplet components. The colors indicate the strength and sign of the pairing kernel. (b),(c) Singlet and triplet components of the pairing kernel. In the case of vanishing SOC, the triplet components are all identical.}
\label{fig:kernel_undoped_U_0p50}
\end{figure}

In Fig.~\ref{fig:kernel_undoped_U_0p50} we show a visualization of the (non-symmetric) pairing kernel $ \mathcal{M}_{b,b^{\prime}}^{\varsigma,\varsigma^{\prime}}({\bf k}_{\mathrm{F}}, {\bf k}_{\mathrm{F}}^{\prime}) $ obtained for $\mu = 0\,$eV, $ T = 0.01\, $eV and interaction parameter $ U = 0.50 \,$eV. The generalized Stoner factor for these parameters, indicating the proximity to an SDW instability, has a value of 0.57 (the instability occurs for a Stoner factor of 1). The color-code in Fig.~\ref{fig:kernel_undoped_U_0p50} indicates the magnitude and sign of the real-valued pairing kernel. Within each of the blocks in singlet-triplet space, which are separated by vertical and horizontal black lines for clarity, each `pixel' corresponds to the value of the pairing kernel for the pair (${\bf k}_{\mathrm{F}}$,${\bf k}_{\mathrm{F}}^{\prime}$). We note that the Fermi momenta are ordered with respect to the Fermi pockets and increasing pocket angle. Zoom-ins for singlet and triplet components are displayed in Fig.~\ref{fig:kernel_undoped_U_0p50}(b) and Fig.~\ref{fig:kernel_undoped_U_0p50}(c). As mentioned above, for vanishing SOC, and thereby full spin-rotational invariance, the pairing kernel is fully block diagonal in singlet-triplet space and all triplet components are identical.

We note that in the case of vanishing SOC, we omit the hat in the notation of the pairing kernel and related quantities to indicate that the pseudospin degree of freedom coincides with physical spin, see Sec.~\ref{sec:gap} for details. At the chosen interaction parameters, the most prominent features of the static RPA spin susceptibility are broad humps around momenta $ {\bf Q}_{1} $ and $ {\bf Q}_{2} $. 

The pairing kernel features a rich momentum structure, corresponding to the possible pairing solutions with increasing degree of momentum space anisotropy. While the singlet component only has negative values, the triplet component exhibits sign changes as required by the antisymmetric nature of the effective pairing interaction in the intraband triplet channel. As the pairing kernel is diagonal in singlet-triplet space, the LGE for the singlet and triplet gap functions decouples and the eigenvalue problems can be solved independently. Solving \Eqref{eq:LGE_main} allows to disentangle the complicated momentum structure of the pairing kernel. The leading Cooper instability is an $s_{+-}$ solution, as expected from the momentum structure of the static spin susceptibility, while we observe subleading solutions with $d$-wave symmetry. 

\begin{figure}[t!]
\centering
\begin{minipage}{0.89\columnwidth}
\centering
\includegraphics[width=1\columnwidth]{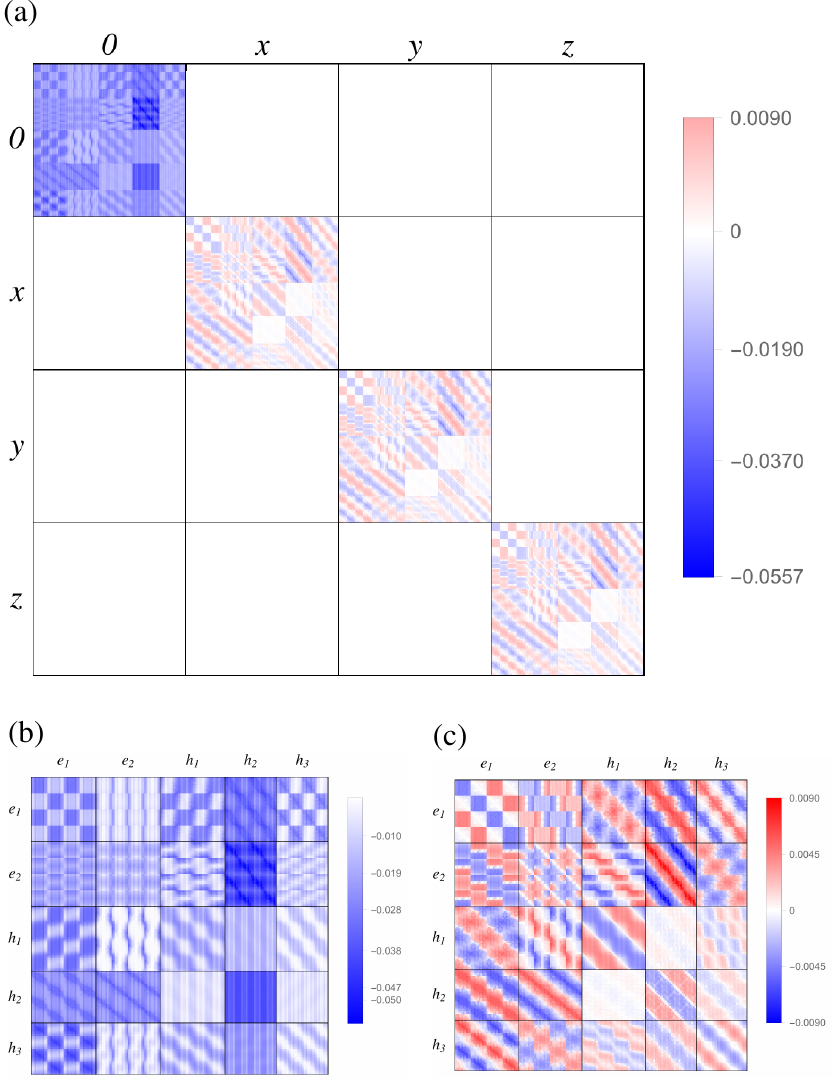}
\end{minipage}
\caption{Visualization of the pairing kernel entering the LGE. The spin-fluctuation mediated pairing interaction was determined for $\mu = 0\,$eV, $ T = 0.01\,$eV, $U = 0.80\,$eV and $ J = U/4 $ for vanishing SOC. (a) Full pairing kernel in singlet-triplet space. The colors indicate the strength and sign of the pairing kernel. (b),(c) Singlet and triplet components of the pairing kernel.}
\label{fig:kernel_undoped_U_0p80}
\end{figure}

To assess the effect of the interaction parameter, and thereby the proximity of the system to an SDW instability on the gap solutions, Fig.~\ref{fig:kernel_undoped_U_0p80} shows the vertex plot for $ U = 0.80 \, $eV, leading to a Stoner factor of 0.91. 

By inspection of Figs.~\ref{fig:kernel_undoped_U_0p50} and~\ref{fig:kernel_undoped_U_0p80}, we observe changes in the momentum structure of the pairing kernel in the singlet and triplet channels. As an increased interaction parameter $ U $ leads to a more pronounced peak structure of the static susceptibility, the changes in the pairing kernel can be attributed to the gain of weight of the peaks at $ {\bf Q}_{1} $ and $ {\bf Q}_{2} $ relative to the `background' susceptibility at other momenta.

\begin{figure}[t!]
\centering
\begin{minipage}{1.00\columnwidth}
\centering
\includegraphics[width=1\columnwidth]{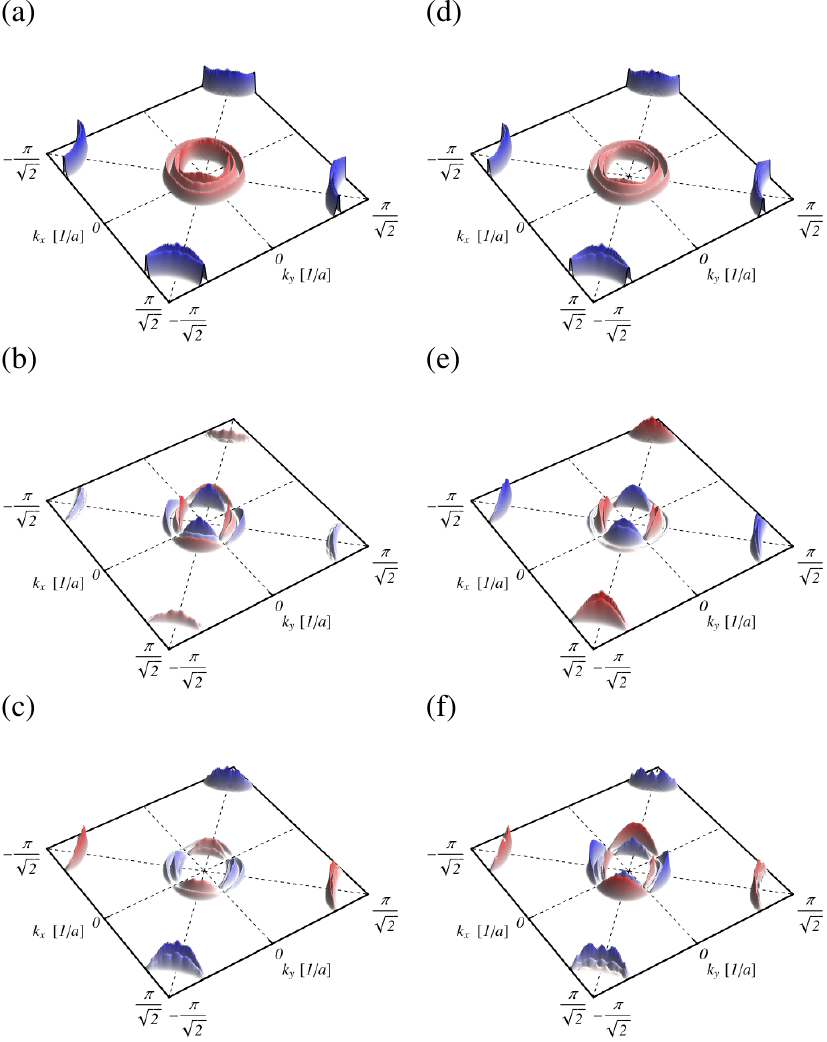}
\end{minipage}
\caption{Leading and first two subleading LGE solutions for (a)-(c) $ U = 0.50\,$eV and (d)-(f) $ U = 0.80\,$eV for vanishing SOC. The LGE solutions are ordered according to their $\lambda$ values, where $\lambda$ decreases from top to bottom in each column. All solutions shown have spin singlet character.}
\label{fig:gap_undoped}
\end{figure}

The leading $s_{+-}$ and subleading $d$-wave solutions obtained for interaction parameters $ U = 0.50 \,$eV and $ U = 0.80\,$eV are shown in Fig.~\ref{fig:gap_undoped}, where we applied the procedure described in Ref.~\onlinecite{maier2009} to obtain gap solutions $ \Delta_{\varsigma}^{b}({\bf k}) $ from the Fermi surface projected LGE that are defined on the entire BZ. The leading solutions for different interaction parameters are displayed in Figs.~\ref{fig:gap_undoped}(a) and (d), respectively. The subleading solutions have $d$-wave character, and display sign changes due to repulsive interpocket interactions. As seen from Fig.~\ref{fig:gap_undoped}, the subleading $d$-wave solutions differ e.g. by in-phase versus out-of-phase gap signs on the hole pockets. As mentioned above, triplet solutions have rather small LGE eigenvalue $ \lambda $ for both small and large interactions and only appear after a series of singlet solutions.

\begin{figure}
\centering
\begin{minipage}{1\columnwidth}
\centering
\includegraphics[width=1\columnwidth]{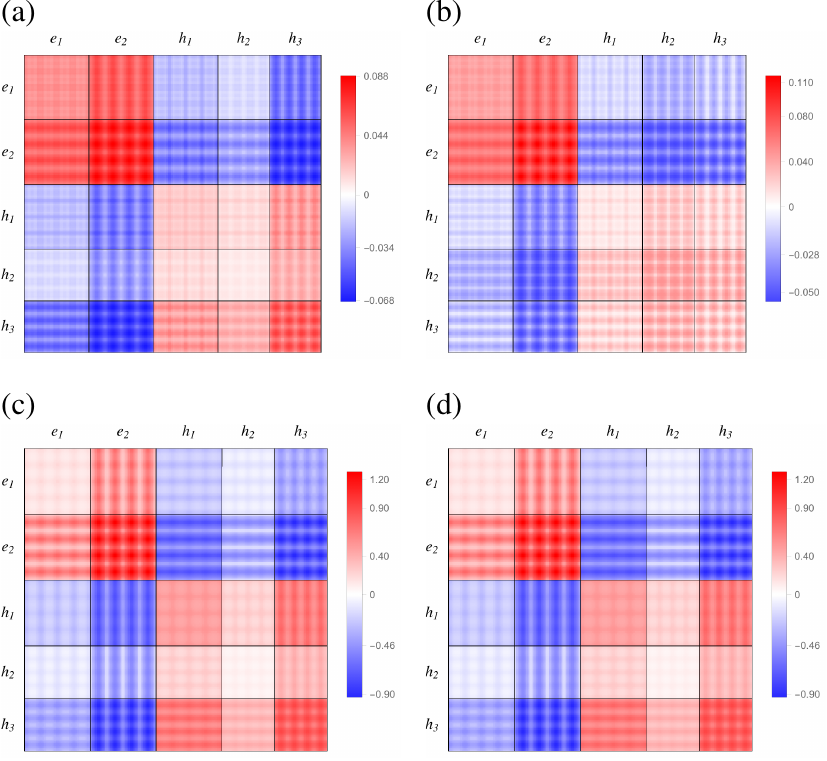}
\end{minipage}
\caption{Reduced kernel corresponding to the leading LGE eigenvalue for (a) $ U = 0.50 \, $eV and (b) $ U = 0.80 \, $eV. (c),(d) Are identical and show $ f^{b}({\bf k}_{\mathrm{F}})  f^{b^{\prime}}({\bf k}_{\mathrm{F}}^{\prime})  v_{\mathrm{F}}({\bf k}_{\mathrm{F}}) v_{\mathrm{F}}({\bf k}_{\mathrm{F}}^{\prime}) $ to facilitate easy comparison with the reduced kernels. The function $f({\bf k}_{\mathrm{F}}) $ implements a sign change between electron and hole pockets. Within our conventions for the kernel, a negative value corresponds to repulsion, while a positive value corresponds to attraction.}
\label{fig:reduced_kernel_undoped}
\end{figure}

To proceed with the analysis of the LGE solutions obtained by diagonalizing the pairing kernel, we note that we can write
\be 
\mathcal{M}_{b,b^{\prime}}^{\varsigma,\varsigma^{\prime}}({\bf k}_{\mathrm{F}}, {\bf k}_{\mathrm{F}}^{\prime}) =
\lambda \,
g_{\mathrm{R},\varsigma}^{b}({\bf k}_{\mathrm{F}}) \,
g_{\mathrm{L},\varsigma^{\prime}}^{b^{\prime}}({\bf k}_{\mathrm{F}}^{\prime}) + \dots,
\ee
where $ \lambda $ denotes the largest attractive LGE eigenvalue (here corresponding to solutions with $s_{+-}$ symmetry), and the functions $ g_{\mathrm{L},\varsigma}^{b}({\bf k}_{\mathrm{F}}) $ and $ g_{\mathrm{R},\varsigma}^{b}({\bf k}_{\mathrm{F}}) $ denote the corresponding left (L) and right (R) eigenfunctions of the pairing kernel. The dots stand for the corresponding terms of subleading LGE eigenvalues $ \lambda $. We observe that a symmetric, reduced kernel can be obtained as
\be 
m_{b,b^{\prime}}^{\varsigma,\varsigma^{\prime}}({\bf k}_{\mathrm{F}}, {\bf k}_{\mathrm{F}}^{\prime}) \equiv
g_{\mathrm{R},\varsigma}^{b}({\bf k}_{\mathrm{F}}) \,
g_{\mathrm{L},\varsigma^{\prime}}^{b^{\prime}}({\bf k}_{\mathrm{F}}^{\prime})
\left(\frac{dk^{\prime}}{v_{\mathrm{F}}({\bf k}_{\mathrm{F}}^{\prime})}\right)^{-1},
\ee
where we factored out the eigenvalue $ \lambda $, as we are mostly concerned with the momentum structure rather than the overall magnitude of the dominant pairing interaction.
The symmetry property and the factorized form of $ m_{b,b^{\prime}}^{\varsigma,\varsigma^{\prime}}({\bf k}_{\mathrm{F}}, {\bf k}_{\mathrm{F}}^{\prime}) $ imply that the reduced kernel can be reconstructed from a function $ \phi_{\varsigma}^{b}({\bf k}_{\mathrm{F}}) $ as
\be 
m_{b,b^{\prime}}^{\varsigma,\varsigma^{\prime}}({\bf k}_{\mathrm{F}}, {\bf k}_{\mathrm{F}}^{\prime}) = \phi_{\varsigma}^{b}({\bf k}_{\mathrm{F}}) \, \phi_{\varsigma^{\prime}}^{b^{}\prime}({\bf k}_{\mathrm{F}}^{\prime}),
\ee
which coincides with the right eigenfunction of the full kernel. The reduced kernels for the interaction parameters $ U = 0.50\, $eV and $ U = 0.80 \, $eV are shown in Fig.~\ref{fig:reduced_kernel_undoped}(a),(b). The reduced kernel contains, of course, the same information as the leading LGE solution itself. Plotting the reduced kernel, however, allows to clearly identify the dominant interactions corresponding to the observed momentum space anisotropy of the gap function. As we have absorbed a global minus sign into the definition of the pairing kernel, a positive value corresponds to attraction, while a negative one to repulsion. As shown in Fig.~\ref{fig:reduced_kernel_undoped}(a), for small interaction parameter, the effective $e_{2} - h_{3}$ interpocket repulsion is strongest, while the intrapocket attraction is strongest on the $ e_{2} $ and $ h_{3} $ pockets. We note, that nesting with $ {\bf Q}_{1} $ and $ {\bf Q}_{2} $ would prefer interpocket interactions of $ e_{1}-h_{1}$ and $e_{2}-h_{1}$ type. The strong repulsion between $ e_{2} $ and $ h_{3} $ is partly explained by the fact that also momenta away from the nesting vectors contribute significantly to the effective pairing interaction.  

The effective interaction features both small and large scale variations, where the large scale variation corresponds to the pocket index. In trying to understand the small scale variation, we plot the product $  f^{b}({\bf k}_{\mathrm{F}})  f^{b^{\prime}}({\bf k}_{\mathrm{F}}^{\prime})  v_{\mathrm{F}}({\bf k}_{\mathrm{F}}) v_{\mathrm{F}}({\bf k}_{\mathrm{F}}^{\prime}) $ in Fig.~\ref{fig:reduced_kernel_undoped}(c) and observe a rather good agreement in both the large and small scale variations of the interaction, cf. Fig.~\ref{fig:reduced_kernel_undoped}(a). Here, the function $ f^{b}({\bf k}_{\mathrm{F}}) $ implements the sign change of the interaction between electron and hole pockets, as required by the interpocket repulsion. The kernel blocks involving the $ h_{2} $ pocket, however, are an exception, in that the maxima of the reduced kernel and the simplified expression are shifted with respect to each other. 

Nevertheless, despite the rather involved procedure of obtaining the 2PI pairing vertex from the multiorbital RPA, the small scale variation of the leading LGE solution with $ s_{+-} $ symmetry can be understood as coming from a simple Fermi surface quantity, namely the Fermi velocity $ v_{\mathrm{F}}({\bf k}_{\mathrm{F}}) $. Concretely, this implies that the maxima of $ | \Delta_{0}^{b}({\bf k}_{\mathrm{F}}) | $ coincide with the maxima of $ v_{\mathrm{F}}({\bf k}_{\mathrm{F}}) $, with exception of the $h_{2}$ pocket, where it is the minima of $ | \Delta_{0}^{b}({\bf k}_{\mathrm{F}}) | $ that coincide with the maxima of $ v_{\mathrm{F}}({\bf k}_{\mathrm{F}}) $, as can be seen from the inversion of the pattern in e.g. the $h_{2}$-$h_{2}$ block when comparing Fig.~\ref{fig:reduced_kernel_undoped}(a) and Fig.~\ref{fig:reduced_kernel_undoped}(c). We note that while the gap amplitude on $ h_{2} $ has a small variation, it is almost constant compared to the variation on e.g. the $ e_{2} $ and $ h_{3} $ pockets. Similarly, the $ e_{1} $ pocket features an almost constant gap. For the plot in Fig.~\ref{fig:reduced_kernel_undoped}(c), we have simply taken $ f^{b}({\bf k}_{\mathrm{F}}) = + 1$ for electron pockets and $ f^{b}({\bf k}_{\mathrm{F}}) = -1 $ for hole pockets, but neglected any further large scale variation in the interaction strength. In the true reduced kernel, we observe a rather strong, pocket dependent modulation of the gap amplitude.

The fact that the gap amplitude in Fig.~\ref{fig:gap_undoped}(a) is largest on the $e_{2}$ and $h_{3}$ pockets can thus be traced back to the particular form of the electronic dispersion around the Fermi level, together with the rather broad humps around momenta $ {\bf Q}_{1} $ and $ {\bf Q}_{2} $ in the static susceptibility, allowing a wider range of momenta to contribute more or less equally to the effective interaction. At this point we can conclude, that there seem to be two mechanisms governing the small scale variation of the leading $ s_{+-} $ gap function. The gaps of the singlet solution indeed approximately follow
\be 
\Delta_{0}^{b}({\bf k}_{\mathrm{f}}) \approx f^{b}({\bf k}_{\mathrm{f}}) v_{\mathrm{F}}({\bf k}_{\mathrm{F}}),
\ee
with an envelope function $ f^{b}({\bf k}_{\mathrm{f}}) $ (which is more complicated than our naive ansatz above) for most of the pockets, while we find an almost constant $ \Delta_{0}^{b}({\bf k}_{\mathrm{f}}) $ on $ e_{1} $ and $ h_{2} $. 

\begin{figure}[t!]
\centering
\begin{minipage}{0.89\columnwidth}
\centering
\includegraphics[width=1\columnwidth]{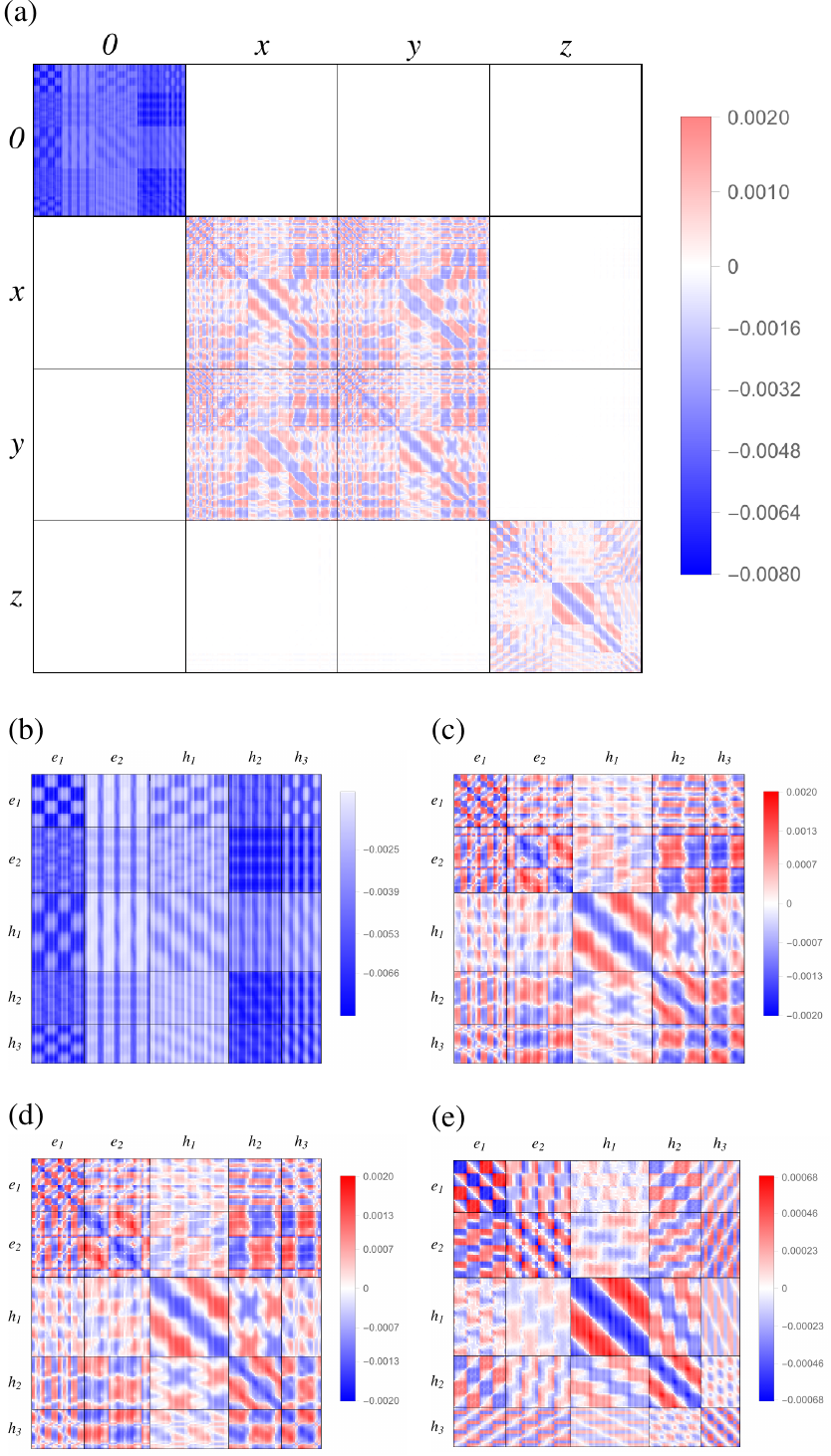}
\end{minipage}
\caption{Visualization of the pairing kernel entering the LGE. The spin-fluctuation mediated pairing interaction was determined for $\mu = 0\,$eV, $ T = 0.01\,$eV, $U = 0.50\,$eV and $ J = U/4 $ and $\lambda_{\mathrm{SOC}} = 75\,$meV. (a) Full pairing kernel in pseudospin singlet-triplet space. The colors indicate the strength and sign of the pairing kernel. (b) Pseudospin singlet and pseudspin triplet (c) $x$, (d) $y$ and (e) $z$ components of the pairing kernel. The blocks coupling pseudospin triplet $x$ and $y$ sectors are not shown separately.}
\label{fig:kernel_undoped_lambda_0p075_U_0p50}
\end{figure}
\begin{figure}[t!]
\centering
\begin{minipage}{0.88\columnwidth}
\centering
\includegraphics[width=1\columnwidth]{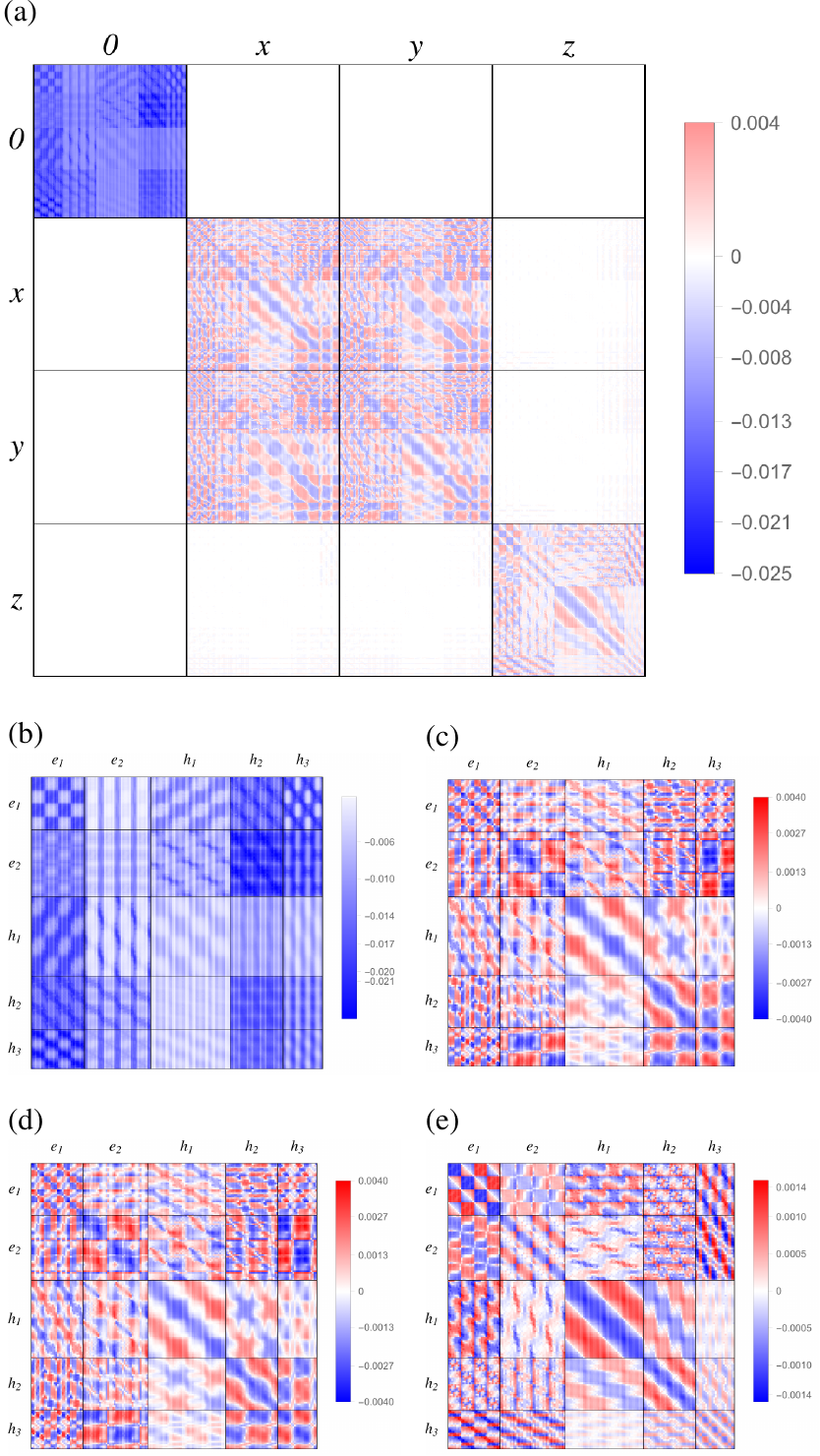}
\end{minipage}
\caption{Visualization of the pairing kernel entering the LGE. The spin-fluctuation mediated pairing interaction was determined for $\mu = 0\,$eV, $ T = 0.01\,$eV, $U = 0.80\,$eV and $ J = U/4 $ and $\lambda_{\mathrm{SOC}} = 75\,$meV. (a) Full pairing kernel in pseudospin singlet-triplet space. The colors indicate the strength and sign of the pairing kernel. (b) Pseudospin singlet and pseudspin triplet (c) $x$, (d) $y$ and (e) $z$ components of the pairing kernel. The blocks coupling pseudospin triplet $x$ and $y$ sectors are not shown separately.}
\label{fig:kernel_undoped_lambda_0p075_U_0p80}
\end{figure}

Performing the same analysis for the larger interaction parameter $ U = 0.80 \, $eV and comparing the corresponding reduced kernel, as shown in Fig.~\ref{fig:reduced_kernel_undoped}(b) with the pocket modulated product of Fermi velocities (for convenience reproduced in Fig.~\ref{fig:reduced_kernel_undoped}(d) to facilitate easy comparison), we observe that we no longer have qualitative agreement between the naive ansatz and the reduced kernel extracted from the leading LGE solution. Indeed, the $ e_{2} - h_{2} $ and $ h_{2}-h_{2} $ interactions have become dominant. We attribute this change in the structure of the interaction to the increased contribution to the static susceptibility from the nesting of the $ d_{xy} $-dominated pockets $e_{2}$ and $h_{2}$ with increased interaction strength. Correspondingly, the increased interaction strength favors a larger gap amplitude on the $ h_{2} $ pocket, while the gap on the electron pockets remains more or less unaffected. The increase in strength of spin fluctuations at wavevectors $ {\bf Q}_{1} $ and $ {\bf Q}_{2} $, which amounts to a decrease in the paramagnon gap for the three different paramagnon polarization branches (with polarizations denoted by $x, y$ and $z$), concomitantly leads to a larger LGE eingevalue $ \lambda $, indicating an enhanced tendency towards a Cooper instability. 

\bigskip

\textit{Finite SOC}, $\lambda_{\mathrm{SOC}} \neq 0$. Having analyzed the case of vanishing SOC to establish a baseline for comparison, we now turn to finite SOC. While we have produced and analyzed numerical data for a range of values for $ \lambda_{\mathrm{SOC}} $ between $25\,$ and $100\,$meV, we will focus on presenting results obtained for $ \lambda_{\mathrm{SOC}} = 75 \,$meV. For the considered range of SOC values, the results turned out to be qualitatively similar, with only small SOC-induced quantitative differences. The Fermi surface of the electronic system with $ \lambda_{\mathrm{SOC}} = 75 \,$meV is shown in Fig.~\ref{fig:FS_undoped}(b). The main effect of SOC is to split the electronic states on the boundary of the 2-Fe BZ. Further, SOC tends to enhance the orbital mixing of eigenstates of the quadratic part of the electronic Hamiltonian. By comparing the Fermi surfaces for vanishing and finite SOC, see Fig.~\ref{fig:FS_undoped}(a) and Fig.~\ref{fig:FS_undoped}(b), it is obvious, that for fixed chemical potential, SOC leads to a deformation of both electron and hole pockets. 

As reported previously by us~\cite{scherer2018}, the changes in the electronic bandstructure together with the SOC-induced alteration of the electronic single-particle states lead to a polarization-dependent differentiation of the energy gaps of the three different paramagnon branches (see Sec.~\ref{sec:orbital} for an example). For the current value of the chemical potential, $\mu_{0} = 0\,$eV, it is the excitations with $x$ polarization which have the smallest gap, and thereby the corresponding component of the static susceptibility has the largest peaks at $ {\bf Q}_{1} $. The gap for $y$-polarized excitations is largest, with the gap for out-of-plane excitations falling in between the two other values. At $ {\bf Q}_{2} $, the hierarchy of excitation gaps and susceptibility peaks of magnetic in-plane excitations is swapped, as dictated by $C_{4}$ symmetry, i.e., rotation of orbital, spin and lattice by an angle of $\pi/2$. We note, that due to finite SOC, symmetry transformations acting on orbital and lattice entail a simultaneous transformation of the spin degree of freedom.

In Figs.~\ref{fig:kernel_undoped_lambda_0p075_U_0p50} and~\ref{fig:kernel_undoped_lambda_0p075_U_0p80} we show the resulting pairing kernels in pseudospin singlet-triplet space for interaction parameters $ U = 0.50\, $eV and $ U = 0.80 \, $eV, respectively. Besides small-scale changes in the momentum-space details of the effective (pseudo-)spin singlet and triplet interactions, the most prominent change is the coupling of pseudospin triplet-$x$ and -$y$ sectors. Without SOC, the triplet solutions feature a six-fold degeneracy (two symmetry-related momentum-space form factors and three independent directions for the $d$-vector). In the presence of SOC, we find this degeneracy to be lifted. The pseudospin triplet-$z$ component and the $x$, $y$ components are no longer identical, see Figs.~\ref{fig:kernel_undoped_lambda_0p075_U_0p50}(c)-(e) and~\ref{fig:kernel_undoped_lambda_0p075_U_0p80}(c)-(e). The additional $x$-$y$ coupling lifts the degeneracy between helical solutions with in-plane $d$-vector. While triplet solutions in the triplet-$z$ channel display a residual degeneracy, possible chiral solutions in the pseudospin triplet-$z$ channel seem to be disfavoured by SOC: While increasing SOC leads to an enhanced LGE eigenvalue of helical solutions, there is no sign of a possible chiral solution among the first eighteen LGE solutions. The helical solutions, however, never become leading, even as the system approaches the SDW instability (which typically leads to an enhanced anisotropy in spin-fluctuations, as only a single paramagnon branch with a specific polarization channel becomes gapless and eventually leads to condensation of magnetic order)~\cite{scherer2018}. 

We note, that while it is possible to transform the Fermi surface projected pairing kernel back to spin singlet-triplet space, this would ultimately blow up the representation of the problem, as due to the presence of SOC, working with physical spin necessitates working in sublattice $ \otimes $ orbital space as well. In such a representation, additional spin singlet-triplet couplings would appear. In this sense, the adaption of the LGE to pseudospin space corresponds to the most economical representation of the Fermi surface projected pairing problem, especially as pseudospin singlet and triplet sectors remain uncoupled, and the LGE problems can, in principle, still be solved independently.

\begin{figure}[t!]
\centering
\begin{minipage}{1.00\columnwidth}
\centering
\includegraphics[width=1\columnwidth]{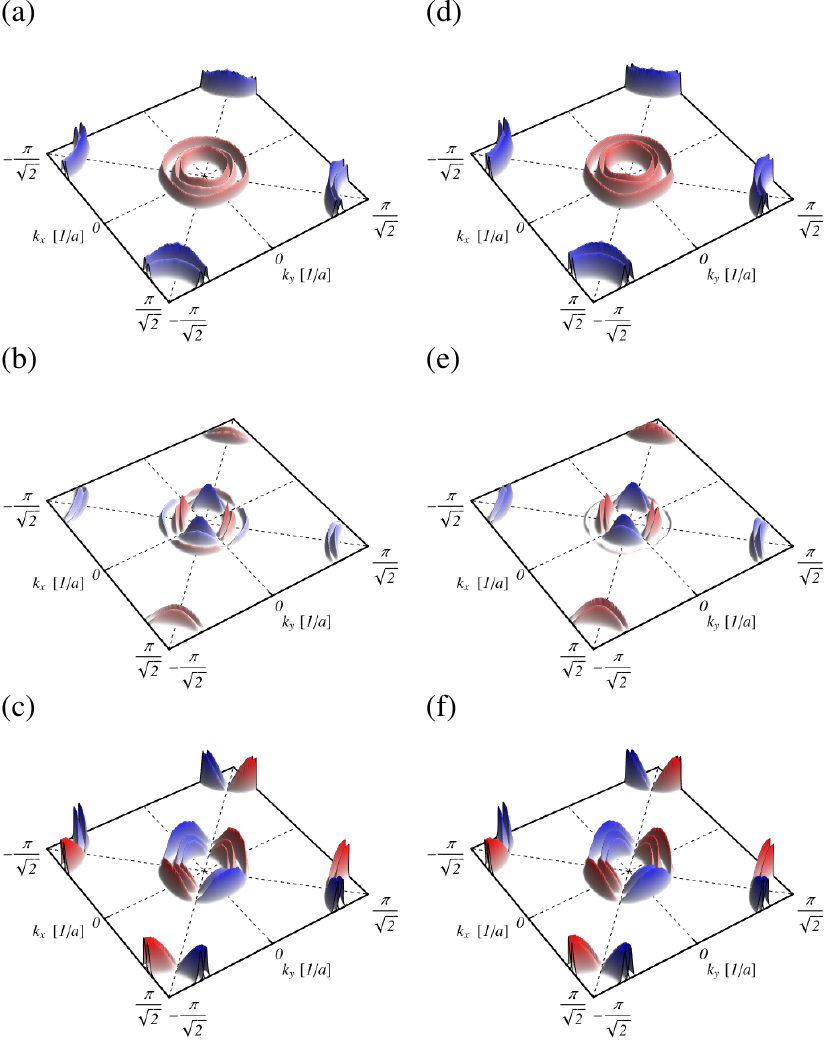}
\end{minipage}
\caption{Leading and first two subleading LGE solutions for (a)-(c) $ U = 0.50\,$eV and (d)-(f) $ U = 0.80\,$eV for $ \lambda_{\mathrm{SOC}} = 75 \, $meV. The LGE solutions are ordered according to their $\lambda$ values, where $\lambda$ decreases from top to bottom in each column. All solutions shown have spin singlet character.}
\label{fig:gap_undoped_SOC}
\end{figure}

The first few associated leading gap solutions (all pseudospin singlet) are shown in Fig.~\ref{fig:gap_undoped_SOC}. For both interaction values, the leading solution is still of $ s_{+-} $ type, now in the pseudospin singlet channel, see Fig.~\ref{fig:gap_undoped_SOC}(a) and (d). In fact, sweeping a rather large window of SOC and interaction parameters, the $s_{+-}$ pseudospin singlet solution always turned out to be the leading solution for the current level of the chemical potential. Increasing the interaction has the same tendency as in the case of vanishing SOC, namely to enhance the gap on the $ d_{xy} $ dominated hole pocket. Overall, however, this effect is less pronounced in the presence of SOC. In fact, two mechanism are at play. We observe that the LGE eigenvalue $ \lambda $ for the $ s_{+-} $ solution is slightly suppressed due to finite SOC for fixed interaction strength. This suppression can be understood by the polarization dependent differentiation of spin fluctuations that mediate the pairing interaction. While the spin fluctuations with $x$ polarization are strongest at $ {\bf Q}_{1} $ (and in fact slightly enhanced compared to the case without SOC), the fluctuations in the other two polarization channels are suppressed, and thereby contribute less to the effective pairing interaction. 

\begin{figure}[t!]
\centering
\begin{minipage}{1.00\columnwidth}
\centering
\includegraphics[width=1\columnwidth]{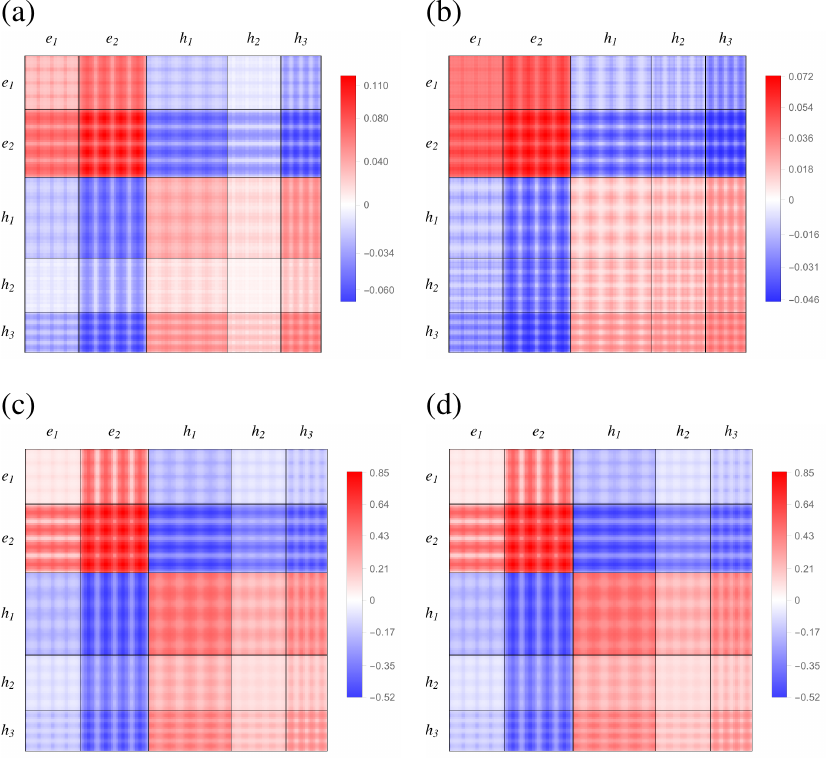}
\end{minipage}
\caption{Reduced kernel corresponding to the leading LGE eigenvalue for (a) $ U = 0.50 \, $eV and (b) $ U = 0.80 \, $eV with finite SOC, $\lambda_{\mathrm{SOC}} = 75\,$meV. (c),(d) Are identical and show $ f^{b}({\bf k}_{\mathrm{F}})  f^{b^{\prime}}({\bf k}_{\mathrm{F}}^{\prime})  v_{\mathrm{F}}({\bf k}_{\mathrm{F}}) v_{\mathrm{F}}({\bf k}_{\mathrm{F}}^{\prime}) $ to facilitate easy comparison with the reduced kernels. The function $f({\bf k}_{\mathrm{F}}) $ implements a sign change between electron and hole pockets. Within our definitions, a negative value corresponds to repulsion, while a positive value corresponds to attraction.}
\label{fig:reduced_kernel_undoped_SOC}
\end{figure}

Further, the enhanced mixing of orbital character of the electronic Fermi surface states due to SOC tends to smear out the strong, interaction enhanced $d_{xy}$ contribution to the pairing interaction across the pockets, and thereby tends to generate a more uniform gap amplitude. As before, the maxima of the gap amplitude on most of the pockets follow the angular variation of the Fermi velocity, see Fig.~\ref{fig:reduced_kernel_undoped_SOC} where we again plot the reduced pairing kernel corresponding to the leading kernel eigenfunction. In line with the changes discussed above, we observe a rather uniform interaction structure, albeit with dominant intra-$e_{2}$, intra-$h_{3}$ and inter $e_{2}$-$h_{3}$ components.
The subleading solutions display $d$-wave character, where SOC seemingly stabilizes the shown form factors (see. Fig.~\ref{fig:gap_undoped_SOC}(b),(c) and (e),(f)) against changes due to increased interaction.

Having obtained the LGE solution, we can in principle construct the corresponding spin singlet and triplet pairing fields in sublattice $ \otimes $ orbital space (see Sec.~\ref{sec:BdG} for the transformation rule). In Sec.~\ref{sec:orbital}, we show a subset of the resulting $ 10 \times 10 $ form factors in spin singlet and triplet channels without and with SOC, as generated from transforming the $s_{+-}$ form factor of the leading pseudospin singlet solution. Analyzing the orbital structure of the spin singlet pairing fields, we find the intraorbital $d_{xz}$, $d_{yz}$ and $ d_{xy} $ elements to be dominant. Given the orbital character of the Fermi surface states (see Fig.~\ref{fig:FS_undoped}), this observation is not overly surprising. Interestingly, the intersublattice components feature a different phase structure than the intrasublattice ones. The physical spin $d$-vector of the spin triplet component is basically purely in-plane. The triplet-$x$ pairing fields are dominated by intra-$d_{xy}$ and inter $d_{xz}$-$d_{xy}$ components, while the triplet-$y$ pairing fields have dominant interorbital components of $d_{yz}$-$d_{xy}$ type. The intra- and intersublattice components again feature subtle differences in the detailed phase structure of the pairing fields. The sizable triplet form factors are rather sparse, but for $ \lambda_{\mathrm{SOC}} = \, 75$meV, the largest triplet pairing amplitudes can be as large as 60 \% of the maximum of the singlet amplitude. We note, that a smearing procedure as described in Ref.~\onlinecite{maier2009} and employed in Ref.~\onlinecite{scherer2019} tends to decrease the triplet component upon increasing the energy cutoff.

To gain a better understanding of the influence of spin-fluctuation anisotropy on the pairing solution, we analyzed the LGE in the presence of SOC with a modified, spin-isotropic pairing kernel. The irreducible particle-hole bubble entering the RPA resummation in the construction of the pairing kernel can be split into spin isotropic and anisotropic contributions, see Ref.~\onlinecite{scherer2018}. The resulting RPA correlation function inherits the spin isotropy from the irreducible bubble and is then dressed by orbital-to-band matrix elements (obtained with finite SOC). Interestingly, the resulting pairing kernel, despite the absence of any spin anisotropy in the spin fluctuations, still features the same overall structure as shown previously in Figs.~\ref{fig:kernel_undoped_lambda_0p075_U_0p50} and~\ref{fig:kernel_undoped_lambda_0p075_U_0p80}. The coupling of pseudospin triplet-$x$ and triplet-$y$ components is thus an effect due to the orbital-to-band matrix elements, which also encode the relation between physical spin and the pseudospin degree of freedom. In the isotropic approximation, we observe the same tendency towards a more uniform gap amplitude as in the calculation with the full pairing kernel.

\subsubsection{Hole-Doped System}
\label{subsubsec:hdoped}

In the following, we will present our numerical results for the case of a
hole-doped system. The case of a lightly doped system will be touched upon only
briefly, as the influence of SOC and interaction parameters on the leading and subleading LGE solutions has the same tendencies as presented in Sec.~\ref{subsubsec:undoped} for the undoped system. The strongly hole-doped case, however, features drastic changes in the Fermi surface topology, and, as we will discuss in more detail below, exhibits strong tendencies toward pseudospin triplet solutions.

\begin{figure}[t!]
\centering
\begin{minipage}{1.00\columnwidth}
\centering
\includegraphics[width=1\columnwidth]{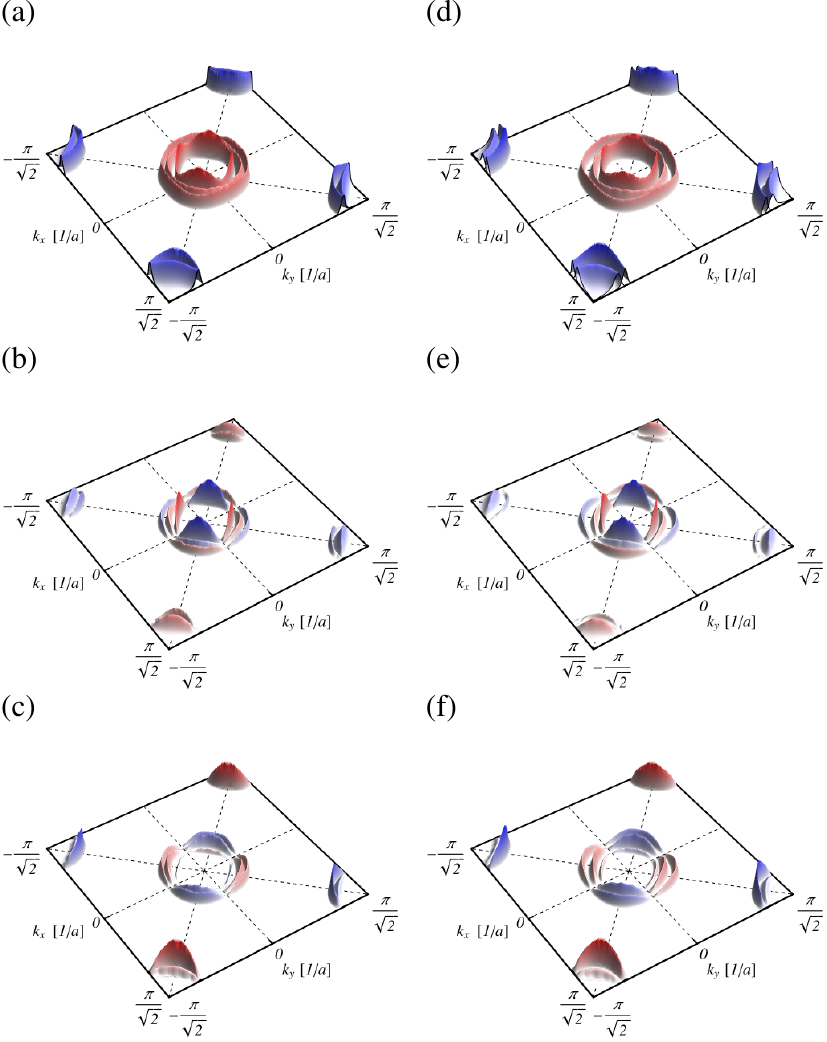}
\end{minipage}
\caption{Leading and first two subleading LGE solutions of the lightly hole-doped system for $ U = 0.70\,$eV with (a)-(c) $ \lambda_{\mathrm{SOC}} = 0 \, $meV and (d)-(f) $ \lambda_{\mathrm{SOC}} = 75 \, $meV, respectively. The LGE solutions are ordered according to their $\lambda$ values, where $\lambda$ decreases from top to bottom in each column. All solutions shown have pseudospin singlet character.}
\label{fig:gap_lightly_hdoped}
\end{figure}

The results for the leading LGE solutions (without and with SOC) for the lightly hole-doped system with a chemical potential $ \mu_{0} = - 45\,$meV and interaction parameter $ U = 0.70 \, $eV are shown in Fig.~\ref{fig:gap_lightly_hdoped}. We chose a smaller interaction parameter as without SOC, since the value of $0.80\,$eV cannot be reached without encountering an SDW instability. The Stoner factor for $ U = 0.70 \, $eV and $ \lambda_{\mathrm{SOC}} = 0 \, $meV is 0.88 and therefore slightly smaller than for the undoped system with $ U = 0.80 \, $eV. As the small shift in chemical potential induces only small changes compared to the undoped case, we do not show the Fermi surface for the lightly doped system. For the leading gap solution, the major difference to the undoped system is an enhanced momentum space anisotropy on the hole pockets. Finite SOC tends to decrease the anisotropy on the hole pockets. We further note, that while hole doping leads to a change in the magnetic anisotropy, where the $z$-axis polarized magnetic fluctuations become dominant, the dominant polarization is in fact rather irrelevant to details of the gap solution. 

The orbital and spin singlet/triplet character of the $s_{+-}$ solution features the same characteristics as in the undoped case. Overall, it seems that the changes in the gap solutions due to a shift of the chemical potential by $ \mu_{0} = - 45\,$meV are rather robust with respect to inclusion of SOC. 

\begin{figure}[t!]
\centering
\begin{minipage}{1\columnwidth}
\centering
\includegraphics[width=1\columnwidth]{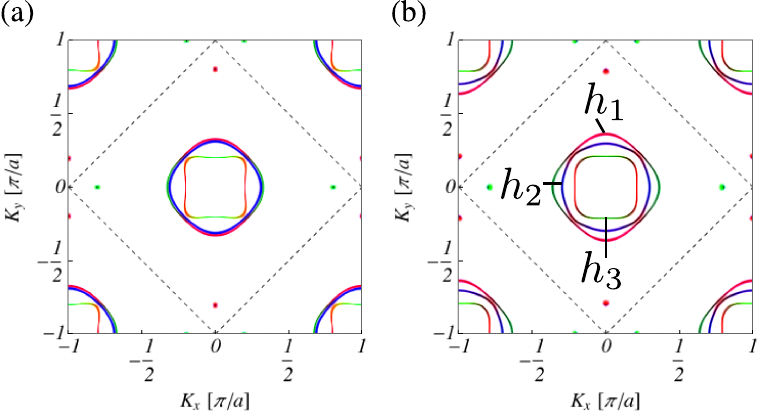}
\end{minipage}
\caption{Normal state Fermi surfaces of the heavily hole-doped generic FeSC band in the 1-Fe BZ ($(K_{x},K_{y})$ denotes momenta in the 1-Fe BZ coordinate system) extracted from the orbitally resolved contributions to the electronic spectral function with $ \mu_{0} = -150 $\,eV for (a) $\lambda = 0\,$meV and (b) $\lambda = 75\,$meV. The 2-Fe BZ is indicated by the dashed square. The colors refer to $d_{xz}$ (red), $d_{yz}$ (green) and $d_{xy}$ (blue) orbital contributions. SOC leads to a splitting of the states at the 2-Fe BZ boundary. As shown in (b), the labels $h_{1}$, $h_{2}$ and $h_{3}$ refer to the outer, middle and inner hole-pocket (as seen from the $\Gamma$ point).}
\label{fig:FS_heavily_hdoped}
\end{figure}

More drastic changes, however, can be observed when moving to the heavily hole-doped case, where we take $ \mu_{0} =  -150 \,$meV, see Fig.~\ref{fig:FS_heavily_hdoped} for the corresponding Fermi surfaces. While the orbitally resolved spectral function at the chemical potential $\mu_{0}$, as displayed in Fig.~\ref{fig:FS_heavily_hdoped}, shows a finite weight close to the corners of the 2-Fe BZ, we neglected these tiny pockets in the treatment of the pairing problem, and focussed only the three hole pockets around the $\Gamma$ point. The outer ($h_{1}$) and middle ($h_{2}$) hole pockets fall almost on top of each other, as can be seen in Fig.~\ref{fig:FS_heavily_hdoped}(a). With finite SOC, the corresponding states repel one another. This repulsion brings about a much more rounded off middle pocket. The orbital composition of the electronic states is qualitatively the same as in the undoped and lightly hole-doped cases, with strong orbital mixing for finite SOC.

\begin{figure}[t!]
\centering
\begin{minipage}{0.88\columnwidth}
\centering
\includegraphics[width=1\columnwidth]{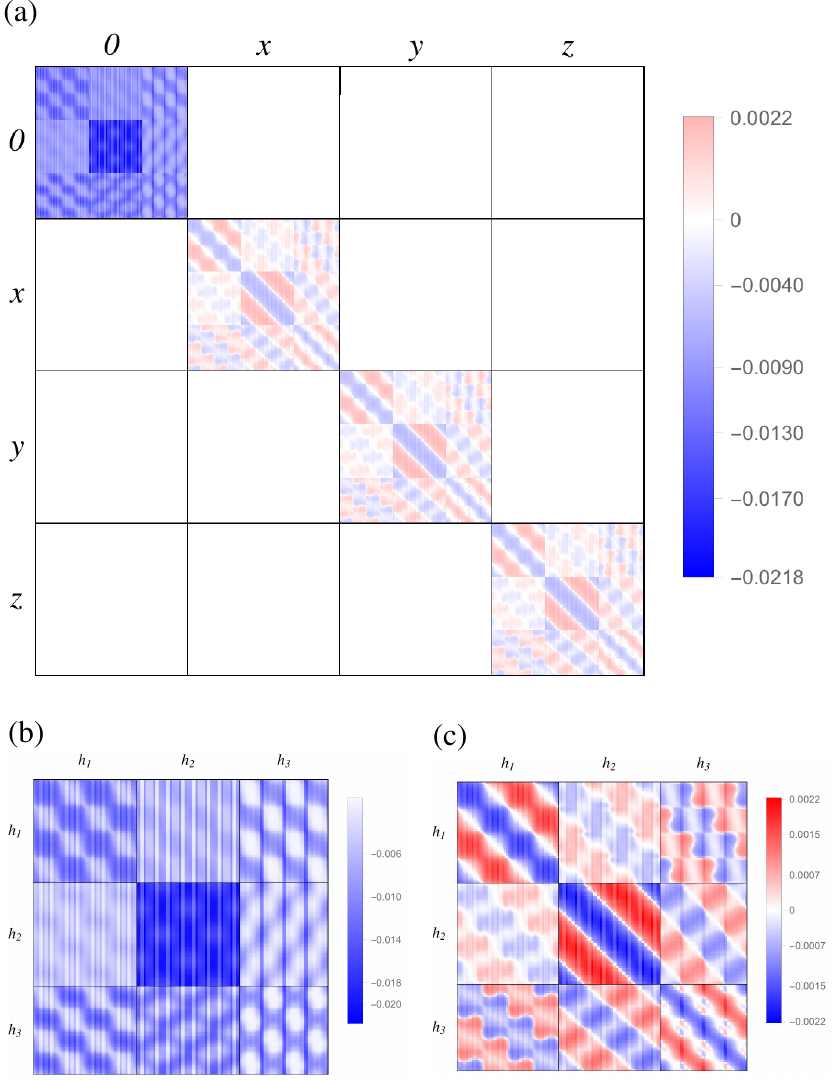}
\end{minipage}
\caption{Visualization of the pairing kernel entering the LGE. The spin-fluctuation mediated pairing interaction was determined for $\mu = -150\,$meV, $ T = 0.01\,$eV, $U = 0.80\,$eV and $ J = U/4 $ and $\lambda_{\mathrm{SOC}} = 0\,$meV. (a) Full pairing kernel in spin singlet-triplet space. The colors indicate the strength and sign of the pairing kernel. (b) Spin singlet and (c) spin triplet components of the pairing kernel.}
\label{fig:kernel_heavily_hdoped_lambda_0p000_U_0p80}
\end{figure}

\begin{figure}[t!]
\centering
\begin{minipage}{0.88\columnwidth}
\centering
\includegraphics[width=1\columnwidth]{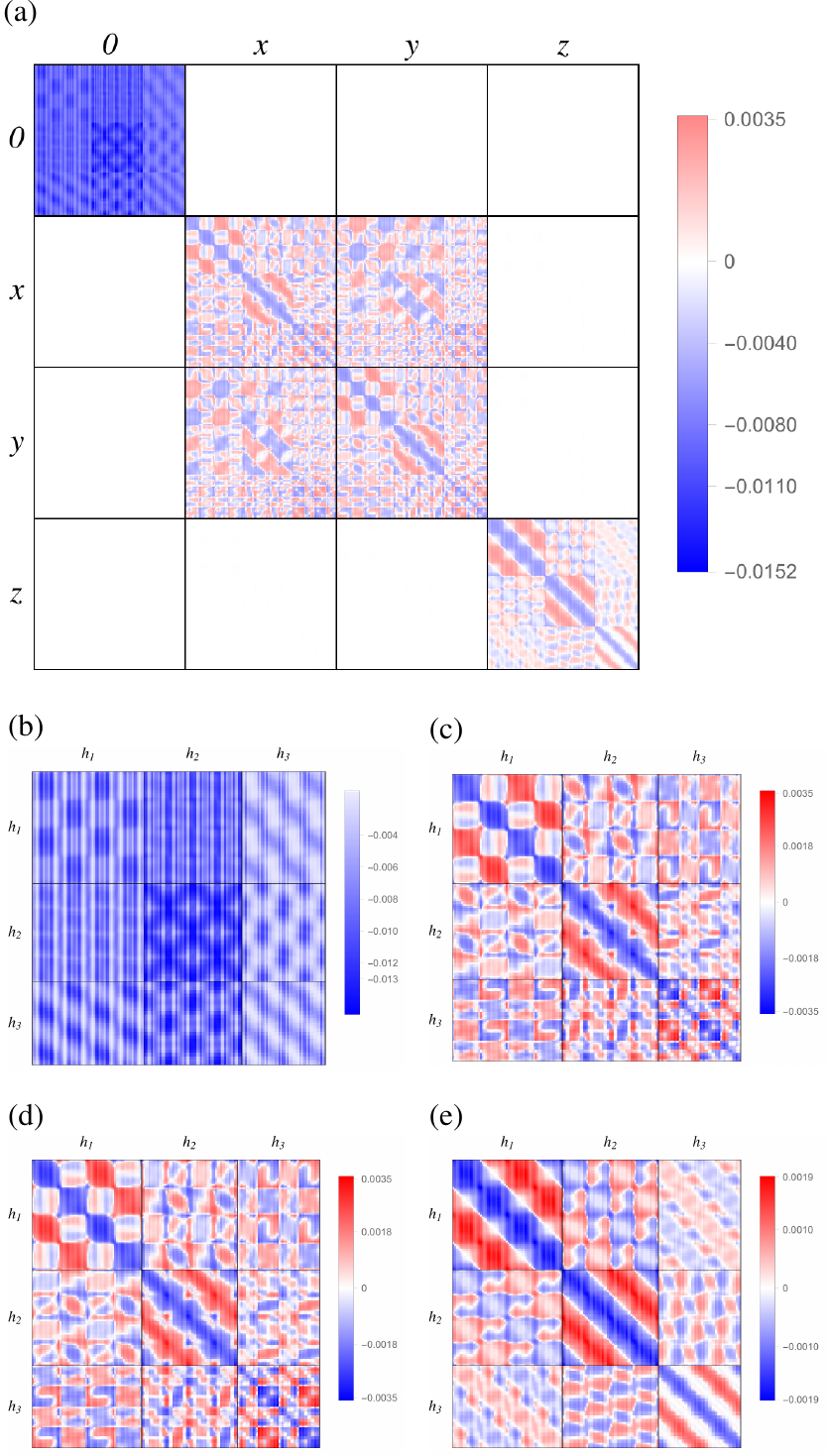}
\end{minipage}
\caption{Visualization of the pairing kernel entering the LGE. The spin-fluctuation mediated pairing interaction was determined for $\mu = -150\,$meV, $ T = 0.01\,$eV, $U = 0.80\,$eV and $ J = U/4 $ and $\lambda_{\mathrm{SOC}} = 75\,$meV. (a) Full pairing kernel in pseudospin singlet-triplet space. The colors indicate the strength and sign of the pairing kernel. (b) Pseudospin singlet and pseudospin triplet (c) $x$, (d) $y$ and (e) $z$ components of the pairing kernel. The blocks coupling pseudospin triplet $x$ and $y$ sectors are not shown separately.}
\label{fig:kernel_heavily_hdoped_lambda_0p075_U_0p80}
\end{figure}

Considering the static spin susceptibility at momenta contributing to the Fermi surface projected LGE, the dominant, contributing spin fluctuations can be identified to be approximately located at the wavevectors $ (2/3 \pi, 0) $ and $ (0, 2/3 \pi) $. We note, that the contributing momenta are concentrated in a small region in momentum space, roughly corresponding to a circle with radius $ 2/3 \pi $. Small momenta contribute relatively strongly to the pairing kernel in the parameter regime under investigation. The resulting pairing kernels without and with SOC are shown in Figs.~\ref{fig:kernel_heavily_hdoped_lambda_0p000_U_0p80} and Figs.~\ref{fig:kernel_heavily_hdoped_lambda_0p075_U_0p80}, respectively. At finite SOC, the $z$-axis polarized magnetic fluctuations dominate, as in the lightly hole-doped system. The resulting magnetic anisotropy, however, is rather small for the chosen interaction parameters.


\begin{figure}[t!]
\centering
\begin{minipage}{1.00\columnwidth}
\centering
\includegraphics[width=1\columnwidth]{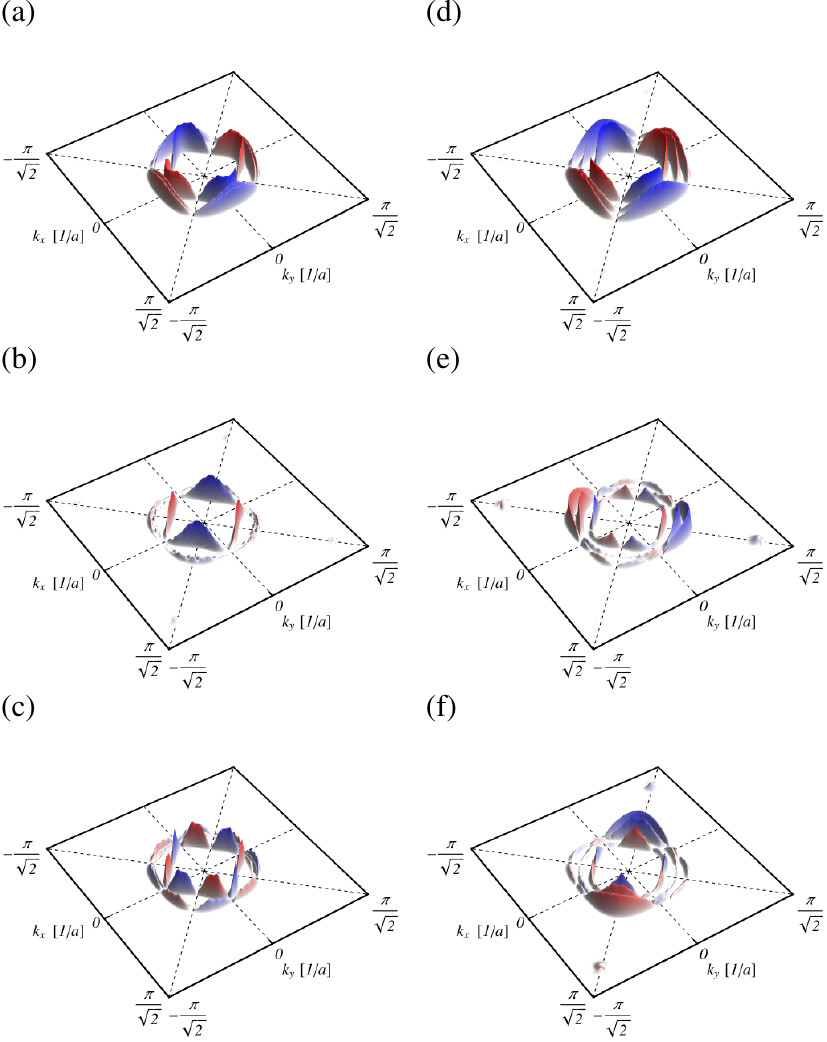}
\end{minipage}
\caption{(a)-(c) Leading and first two subleading LGE solutions with pseudospin singlet character of the heavily hole-doped system for $ U = 0.80\,$eV with $ \lambda_{\mathrm{SOC}} = 0 \, $meV. (d) Leading $d$-wave solution and (e),(f) pseudospin triplet-$x$ and $y$ components of the subleading helical triplet LGE solution for $ \lambda_{\mathrm{SOC}} = 75 \, $meV. The form factors of the two triplet components are related by a $C_{4}$ transformation. The interpolation scheme generates small contributions of the gap on the neglected electron pockets in the corners of the 2-Fe BZ.}
\label{fig:gap_heavily_hdoped}
\end{figure}

Both pairing kernels feature strong intra $h_{2}$ interactions in the (pseudo-)spin singlet channel, as well as strong intra $h_{2}$ and intra $h_{3}$ interactions in the (pseudo-)spin triplet channel. In the case with finite SOC, the usual pseudospin triplet-$x$ triplet-$y$ coupling appears. In both cases, the leading LGE solutions are of $d$-wave type, as shown in Fig.~\ref{fig:gap_heavily_hdoped}(a) and (d). The subleading solutions, however, are much more sensitive to finite SOC as in the undoped and lightly doped cases. While for $ \lambda_{\mathrm{SOC}} = 0\, $meV the first few subleading solutions are all of spin singlet type, finite SOC $ \lambda_{\mathrm{SOC}} \geq 50\, $meV and strong local repulsion $ U \geq 0.80 \,$eV put a strongly anisotropic, helical triplet solution in the place with 2nd largest LGE eigenvalue, see Fig.~\ref{fig:gap_heavily_hdoped}(e),(f). While our analysis is quantitatively not reliable with respect to the LGE eigenvalues $\lambda$, we note that the subleading triplet solution starts out with a $\lambda$ which is smaller than the leading LGE eigenvalue by a factor of two. Sweeping a large window of SOC strength and interaction parameter values, we observe that the pseudospin triplet solution can become the leading instability, but in a rather narrow parameter window close to an SDW instability for $ \lambda_{\mathrm{SOC}} = 75 \, $meV. Upon further increase of the interaction parameter, the pseudospin triplet solution is rendered subleading again. Judging from the observed dependence of the leading solution on the interaction parameter, it appears as if the triplet solution becomes leading by accident.
It is possible, however, that the pseudospin triplet solution can be stabilized by further decrease of temperature or finite-energy and inter-band contributions to the pairing problem, which are neglected in our Fermi surface projected version of the LGE. We finally note, that due to SOC, the degeneracies of solutions with triplet character are lifted (with a residual twofold degeneracy in the pseudo-spin triplet-$z$ channel). On the level of the static spin susceptibility, we trace the enhanced tendency toward an odd parity solution back to the increased weight of small momentum spin fluctuations.

Returning to the (pseudo-)spin singlet solution of $d$-wave type, we note that the maxima of the gap amplitudes on both the $h_{1}$ and the $h_{2}$ pocket follow the maxima of the Fermi velocity as a function of the pocket angle, while on $ h_{3} $ the pocket angles for amplitude maxima coincide with those of Fermi velocity minima. The $d$-wave solution and the described gap amplitude appear to be robust with respect to the inclusion of SOC and variations in the interaction parameter. Analyzing the orbital and spin structure of the $d$-wave solution (see Sec.~\ref{sec:orbital}), we again find large spin-triplet components. The spin singlet is dominated by intra-orbital $d_{xz}$, $d_{yz}$ and $d_{3z^2 - r^2}$ contributions, with intra-$d_{xy}$ contributions being subleading. The largest interorbital components can be found for $d_{xz}$/$d_{yz}$ - $d_{3z^2 - r^2}$ components. The orbital structure in the spin-triplet component is similar to the one in the singlet component.

Recently, another pairing scenario for systems with similar Fermi surface topology as for the heavily hole-doped system was suggested and analyzed~\cite{vafek}. It was argued, that under certain conditions, the effective pairing interaction can be understood to arise from renormalized interaction parameters. Upon inclusion of SOC, the authors of Ref.~\onlinecite{vafek} found from an analysis of the gap equation, that in a certain parameter window, a superconducting order parameter with a finite and dominant triplet component is stabilized. Interband contributions to the gap equation were argued to be relevant for the generation of a finite $T_{\mathrm{c}}$ (at the level of the involved approximations). While our analysis at the level of the Fermi surface projected LGE can, according to the arguments presented in Ref.~\onlinecite{vafek}, not produce a finite $ T_{\mathrm{c}} $, we found it worthwhile to generate a pairing kernel from renormalized interactions, as described in Ref.~\onlinecite{vafek}, and analyze the corresponding pseudospin triplet solution in our multiorbital formalism. The result can then be compared to the prediction for the order parameter obtained from the spin-fluctuation mechanism at the RPA level, see Fig.~\ref{fig:gap_heavily_hdoped}. 

By adjusting the value of $ U^{\prime} $, while keeping $U$, $J$ and $J^{\prime}$ fixed, we can set up the LGE with the pairing kernel arising from a renormalized Hubbard-Hund type interaction, where the spin-fluctuation contribution is dropped completely, see Sec.~\ref{sec:vertex}. We parameterize the interorbital repulsion as $ U^{\prime} = J - \delta $, where an attractive interaction is possible for $\delta > 0$. Solving the LGE, we indeed find a leading triplet solution with small but positive $\lambda$, signalling a potential Cooper instability. We note, that even at $ \lambda_{\mathrm{SOC}} = 0 \,$meV, the LGE solutions supported by the pairing vertex described above have spin triplet character. Increasing $ \delta $ for otherwise fixed parameters increases the LGE eigenvalues, but has otherwise no effect on the gap solutions.

\begin{figure}[t!]
\centering
\begin{minipage}{1.00\columnwidth}
\centering
\includegraphics[width=1\columnwidth]{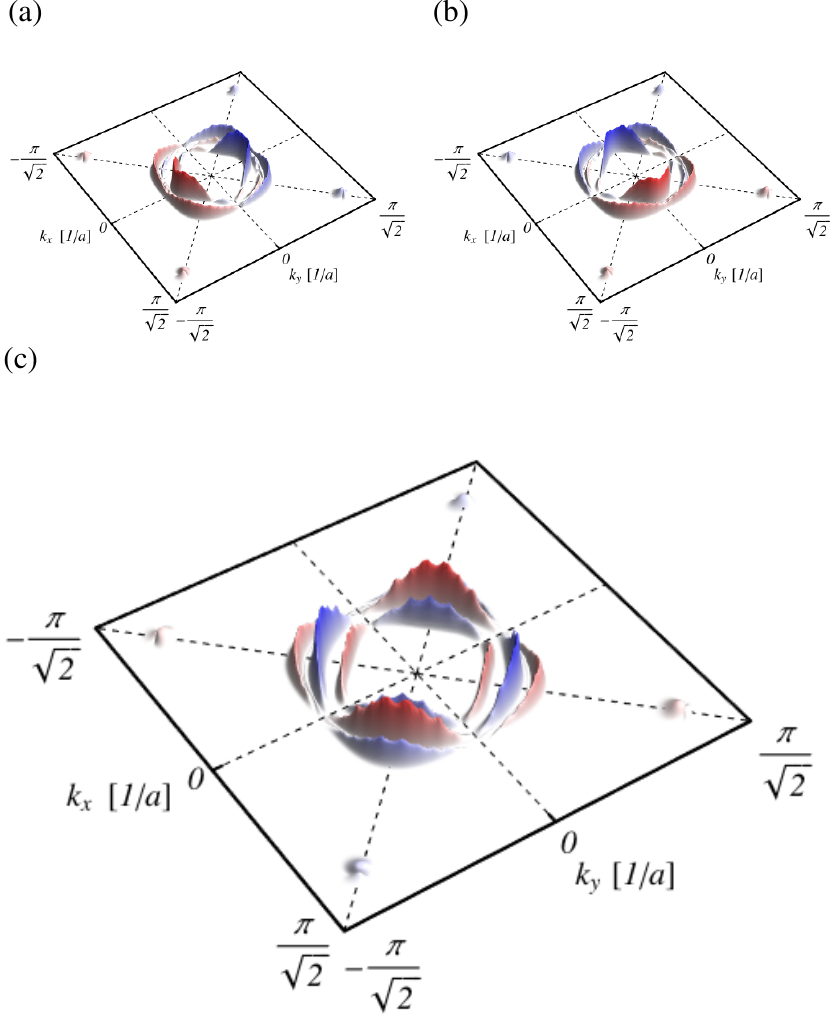}
\end{minipage}
\caption{(a),(b) Leading, degenerate gap solutions in pseudospin triplet-$z$ channel obtained for Vafek-Chubukov-type interactions with $ \delta = 15\, $meV, $ T = 0.01\,$eV, $\mu_{0} = -150\,$meV, $ U = 0.50\,$eV, $ J = U/4$ for $ \lambda_{\mathrm{SOC}} = 50 \,$meV. (c) Leading pseudospin singlet solution for $ \lambda_{\mathrm{SOC}} = 75 \,$meV and otherwise same parameters as in (a),(b). The interpolation scheme generates small contributions of the gap on the neglected electron pockets in the corners of the 2-Fe BZ.}
\label{fig:gap_heavily_hdoped_triplet}
\end{figure}

Switching on SOC, we again observe the splitting of triplet solutions and removal of degeneracies for solutions with in-plane $d$-vector, while the triplet-$z$ channel still features a residual twofold degeneracy. Indeed, the leading pseudospin triplet solution has a $d$-vector aligned with the $z$-axis in pseudospin space. The momentum space structure of the degenerate triplet-$z$ solutions is shown in Fig.~\ref{fig:gap_heavily_hdoped_triplet}(a),(b). Interestingly, increasing both SOC and $ \delta $ eventually favor a pseudospin singlet solution of $d$-wave type with an interpocket phase structure corresponding to interpocket repulsion, see Fig.~\ref{fig:gap_heavily_hdoped_triplet}(c), while the triplet solutions are rendered subleading.

To conclude, the spin-fluctuation mediated pairing scenario and the renormalized interaction parameter scenario produce rather different LGE solutions in both singlet and triplet channels.

\subsubsection{Electron-Doped System}
\label{subsubsec:edoped}

Having presented results for undoped and hole-doped systems in Sec.~\ref{subsubsec:undoped} and Sec.~\ref{subsubsec:hdoped}, respectively, we now move on to our results obtained for the electron doped case. Here, we skip the discussion of the lightly electron-doped case, as neither the magnetic anisotropy nor the Fermi surface
display notable changes compared to the undoped case. Considering a chemical potential
of $ \mu_{0} = 20 \,$meV, however, leads to a Fermi surface with (almost overlapping) small hole and large
electron pockets, where the $d_{xy}$ dominated hole pocket centered around
$\Gamma$ has been pushed below the chemical potential.

\begin{figure}[t!]
\centering
\begin{minipage}{1\columnwidth}
\centering
\includegraphics[width=1\columnwidth]{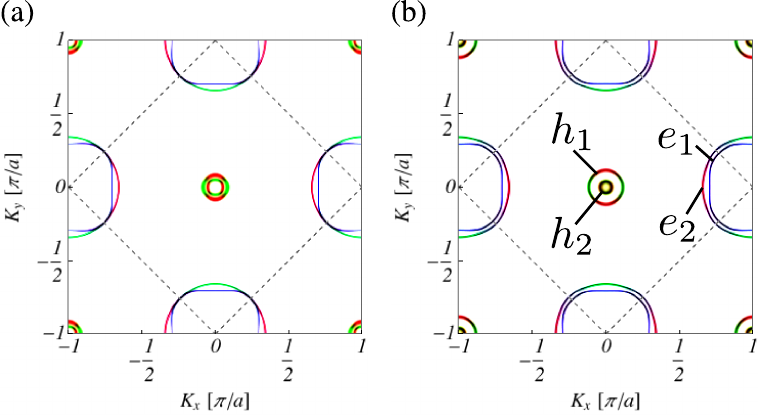}
\end{minipage}
\caption{Normal state Fermi surfaces of the electron doped generic FeSC band in the 1-Fe BZ ($(K_{x},K_{y})$ denotes momenta in the 1-Fe BZ coordinate system) extracted from the orbitally resolved contributions to the electronic spectral function with $ \mu_{0} = 20 $\,eV for (a) $\lambda = 0\,$meV and (b) $\lambda = 75\,$meV. The 2-Fe BZ is indicated by the dashed square. The colors refer to $d_{xz}$ (red), $d_{yz}$ (green) and $d_{xy}$ (blue) orbital contributions. SOC leads to a splitting of the states at the 2-Fe BZ boundary. The labels $h_{1}$, $h_{2}$ and $h_{3}$ refer to the outer, middle and inner hole-pocket (as seen from the $\Gamma$ point).}
\label{fig:FS_heavily_edoped}
\end{figure}

The Fermi surfaces without and with SOC are displayed in Fig.~\ref{fig:FS_heavily_edoped}. As can be seen in  
Fig.~\ref{fig:FS_heavily_edoped}(a), for vanishing SOC, the hole pockets (outer: $h_{1}$, inner: $h_{2}$) are located very close to each other in momentum space. The electron pockets centered around the corners of the 2-Fe BZ feature a squarish $d_{xy}$ dominated inner pocket ($e_{1}$), and almost circular $d_{xz}$/$d_{yz}$ dominated outer pockets ($e_{2}$). Finite SOC splits the bands giving rise to the hole pockets and thereby generates a larger momentum space separation, cf. Fig~\ref{fig:FS_heavily_edoped}(b). The usual splitting of the electronic states on the 2-Fe BZ boundary leads to a mixed orbital character close to the $ \lambda_{\mathrm{SOC}} = 0 \,$eV degeneracy points.

Due to the changes in the shape of the Fermi surface, the static spin susceptibility for momenta relevant for the Fermi surface projected LGE now approximately peaks at wavevector $ (\pi/2, \sqrt{3}\pi /2) $ and the three other wavevectors related by $C_{4}$ rotations, see Sec.~\ref{sec:orbital}. For the interaction parameter $ U = 0.80 \, $eV, the system is still not particularly close to a magnetic instability (at least for an SDW instability with one of the wavevectors entering the Fermi surface projected LGE), as indicated by a Stoner factor of about 0.69. Correspondingly, the static susceptibility at other momenta can still be considered to be comparatively large. The momentum structure of the static susceptibility does not change drastically upon switching on SOC, but the peak position is slightly shifted. The ensuing magnetic anisotropy favors magnetic fluctuations with in-plane polarization with equal components in $x$ and $y$ polarization in a star-shaped momentum space region, while the remaining regions of momentum space have dominant $z$-axis polarization, see Sec.~\ref{sec:orbital}.

\begin{figure}[t!]
\centering
\begin{minipage}{0.88\columnwidth}
\centering
\includegraphics[width=1\columnwidth]{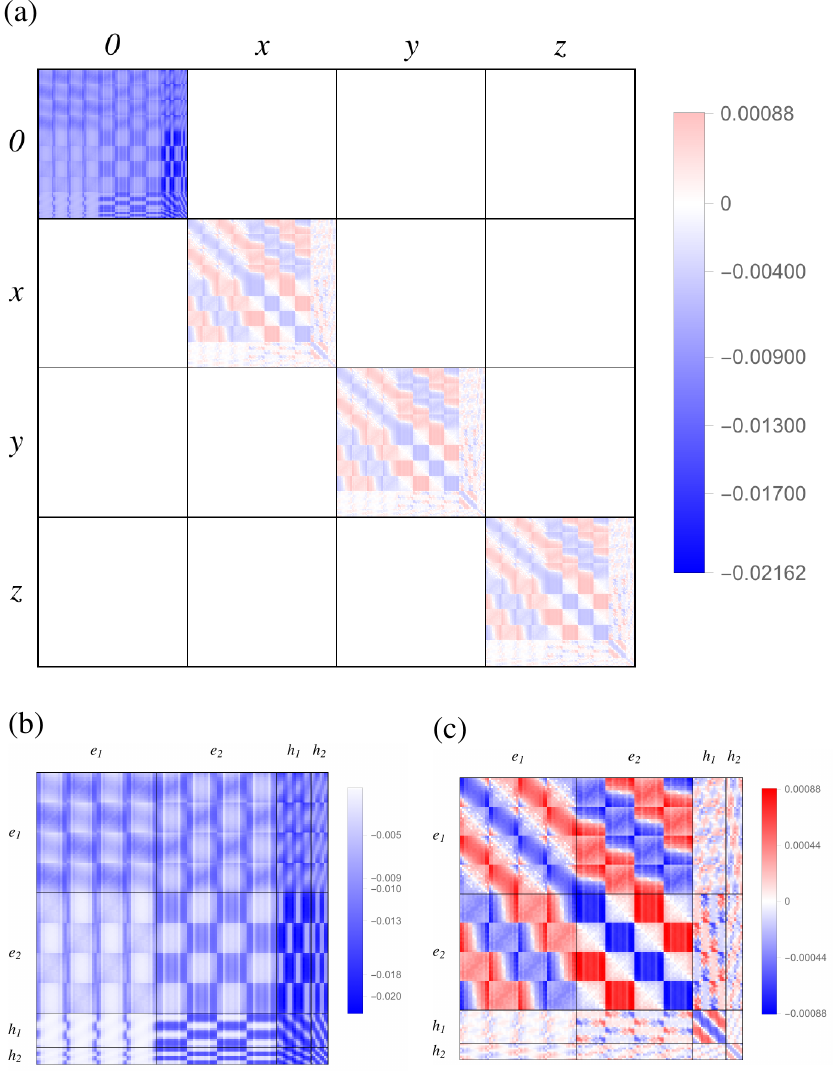}
\end{minipage}
\caption{Visualization of the pairing kernel entering the LGE. The spin-fluctuation mediated pairing interaction was determined for $\mu = 20\,$meV, $ T = 0.01\,$eV, $U = 0.80\,$eV and $ J = U/4 $ and $\lambda_{\mathrm{SOC}} = 0\,$meV. (a) Full pairing kernel in spin singlet-triplet space. The colors indicate the strength and sign of the pairing kernel. (b) Spin singlet and (c) spin triplet components of the pairing kernel.}
\label{fig:kernel_heavily_edoped_lambda_0p000_U_0p80}
\end{figure}

\begin{figure}[t!]
\centering
\begin{minipage}{0.88\columnwidth}
\centering
\includegraphics[width=1\columnwidth]{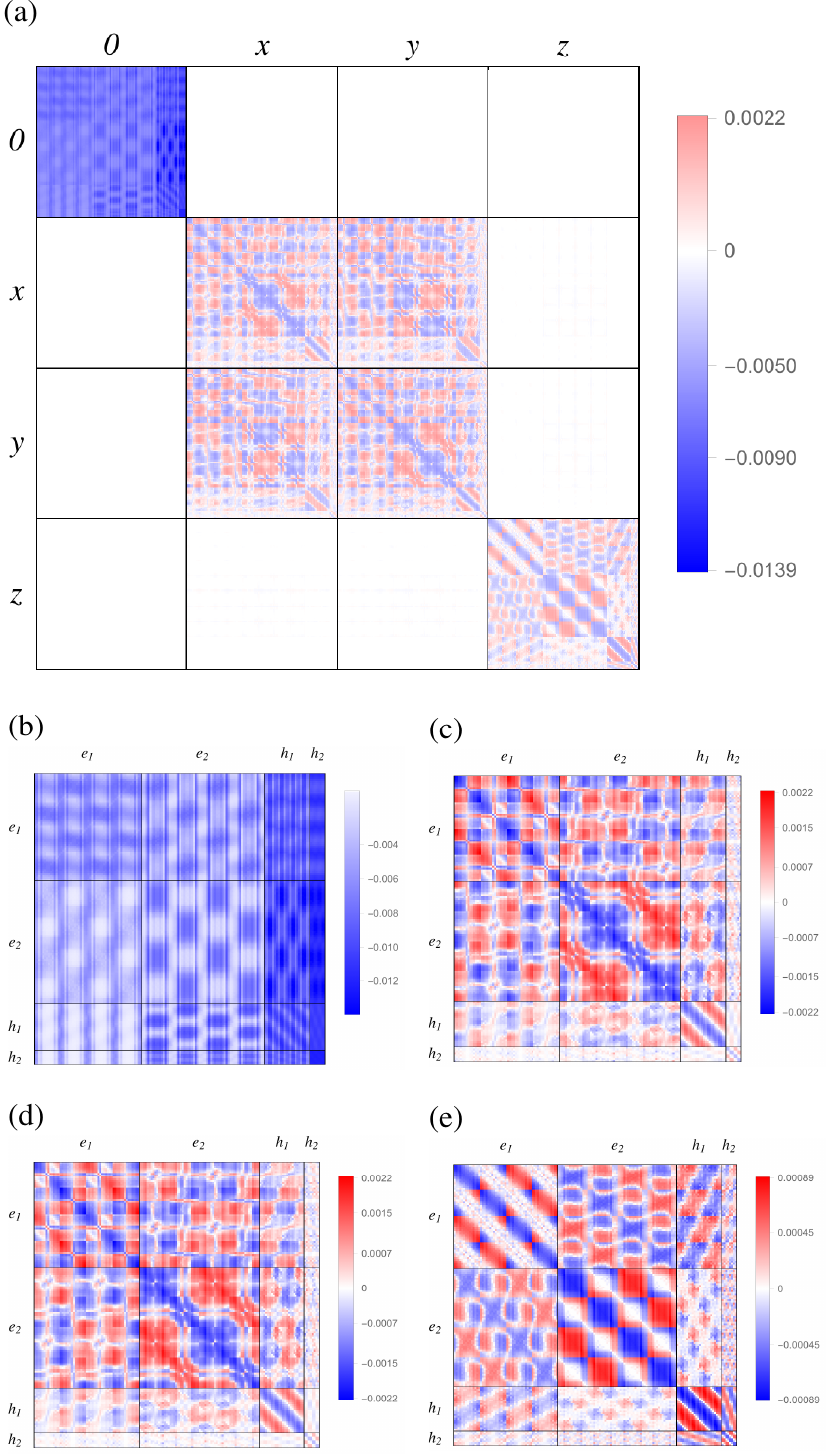}
\end{minipage}
\caption{Visualization of the pairing kernel entering the LGE. The spin-fluctuation mediated pairing interaction was determined for $\mu = 20\,$meV, $ T = 0.01\,$eV, $U = 0.80\,$eV and $ J = U/4 $ and $\lambda_{\mathrm{SOC}} = 75\,$meV. (a) Full pairing kernel in pseudospin singlet-triplet space. The colors indicate the strength and sign of the pairing kernel. (b) Pseudospin singlet and pseudspin triplet (c) $x$, (d) $y$ and (e) $z$ components of the pairing kernel. The blocks coupling pseudospin triplet $x$ and $y$ sectors are not shown separately.}
\label{fig:kernel_heavily_edoped_lambda_0p075_U_0p80}
\end{figure}

\begin{figure}[t!]
\centering
\begin{minipage}{1.00\columnwidth}
\centering
\includegraphics[width=1\columnwidth]{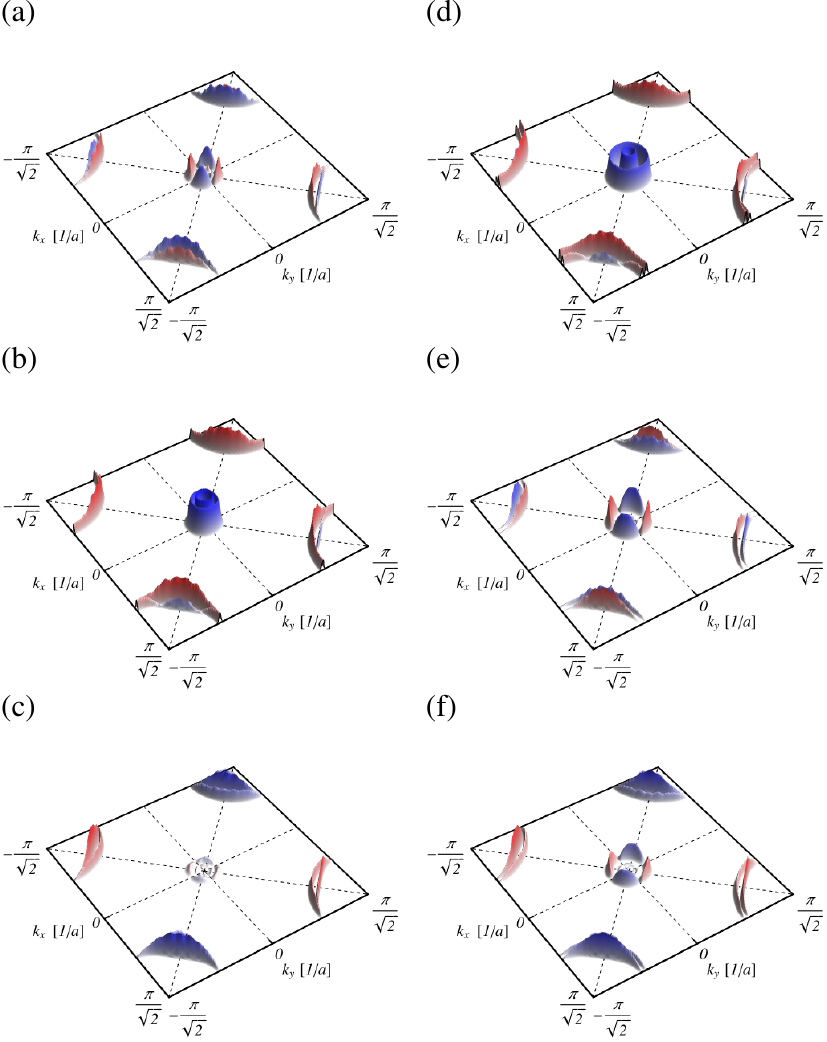}
\end{minipage}
\caption{Leading and first two subleading LGE solutions of the heavily electron-doped system for $ U = 0.80\,$eV with (a)-(c) $ \lambda_{\mathrm{SOC}} = 0 \, $meV and (d)-(f) $ \lambda_{\mathrm{SOC}} = 75 \, $meV, respectively. The LGE solutions are ordered according to their $\lambda$ values, where $\lambda$ decreases from top to bottom in each column. All solutions shown have pseudospin singlet character.}
\label{fig:gap_heavily_edoped}
\end{figure}

The pairing kernels for the electron-doped system without and with SOC are shown in Fig.~\ref{fig:kernel_heavily_edoped_lambda_0p000_U_0p80} and Fig.~\ref{fig:kernel_heavily_edoped_lambda_0p075_U_0p80} for interaction parameter $ U = 0.80\, $eV, respectively. The (pseudo-)spin singlet components of the kernels do not appear to be drastically different.
The gap solutions corresponding to these kernels are collected in Fig.~\ref{fig:gap_heavily_edoped}. Both for
vanishing and finite SOC, the first few attractive LGE solutions have (pseudo-)spin singlet character.
We observe that finite SOC has the effect of pushing the pseudospin triplet solution slightly higher in the hierarchy
of LGE solutions, but even for $ \lambda_{\mathrm{SOC}} = 100 \, $meV and $ U = 0.80 \, $eV, the pseudospin triplet
solution corresponds to the 5th LGE eigenvalue.

For vanishing SOC, the leading solution displays a $d$-wave symmetry, with sign changes between the two electron and the two hole pockets, where the sign of the superconducting order parameter on the outer (inner) hole pocket is the same 
as on the outer (inner) electron pockets. The $d$-wave solution is followed by a subleading solution with $s_{+-}$ character, where, however, the inner electron pocket $e_{1}$ features accidental nodes. The gap on the small hole pockets is basically constant. We further observe, that at finite SOC these two solutions are swapped, and the $s_{+-}$ solution is rendered the leading Cooper instability in the pseudospin singlet channel. The swap can be read off from comparing the two columns of Fig.~\ref{fig:gap_heavily_edoped}, where the first column (Fig.~\ref{fig:gap_heavily_edoped}(a)-(c)) shows the first three solutions for vanishing SOC, and the second column (Fig.~\ref{fig:gap_heavily_edoped}(d)-(f)) the gap solutions obtained for $ \lambda_{\mathrm{SOC}} = 75 \, $meV. The gap amplitude on the pocket $ e_{1} $ has maxima that correspond to the maxima of the Fermi velocity, while on $ e_{2} $ the maxima in the gap amplitude coincide with the Fermi velocity minima. The gap on the hole pockets $ h_{1}$ and $ h_{2} $ is, as mentioned above, constant.

We note, that the same behavior as a function of SOC is observed for a higher level of electron doping. We additionally investigated in detail the gap solutions as a function of SOC and interaction strength for a chemical potential of $ \mu_{0} = 25 \, $eV, which, for vanishing SOC, realizes a Fermi surface comprised of only electron pockets. As SOC increases, small hole pockets open up around the $ \Gamma $ point and the leading $d$-wave solution is replaced by the $s_{+-}$ solution.

In terms of orbital composition, for vanishing SOC, the spin singlet $ d $-wave solution is dominated by intra-$d_{xz}$ and intra $d_{xz}$ components, with intra $d_{xy}$ and inter $d_{xz}$/$d_{yz}$-$d_{xy}$ being subleading components. For SOC strength $ \lambda_{\mathrm{SOC}} = 75 \,$meV, the intra $d_{xy}$ and inter $d_{xz}$/$d_{yz}$-$d_{xy}$ components of the spin singlet part are weakened compared to the intra $d_{xz}$ and $d_{yz}$ components. For the electron-doped system, the largest amplitudes of the spin triplet part grow only as large as 25 \% of the largest amplitudes of the spin singlet part. This suggests that undoped and hole-doped systems have a larger propensity to the formation of a sizable spin triplet component in the presence of SOC. The orbital composition of the spin triplet component has the same characteristics as encountered for the undoped and hole-doped systems, namely dominant inter $d_{xz}$ - $d_{xy}$ components in the triplet-$x$ and inter $d_{yz}$-$d_{xy}$ components in the triplet-$y$ channel. 

\subsection{FeSe}
\label{subsec:FeSe}

Despite its structural simplicity and structural similarity to the pnictides, the chalcogenide FeSe has rather distinct electronic properties~\cite{Coldeareview,Bohmerreview}. Several experimental and theoretical studies indicate the importance of both local and non-local correlation effects beyond a DFT treatment in this material~\cite{Watson_Hub,Evtushinsky_Hub}. While local correlations (as described by DMFT) are believed to be mainly responsible for orbital selective mass- and quasiparticle renormalizations~\cite{luca}, non-local correlations, as arising from the self-energy feedback of collective excitations, might play a role in the rather drastic Fermi surface renormalization observed in FeSe relative to shape and size of the Fermi surface as predicted by DFT+LDA calculations~\cite{Jiang,scherer2017}. To establish a weak-coupling baseline for further research on the interplay of SOC and correlations in FeSe, we will briefly present our numerical results obtained in the weak-coupling RPA approach for the spin-fluctuation mediated pairing problem, even though we find a low-energy spin fluctuation spectrum that is incompatible with experiment. Further, we will analyze the pairing problem in the non-nematic, tetragonal state of the system, while FeSe features a transition to a nematic, orthorhombic state, from which superconductivity eventually emerges upon a lowering temperature.

In order to incorporate the aforementioned Fermi surface renormalization, we apply the self-consistent procedure described in Ref.~\onlinecite{scherer2017}, in order to produce hopping renormalizations arising from the exchange self-energy in a mean-field treatment of a non-local, nearest-neighbor repulsion. The corresponding mean-field Hamiltonian is obtained for vanishing SOC. For finite SOC, we simply add $ H_{\mathrm{SOC}} $ (see \Eqref{eq:SOC}) to the hopping part of the Hamiltonian containing the mean-field corrections.

\begin{figure}[t!]
\centering
\begin{minipage}{1\columnwidth}
\centering
\includegraphics[width=1\columnwidth]{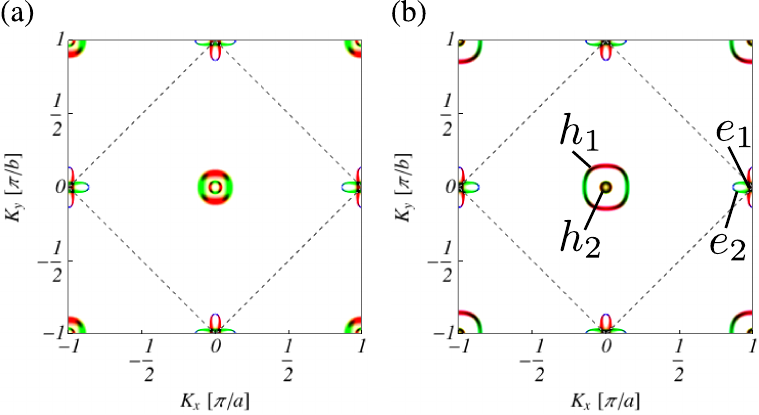}
\end{minipage}
\caption{Normal state Fermi surfaces of the FeSe model in the 1-Fe BZ ($(K_{x},K_{y})$ denotes momenta in the 1-Fe BZ coordinate system) extracted from the orbitally resolved contributions to the electronic spectral function with $ \mu_{0} = 0 $\,eV for (a) $\lambda = 0\,$meV and (b) $\lambda = 75\,$meV. The 2-Fe BZ is indicated by the dashed square. The colors refer to $d_{xz}$ (red), $d_{yz}$ (green) and $d_{xy}$ (blue) orbital contributions. SOC leads to a splitting of the states at the 2-Fe BZ boundary. As shown in (b), the labels $h_{1}$, $h_{2}$ refer to the outer and inner hole-pocket (as seen from the $\Gamma$ point), while $e_{1}$ and $e_{2}$ refer to the inner and outer electron pocket (as seen from the $M$ point).}
\label{fig:FS_FeSe}
\end{figure}

The resulting Fermi surfaces of our FeSe model are displayed in Fig.~\ref{fig:FS_FeSe}. For both vanishing and finite SOC, the Fermi surfaces feature two hole pockets around the $ \Gamma $ point and the FeSe-typical clover-leaf shaped electron pockets in the corners of the 2-Fe BZ, see Fig.~\ref{fig:FS_FeSe}(a). The tips of the clover-leafs have dominant $d_{xy}$ orbital character, while the hole pockets and the remaining clover-leaf structure are dominated by $d_{xz}$ and $d_{yz}$ orbitals. As described already above for the LaFeAsO model, the SOC-induced splitting at the $ \Gamma $ point pushes the hole pockets apart, see Fig.~\ref{fig:FS_FeSe}(b).

Analyzing the static susceptibility for the FeSe model exposes the known weakness of weak-coupling calculations for FeSe\cite{kreisel_2015}: The peak of the susceptibility (at momenta contributing to the Fermi surface projected LGE) is located in the vicinity of $ (\pi,\pi) $ (in 1-Fe notation), see Sec.~\ref{sec:orbital_FeSe}, rather than $ {\bf Q}_{1} = (\pi,0) $ and $ {\bf Q}_{2} = (0,\pi) $ as observed in inelastic neutron scattering experiments~\cite{rahn,Wang_NS1,Wang_NS2,Tong}. Nevertheless, for the purpose of understanding the role of SOC on pairing in FeSe from a weak-coupling perspective, we produce the corresponding pairing kernels for vanishing and finite SOC and solve the resulting LGE. Here, we limit the discussion to interaction parameter $U = 1.30\,$eV. This choice for the local repulsion, yielding a Stoner factor of 0.94, brings the system close to an SDW. For smaller values of the interaction parameter $ U $, the peak close to $ (\pi,\pi) $ is reduced in height, and becomes comparable to the peaks in the vicinity of wavevectors $ {\bf Q}_{1} $ and $ {\bf Q}_{2} $. When switching on SOC, the most prominent effect on the pseudospin singlet pairing kernel is the change in the pocket sizes (with the small pocket $e_{1}$ being absent for $\lambda_{\mathrm{SOC}} = 75 \,$meV), the momentum-space pattern of the kernel, however, remains more or less the same (not shown).

\begin{figure}[t!]
\centering
\begin{minipage}{1.00\columnwidth}
\centering
\includegraphics[width=1\columnwidth]{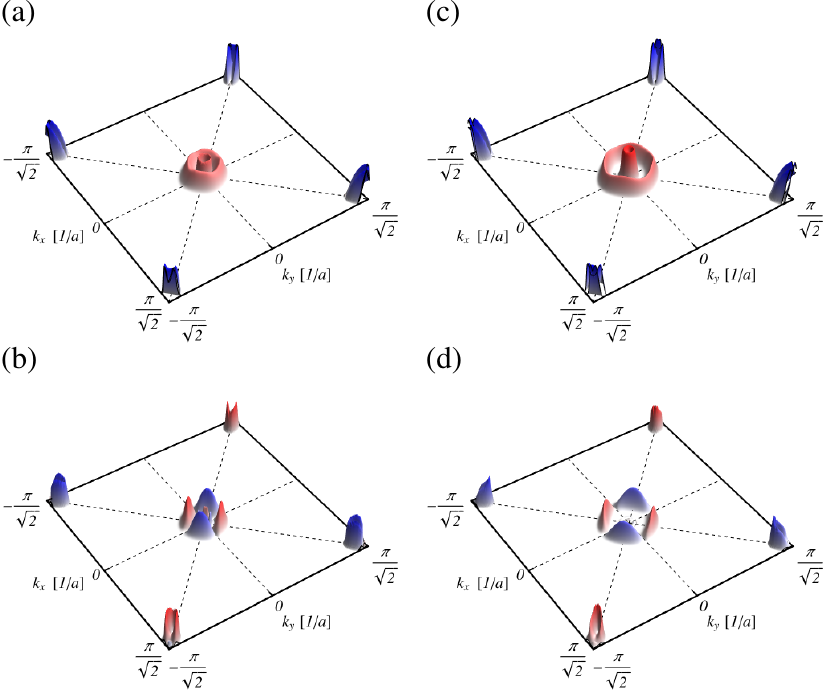}
\end{minipage}
\caption{Leading and first subleading LGE solutions of the FeSe model for $ U = 1.30\,$eV with (a)-(c) $ \lambda_{\mathrm{SOC}} = 0 \, $meV and (d)-(f) $ \lambda_{\mathrm{SOC}} = 75 \, $meV, respectively. The LGE solutions are ordered according to their $\lambda$ values, where $\lambda$ decreases from top to bottom in each column. All solutions shown have pseudospin singlet character.}
\label{fig:gap_FeSe_U_1p30}
\end{figure}

The leading and first subleading LGE solution for interaction parameter $ U = 1.30 \, $eV without (first column) and with SOC (second column) are shown in Fig.~\ref{fig:gap_FeSe_U_1p30}. The leading solution has $s_{+-}$ character in both cases, with the subleading solution being of $d$-wave symmetry. For vanishing SOC, the $ s_{+-} $ solution has accidental nodes close to the clover-leaf tips, with the gap maxima on the electron pockets being located at the base of the leafs, see Fig.~\ref{fig:gap_FeSe_U_1p30}(a). SOC has the effect of moving the nodes closer to the tip, while the positions of the gap maxima remain unaltered, see Fig.~\ref{fig:gap_FeSe_U_1p30}(b). In fact, the nodes are so close to each other, and the gap amplitude between the nodes is so small, that the gap region of opposite sign on the electron pocket is basically invisible in Fig.~\ref{fig:gap_FeSe_U_1p30}(b). We note that for smaller interaction parameter $ U $, the location of the nodes on the electron pockets move slightly outward in the case of vanishing SOC. The shrinking of the internodal region due to SOC appears to be robust with respect to variations in the interaction parameter. For our FeSe model, the gap variation in terms of the position of amplitude maxima for the $s_{+-}$ solution does not follow the maxima of the Fermi velocity. Rather, it is the internodal region that coincides with the maximum of the Fermi velocity on the electron pocket. Both without and with SOC, the gap amplitude on the hole pockets is constant, with the larger amplitude being found on the inner pocket $ h_{2} $. The most striking SOC-induced change for the subleading $d$-wave solution is the small amplitude of the gap on the inner hole pocket.

In terms of its orbital and spin structure, the $s_{+-}$ solution features dominant intra-orbital $d_{xz}$ and $d_{yz}$ components in the singlet part. The triplet-$x$ and -$y$ parts are dominated by inter $d_{xz}$-$d_{yz}$ components, with strong but subleading inter $d_{xz}$-$d_{xy}$ components for the triplet-$x$ and inter $d_{yz}$-$d_{xy}$ components to the triplet-$y$ part, respectively. The largest spin triplet amplitudes reach about 30\% of the largest spin singlet amplitudes.

The main conclusion we draw from the results shown in Fig.~\ref{fig:gap_FeSe_U_1p30} is that, despite the presence of only tiny Fermi surface pockets, SOC does not induce qualitatively new gap features. While the analysis above was not as material-specific as several other recent theoretical investigations~\cite{onari,kreisel2017,she,kang2018,benfatto,rhodes,kreisel2018}, it highlights the role of unusual electronic effects (beyond SOC) generating the experimentally detected peculiar gap structure of FeSe~\cite{sprau}. Current possibilities include, for example, orbital selective effects driven by strong electron interactions,\cite{kreisel2018} or severe Fermi surface modifications imposed by the nematic order at low temperatures~\cite{MingYi}.  

\section{Summary}
\label{sec:summary}

In this work, we have analyzed the influence of SOC on the leading and subleading solutions
of the Fermi surface-projected linearized gap equation for superconducting instabilities relevant iron-based superconductors, assuming a weak-coupling pairing kernel arising from spin fluctuations.
We have carried out an extensive parameter scan in terms of bandstructure, the strength of SOC, and the interaction parameter $U$, 
parameterizing the strength of the local intraorbital repulsion. Here, we presented a selection of our
numerical results. For the generic FeSCs band model, we additionally analyzed the doping dependence of the gap solutions and their momentum space anisotropy and the interplay of changes in the Fermi surface and the concomitant changes in the behavior of low-energy spin fluctuations. 

Not unexpectedly, we found that the momentum structure of the static spin susceptibility determines
the symmetry of the Cooper instability. Despite the complexity of the pairing kernel, as evidenced by
its fine grained momentum structure, we found that certain details of the gap anisotropy seem to be simply 
related to the variation of the Fermi velocity along the hole and electron pockets. We could identify three cases: 1)
the location of maxima in the gap amplitude coincides with the location of maxima in the Fermi velocity, 2) the location of maxima in the gap amplitude coincides with the location of minima in the Fermi velocity and 3) the gap is constant and features no variation as a function of the pocket angle.

It is known that at the level of the electronic structure, SOC basically entangles spin, orbital and momentum degrees of freedom. At the level of the collective spin fluctuations of the electronic system, these SOC-induced changes in the electronic structure translate to a momentum dependent magnetic anisotropy, i.e., paramagnon branches with different polarization have different dispersion relation due to SOC. Analyzing the LGE with SOC included, we found that while the magnetic anisotropy has an influence on the size of the values of the LGE eigenvalues and thereby the value of $ T_{\mathrm{c}} $, the magnetic anisotropy entering through spin fluctuations has no decisive influence on the overall structure of the pairing kernel, nor the resulting gap solutions. Rather, it is the orbital-to-band matrix elements, which encode the entangled structure of spin, orbital and momentum degrees of freedom, which are crucial for the structure of the pairing kernel. We note, that the situation would indeed be more complicated, if the anisotropic part of the susceptibility featured an additional, pronounced momentum structure not present in the isotropic case. We find, however, that no additional peaks are generated in the presence of SOC. Small shifts in the peak positions due to SOC-induced changes in the Fermi surface are not sufficient to generate drastic changes in the pairing kernel that could possibly alter the leading Cooper instability.

Most interestingly, perhaps, we found that the spin triplet components of the leading $ s_{+-} $ pseudospin singlet solutions can acquire quite sizable gap amplitude, where the strength of the spin triplet component depends on details such as the doping level. In addition, we conclude that SOC can generate a helical triplet pairing state for realistic parameters in the strongly hole-doped case. SOC was also shown to favor $s_{+-}$ superconductivity over $d$-wave pairing in the case of strongly electron-doped systems.

\begin{acknowledgments}
We are grateful for valuable discussions with P. J. Hirschfeld and A. T. R\o mer.
We acknowledge financial support from the Carlsberg Foundation. 
\end{acknowledgments}




\begin{widetext}

\newpage

\setcounter{equation}{0}
\setcounter{section}{0}
\setcounter{figure}{0}
\setcounter{table}{0}
\setcounter{page}{1}
\makeatletter
\renewcommand{\theequation}{S\arabic{equation}}
\renewcommand{\thesection}{S\arabic{section}}
\renewcommand{\thefigure}{S\arabic{figure}}
\renewcommand{\bibnumfmt}[1]{[S#1]}
\renewcommand{\citenumfont}[1]{S#1}

\begin{center}
\textbf{\large Supplementary Material: ``Effects of spin-orbit coupling on spin-fluctuation induced pairing in iron-based superconductors''}
\end{center}

\section{BdG Hamiltonian and Nambu-Gorkov Greens Function}
\label{sec:BdG}

In order to provide a complete and consistent account of all conventions
employed to obtain the results described in the main text, we start out with
providing a definition of the the Bogoliubov-de-Gennes (BdG) Hamiltonian for a spin-orbit coupled superconductor and the corresponding Nambu-Gorkov Greens function. We begin with representing the quadratic part of the normal-state Hamiltonian, $H_{0}$, in momentum space, where we define the Fourier-transformed fermionic operators as
\be 
c_{{\bf k}l\mu\sigma} = \frac{1}{\sqrt{\mathcal{N}}} \sum_{i} \mathrm{e}^{\mathrm{i}{\bf k}\cdot({\bf r}_{i}+\boldsymbol{\delta}_{l})} \, c_{li\mu\sigma}, \quad c_{{\bf k}l\mu\sigma}^{\dagger} = \frac{1}{\sqrt{\mathcal{N}}} \sum_{i} \mathrm{e}^{-\mathrm{i}{\bf k}\cdot({\bf r}_{i}+\boldsymbol{\delta}_{l})} \, c_{li\mu\sigma}^{\dagger}.
\ee
In the equations above, the sum runs over lattice sites, ${\bf r}_{i}$ denotes the lattice vector corresponding to the $i$th unit cell and $\boldsymbol{\delta}_{l}$ denotes an intra-unitcell vector referencing the relative positions of the two Fe atoms. The non-interacting Hamiltonian $ H_{0} + H_{\mathrm{SOC}} $, see \Eqref{eq:hopping} and \Eqref{eq:SOC}, can then be written as
\be 
\label{eq:hamiltonian}
H_{0} + H_{\mathrm{SOC}} = \sum_{{\bf k},l,l^{\prime},\sigma,\sigma^{\prime}} 
c_{{\bf k}l\mu\sigma}^{\dagger} \,
[h({\bf k})]_{l\mu\sigma,l^{\prime}\mu^{\prime}\sigma^{\prime}} \,
c_{{\bf k}l^{\prime}\mu^{\prime}\sigma^{\prime}},
\ee
with $ h({\bf k}) $ denoting a $20 \times 20$ ${\bf k}$-dependent Bloch-matrix acting in the single-particle Hilbert space of sublattice $\otimes$ orbital $\otimes$ spin. The Bloch-matrix $ h({\bf k}) $ is diagonalized by a unitary transformation
\be 
\label{eq:transformation}
c_{{\bf k}l\mu\sigma} = \sum_{b,\kappa} [\mathcal{U}({\bf k})]_{l\mu\sigma,b\kappa} \Phi_{{\bf k}b\kappa}, \quad
c_{{\bf k}l\mu\sigma}^{\dagger} = \sum_{b,\kappa} [\mathcal{U}({\bf k})]_{l\mu\sigma,b\kappa}^{\ast} \Phi_{{\bf k}b\kappa}^{\dagger}
\ee
with the unitary matrix $ \mathcal{U}({\bf k}) $ comprising the eigenvectors of $ h({\bf k}) $, with corresponding eigenvalues $ \epsilon_{b\kappa}({\bf k}) $. Here, the index $ b $ labels electronic bands, while $ \kappa \in \{ +, - \}$ labels a pair of degenerate states, that is guaranteed to exist at a given ${\bf k}$ for a time-reversal-invariant and inversion-symmetric system. In the solution of the pairing problem with spin-orbit coupling, the $\kappa$-degree of freedom plays the role of a pseudospin. With a judicious choice of the eigenbasis in the degenerate subspaces, the pseudospin indeed has the transformation properties of a spin-1/2 degree of freedom. The construction of the pseudospin basis is described in Sec.~\ref{sec:pseudospin} below.   

Moving to the superconducting state by adding a Hamiltonian describing the coupling of electrons to the pairing field, $H_{\Delta}$, the full BCS Hamiltonian containing the superconducting order parameter can be written as
\be 
\label{eq:BCS}
H_{\mathrm{BCS}} = H_{0} - \mu_{0} N + H_{\mathrm{SOC}} + H_{\Delta} = \frac{1}{2}\sum_{{\bf k}} \Psi_{{\bf k}l\mu}^{\dagger} [ h_{\mathrm{BdG}}({\bf k}) ]_{l\mu,l^{\prime}\mu^{\prime}} \Psi_{{\bf k}l^{\prime}\mu^{\prime}},
\ee
where we introduced the Nambu-Gorkov spinors $\Psi_{{\bf k}l\mu}$, $\Psi_{{\bf k}l\mu}^{\dagger}$. In terms of the original electron operators in the orbital Wannier basis, these are defined as
\be 
\Psi_{{\bf k}l\mu} = 
\left( c_{{\bf k}l\mu\uparrow}, c_{{\bf k}l\mu\downarrow}, c_{-{\bf k}l\mu\uparrow}^{\dagger}, c_{-{\bf k}l\mu\downarrow}^{\dagger} \right)^{T}, \quad
\Psi_{{\bf k}l\mu}^{\dagger} = 
\left( c_{{\bf k}l\mu\uparrow}^{\dagger}, c_{{\bf k}l\mu\downarrow}^{\dagger}, c_{-{\bf k}l\mu\uparrow}, c_{-{\bf k}l\mu\downarrow} \right),
\ee
where the transposition turns the row- into a columnvector, but acts trivially on the Fock-space operators. For each pair of indices $ l, \mu $ and $ l^{\prime }, \mu ^{\prime} $, $ [h_{\mathrm{BdG}}({\bf k}) ]_{l\mu,l^{\prime}\mu^{\prime}} $ is a $ 4 \times 4 $ matrix acting in particle-hole $\otimes$ spin-space. We can compactly write
\be 
\label{eq:BdG}
h_{\mathrm{BdG}}({\bf k})  = 
\mathcal{P}_{+} \otimes  \left( h({\bf k}) - \mu_{0} \mathbbm{1} \right)+
\mathcal{P}_{-} \otimes \left( -h^{T}(-{\bf k}) + \mu_{0}\mathbbm{1} \right)  +
\tau_{+} \otimes  \Delta({\bf k}) + \tau_{-}  \otimes  \bar{\Delta}({\bf k}),
\ee
where $ \mathcal{P}_{\pm} = \frac{1}{2} \left( \mathbbm{1} \pm \tau_{z} \right) $, 
$ \tau_{\pm} = \frac{1}{2} \left( \tau_{x} \pm \mathrm{i}\tau_{y} \right) $ with $ \tau_{i} $, $i = x,y,z$ Pauli matrices acting in particle-hole space. We keep a general orbital and spin structure for the pairing fields $ \Delta({\bf k}) $, $ \bar{\Delta}({\bf k}) $. The spin structure can be decomposed into spin-singlet ($\varsigma = 0$) and spin-triplet ($\varsigma = x, y, z$) as
\be
\label{eq:spin_struct}
[\Delta({\bf k})]_{l\mu\sigma,l^{\prime}\mu^{\prime}\sigma^{\prime}} = 
\sqrt{2} \sum_{\varsigma} s_{\varsigma}  [\Delta_{\varsigma}({\bf k})]_{l\mu,l^{\prime}\mu^{\prime}} [\Gamma_{\varsigma}]_{\sigma,\sigma^{\prime}}, \quad
[\bar{\Delta}({\bf k})]_{l\mu\sigma,l^{\prime}\mu^{\prime}\sigma^{\prime}} = 
\sqrt{2} \sum_{\varsigma} [\bar{\Delta}_{\varsigma}({\bf k})]_{l\mu,l^{\prime}\mu^{\prime}} [\Gamma_{\varsigma}]_{\sigma,\sigma^{\prime}},
\ee
where we introduced the Balian-Werthamer (BH) spin matrices 
\be 
\Gamma_{0} = \frac{1}{\sqrt{2}}\mathbbm{1}\mathrm{i}\sigma_{y}, \quad
\Gamma_{x} = \frac{1}{\sqrt{2}}\sigma_{x}\mathrm{i}\sigma_{y}, \quad
\Gamma_{y} = \frac{1}{\sqrt{2}}\sigma_{y}\mathrm{i}\sigma_{y}, \quad
\Gamma_{z} = \frac{1}{\sqrt{2}}\sigma_{z}\mathrm{i}\sigma_{y},
\ee
characterizing the spin structure of singlet and triplet Cooper pairs, and $ s_{\varsigma} = +1 $ for $ \varsigma \in \{ x,z \} $ and $ s_{\varsigma} = -1 $ for $ \varsigma \in \{ 0, y \} $. In a self-consistent treatment of pairing, the pairing fields need to be defined in terms of the operators describing the microscopic electronic degrees of freedom, glued together by a suitable, attractive interaction encoded in the pairing vertex. In our conventions, the definitions for the singlet and triplet pairing fields read as
\be 
[\Delta_{\varsigma}({\bf k})]_{l_{1}\mu_{1},l_{2}\mu_{2}} & = &
- \frac{1}{\sqrt{2} \mathcal{N}} 
\sum_{\varsigma^{\prime},{\bf k}^{\prime}}
\sum_{l_{3}\mu_{3},l_{4}\mu_{4}}
[\Gamma^{\varsigma,\varsigma^{\prime}}({\bf k},{\bf k}^{\prime})]^{l_{1}\mu_{1};l_2\mu_{2}}_{l_{3}\mu_{3};l_4\mu_{4}}
\left\langle 
\sum_{\sigma,\sigma^{\prime}}
c_{-{\bf k}l_{3}\mu_{3}\sigma} 
[\Gamma_{\varsigma^{\prime}}]_{\sigma\sigma^{\prime}}  
c_{{\bf k}l_{4}\mu_{4}\sigma^{\prime}} 
\right\rangle_{\mathrm{BCS}}
\\[0.5em]
 [\bar{\Delta}_{\varsigma}({\bf k})]_{l_{1}\mu_{1},l_{2}\mu_{2}} & = &
- \frac{1}{\sqrt{2} \mathcal{N}} 
\sum_{\varsigma^{\prime},{\bf k}^{\prime}}
\sum_{l_{3}\mu_{3},l_{4}\mu_{4}}
[\Gamma^{\varsigma^{\prime},\varsigma}({\bf k}^{\prime},{\bf k})]^{l_{3}\mu_{3};l_4\mu_{4}}_{l_{1}\mu_{1};l_2\mu_{2}}
\left\langle 
\sum_{\sigma,\sigma^{\prime}}
s_{\varsigma^{\prime}}
c_{{\bf k}l_{3}\mu_{3}\sigma}^{\dagger}
[\Gamma_{\varsigma^{\prime}}]_{\sigma\sigma^{\prime}}  
c_{-{\bf k}l_{4}\mu_{4}\sigma^{\prime}}^{\dagger}
\right\rangle_{\mathrm{BCS}},
\ee
where the static 2-PI vertex in the pairing channel is denoted by $ [\Gamma^{\varsigma,\varsigma^{\prime}}({\bf k},{\bf k}^{\prime})]^{l_{1}\mu_{1};l_2\mu_{2}}_{l_{3}\mu_{3};l_4\mu_{4}} $. We note that in the presence of SOC, a mixing of triplet components is possible in general. Therefore, the pairing vertex acts as a matrix in singlet-triplet space, as indicated by the labels $ \varsigma $ and $ \varsigma^{\prime} $. Details on the construction of the spin-fluctuation mediated pairing vertex in the presence of SOC can be found in Sec.~\ref{sec:vertex} below. The expectation value above is taken with respect to a Gibbs state of the BCS mean-field Hamiltonian \Eqref{eq:BCS}.

By applying the transformation \Eqref{eq:transformation} to \Eqref{eq:BdG}, the BdG Hamiltonian can be brought into the band-space representation. Here, we simply note the following transformation rules that connect the orbital- and band-space representations of the pairing fields:
\be 
\label{eq:transformation_1}
[\hat{\Delta}({\bf k})]_{b\kappa,b^{\prime}\kappa^{\prime}} & = & 
\sum_{l,l^{\prime}}\sum_{\mu,\mu^{\prime}}\sum_{\sigma,\sigma^{\prime}}
[\mathcal{U}({\bf k})]_{l\mu\sigma,b\kappa}^{\ast}
[\mathcal{U}(-{\bf k})]_{l^{\prime}\mu^{\prime}\sigma^{\prime},b^{\prime}\kappa^{\prime}}^{\ast}
[\Delta({\bf k})]_{l\mu\sigma,l^{\prime}\mu^{\prime}\sigma^{\prime}}, \\[0.5em]
\label{eq:transformation_2}
[\bar{\hat{\Delta}}({\bf k})]_{b\kappa,b^{\prime}\kappa^{\prime}} & = & 
\sum_{l,l^{\prime}}\sum_{\mu,\mu^{\prime}}\sum_{\sigma,\sigma^{\prime}}
[\mathcal{U}(-{\bf k})]_{l\mu\sigma,b\kappa}
[\mathcal{U}({\bf k})]_{l^{\prime}\mu^{\prime}\sigma^{\prime},b^{\prime}\kappa^{\prime}}
[\Delta({\bf k})]_{l\mu\sigma,l^{\prime}\mu^{\prime}\sigma^{\prime}}.
\ee
The band-space pairing fields in turn can be decomposed into pseudospin-singlet and pseudospin-triplet components:
\be 
\label{eq:pseudospin_struct}
[\hat{\Delta}({\bf k})]_{b\kappa,b^{\prime}\kappa^{\prime}} = 
\sqrt{2} \sum_{\varsigma} s_{\varsigma}  [\hat{\Delta}_{\varsigma}({\bf k})]_{b,b^{\prime}} [\hat{\Gamma}_{\varsigma}]_{\kappa,\kappa^{\prime}}, \quad
[\bar{\hat{\Delta}}({\bf k})]_{b\kappa,b^{\prime}\kappa^{\prime}} = 
\sqrt{2} \sum_{\varsigma} [\bar{\hat{\Delta}}_{\varsigma}({\bf k})]_{b,b^{\prime}} [\hat{\Gamma}_{\varsigma}]_{\kappa,\kappa^{\prime}},
\ee
where we notationally distinguish BH matrices in pseudospin space by a hat. The pseudospin basis is constructed such that for vanishing SOC it coincides with the physical spin. Correspondingly, in this limit the unitary transformation diagonalizing \Eqref{eq:hamiltonian} factorizes as
$ [\mathcal{U}({\bf k})]_{l\mu\sigma,b\kappa} = [\mathcal{U}({\bf k})]_{l\mu,b^{\prime}}\delta_{\sigma,\kappa} $. For finite SOC, the discussion of inter- and intraband pairing is thus necessarily tied to the pseudospin degree of freedom.

Having defined the BdG Hamiltonian in \Eqref{eq:BdG}, the corresponding imaginary-time Nambu-Gor'kov Green function can be obtained as 
\be 
\mathcal{G}(\mathrm{i}\omega_{n},{\bf k}) = - \int_{0}^{\beta} \! d\tau \mathrm{e}^{\mathrm{i} \omega_{n} \tau} \langle \mathcal{T}_{\tau} \Psi_{{\bf k}}(\tau) \Psi_{{\bf k}}^{\dagger}(0) \rangle_{\scriptsize\mathrm{BCS}},
\ee
where the expectation value is evaluated with respect to a thermal Gibbs state of inverse temperature $\beta = 1/k_{\mathrm{B}}T$ of the BCS Hamiltonian \Eqref{eq:BCS} and $ \omega_{n} = \frac{2\pi}{\beta} (n + 1/2) $, $n \in \mathbbm{Z}$ denotes a fermionic Matsubara frequency. We then decompose the Nambu-Gor'kov Green function in the same way as the BdG Hamiltonian to obtain
\be   
\mathcal{G}(\mathrm{i}\omega_{n},{\bf k}) =
\mathcal{P}_{+} \otimes G_{+}(\mathrm{i}\omega_{n},{\bf k})  - 
\mathcal{P}_{-} \otimes  G_{-}(\mathrm{i}\omega_{n},{\bf k})  +
\tau_{+} \otimes F(\mathrm{i}\omega_{n},{\bf k})  + \tau_{-}  \otimes \bar{F}(\mathrm{i}\omega_{n},{\bf k}).
\ee
and we have $ G_{-}(\mathrm{i}\omega_{n},{\bf k}) = [G_{+}(-\mathrm{i}\omega_{n},-{\bf k})]^T $. Below, we will formulate the non-linear gap equation for the pairing fields in terms 
of the Nambu-Gor'kov Greens function. Upon linearization in the pairing fields, we obtain the linearized gap equation, which forms the basis for the numerical results presented in the main text.

\section{Gap Equation}
\label{sec:gap}

Having set up the BdG Hamiltonian and the corresponding Nambu-Gor'kov Greens function,
we can formulate the non-linear gap equation for the pairing fields in terms of the Nambu-Gor'kov Greens function. The resulting self-consistency equation for the static pairing fields reads as
\be 
[\Delta_{\varsigma}({\bf k})]_{l_{1}\mu_{1},l_{2}\mu_{2}} & = &
- \frac{1}{\sqrt{2} \beta \mathcal{N}} 
\sum_{\varsigma^{\prime}, n, {\bf k}^{\prime}}
\sum_{l_{3}\mu_{3},l_{4}\mu_{4}}
[\Gamma^{\varsigma,\varsigma^{\prime}}({\bf k},{\bf k}^{\prime})]^{l_{1}\mu_{1};l_2\mu_{2}}_{l_{3}\mu_{3};l_4\mu_{4}} \,
\mathrm{tr} 
\Bigl\{ 
[\mathcal{G}(\mathrm{i}\omega_{n},{\bf k}^{\prime})]_{l_{4}\mu_{4},l_{3}\mu_{3}} \,
\tau_{-}\otimes \Gamma_{\varsigma^{\prime}} 
\Bigr\},
\\[0.5em]
 [\bar{\Delta}_{\varsigma}({\bf k})]_{l_{1}\mu_{1},l_{2}\mu_{2}} & = &
- \frac{1}{\sqrt{2} \beta \mathcal{N}} 
\sum_{\varsigma^{\prime}, n, {\bf k}^{\prime}}
\sum_{l_{3}\mu_{3},l_{4}\mu_{4}}
[\Gamma^{\varsigma^{\prime},\varsigma}({\bf k}^{\prime},{\bf k})]^{l_{3}\mu_{3};l_4\mu_{4}}_{l_{1}\mu_{1};l_2\mu_{2}} \, s_{\varsigma^{\prime}} \,
\mathrm{tr} 
\Bigl\{ 
[\mathcal{G}(\mathrm{i}\omega_{n},{\bf k}^{\prime})]_{l_{4}\mu_{4},l_{3}\mu_{3}} \,
\tau_{+}\otimes \Gamma_{\varsigma^{\prime}} 
\Bigr\},
\ee
where the trace runs over particle-hole and spin indices and the pairing vertex $ [\Gamma^{\varsigma,\varsigma^{\prime}}({\bf k},{\bf k}^{\prime})]^{l_{1}\mu_{1};l_2\mu_{2}}_{l_{3}\mu_{3};l_4\mu_{4}} $ is defined below (see Sec.~\ref{sec:vertex}). Focussing on the expression for $ [\Delta_{\varsigma}({\bf k})]_{l\mu,l^{\prime}\mu^{\prime}} $ and linearizing the right-hand side in the pairing fields yields
\be 
\label{eq:LGE_1}
[\Delta_{\varsigma}({\bf k})]_{l_{1}\mu_{1},l_{2}\mu_{2}} & = &
- \frac{1}{\beta \mathcal{N}} 
\sum_{n, {\bf k}^{\prime}}
\sum_{\varsigma^{\prime},\varsigma^{\prime\prime}}
\sum_{l_{3}\mu_{3},l_{4}\mu_{4}}
\sum_{m_{1}\nu_{1},m_{2}\nu_{2}}
\sum_{\sigma_{1},\dots,\sigma_{4}}
[\Gamma^{\varsigma,\varsigma^{\prime}}({\bf k},{\bf k}^{\prime})]^{l_{1}\mu_{1};l_2\mu_{2}}_{l_{3}\mu_{3};l_4\mu_{4}} 
\times \nn \\
& & 
[G(-\mathrm{i}\omega_{n},-{\bf k}^{\prime})]_{l_{3}\mu_{3}\sigma_{2},m_{2}\nu_{2}\sigma_{1}} 
[\Gamma_{\varsigma^{\prime}}]_{\sigma_{2}\sigma_{3}}
[G(\mathrm{i}\omega_{n},{\bf k}^{\prime})]_{l_{4}\mu_{4}\sigma_{3},m_{1}\nu_{1}\sigma_{4}}
[\Gamma_{\varsigma^{\prime\prime}}]_{\sigma_{4}\sigma_{1}}
s_{\varsigma^{\prime\prime}}
[\Delta_{\varsigma^{\prime\prime}}({\bf k}^{\prime})]_{m_{1}\nu_{1},m_{2}\nu_{2}},
\ee
which can be recognized as the Bethe-Salpeter equation in the pairing channel, formulated for singlet and triplet pairing fields. In our approximation, the irreducible particle-particle bubble is built from the bare normal-state electronic Greens function $ G(\mathrm{i}\omega_{n},{\bf k}) $, and we exploited that $ G_{-}(\mathrm{i}\omega_{n},{\bf k}) = [G_{+}(-\mathrm{i}\omega_{n},-{\bf k})]^T $ and $ G_{+}(\mathrm{i}\omega_{n},{\bf k}) = G(\mathrm{i}\omega_{n},{\bf k}) $ for vanishing pairing fields. The normal-state Greens function can be expressed as
\be 
[G(\mathrm{i}\omega_{n},{\bf k})]_{l_{1}\mu_{1}\sigma_{1},l_{2}\mu_{2}\sigma_{2}} = 
\sum_{b,\kappa} 
[\mathcal{U}({\bf k})]_{l_{1}\mu_{1}\sigma_{1},b\kappa} 
[\mathcal{U}({\bf k})]_{l_{2}\mu_{2}\sigma_{2},b\kappa}^{\ast} 
\frac{1}{\mathrm{i}\omega_{n} - E_{b}({\bf k})},
\ee
with $ E_{b}({\bf k}) = \epsilon_{b}({\bf k}) - \mu_{0} $ and we dropped the index $ \kappa $ on the energies, as $ \epsilon_{b,+}({\bf k}) = \epsilon_{b,-}(\bf k) $. Plugging in this representation for the Greens function and evaluating the Matsubara sum, the summed irreducible particle-particle bubble becomes
\be 
\label{eq:irrbub}
\frac{1}{\beta}\sum_{n}\sum_{\sigma_{1},\dots,\sigma_{4}}
[G(-\mathrm{i}\omega_{n},-{\bf k}^{\prime})]_{l_{3}\mu_{3}\sigma_{2},m_{2}\nu_{2}\sigma_{1}} 
[\Gamma_{\varsigma^{\prime}}]_{\sigma_{2}\sigma_{3}}
[G(\mathrm{i}\omega_{n},{\bf k}^{\prime})]_{l_{4}\mu_{4}\sigma_{3},m_{1}\nu_{1}\sigma_{4}}
[\Gamma_{\varsigma^{\prime\prime}}]_{\sigma_{4}\sigma_{1}} & = & \nn \\
& & \hspace{-11cm} \sum_{b,b^{\prime}}\sum_{\kappa,\kappa^{\prime}}\sum_{\sigma_{1},\dots,\sigma_{4}}
\left( 
[\mathcal{U}(-{\bf k}^{\prime})]_{l_{3}\mu_{3}\sigma_{2},b\kappa} 
[\Gamma_{\varsigma^{\prime}}]_{\sigma_{2}\sigma_{3}} 
[\mathcal{U}({\bf k}^{\prime})]_{l_{4}\mu_{4}\sigma_{3},b^{\prime}\kappa^{\prime}} 
\right)
\left( 
[\mathcal{U}({\bf k}^{\prime})]_{m_{1}\nu_{1}\sigma_{4},b^{\prime}\kappa^{\prime}}^{\ast}
[\Gamma_{\varsigma^{\prime\prime}}]_{\sigma_{4}\sigma_{1}} 
[\mathcal{U}(-{\bf k}^{\prime})]_{m_{2}\nu_{2}\sigma_{1},b\kappa}^{\ast}
\right) \times \nn \\
\frac{1 - n_{\mathrm{F}}(E_{b}({\bf k}^{\prime})) - n_{\mathrm{F}}(E_{b^{\prime}}({\bf k}^{\prime}))}{E_{b}({\bf k}^{\prime}) + E_{b^{\prime}}({\bf k}^{\prime})}.
\ee
Above, we defined the Fermi-Dirac distribution function $ n_{\mathrm{F}}(E) = 1/(\mathrm{exp}(\beta E) + 1) $. In order to simplify the irreducible bubble further, we restrict to intraband pairing with $ b = b^{\prime} $. Indeed, intraband contributions close to the Fermi level are expected to dominate the last factor in \Eqref{eq:irrbub}, as then both energies in the denominator are small (relative to the Fermi level). We thus perform the replacement
\be 
\label{eq:factor}
\frac{1 - n_{\mathrm{F}}(E_{b}({\bf k}^{\prime})) - n_{\mathrm{F}}(E_{b^{\prime}}({\bf k}^{\prime}))}{E_{b}({\bf k}^{\prime}) + E_{b^{\prime}}({\bf k}^{\prime})} \to 
\frac{1 - 2 n_{\mathrm{F}}(E_{b}({\bf k}^{\prime}))}{2 E_{b}({\bf k}^{\prime})} \delta_{b,b^{\prime}} =
\frac{\mathrm{tanh}(\beta E_{b}({\bf k}^{\prime}) / 2)}{2 E_{b}({\bf k}^{\prime})}
\delta_{b,b^{\prime}},
\ee
in order to carry out the projection of the linearized gap equation onto the Fermi surface. Inserting $ 1 = \int_{-\infty}^{+\infty} \! dE \, \delta( E - E_{b}({\bf k}) )$ into \Eqref{eq:LGE_1}, exploiting that $\mathrm{tanh}(\beta E / 2) / 2 E $ is sharply peaked around $ E = 0 $ and subsequently rewriting the momentum sum as an integration over the Fermi surface (which for a 2D system is given by 1D sub-manifolds) allows to express the LGE purely in terms of quantities defined at Fermi momenta ${\bf k}_{\mathrm{F}}$ and an energy-cutoff and temperature dependent factor. In order to further simplify the problem, we bring the LGE into the band representation. Due to the restriction to intraband pairing, only pairing fields with $ b = b^{\prime} $ have to be considered. Using the decomposition of the pairing fields into spin-singlet/-triplet \Eqref{eq:spin_struct} and pseudospin-singlet/-triplet \Eqref{eq:pseudospin_struct} together with the transformation rules \Eqref{eq:transformation_1} and \Eqref{eq:transformation_2}, we obtain the conversion formulas
\be 
[\hat{\Delta}_{\varsigma}({\bf k})]_{b,b^{\prime}} & = & 
\sum_{\varsigma^{\prime}}
\sum_{l\mu,l^{\prime}\mu^{\prime}}
\sum_{\sigma,\sigma^{\prime}}
\sum_{\kappa,\kappa^{\prime}}
[\mathcal{U}({\bf k})]_{l\mu\sigma,b\kappa}^{\ast}
[\mathcal{U}(-{\bf k})]_{l^{\prime}\mu^{\prime}\sigma^{\prime},b^{\prime}\kappa^{\prime}}^{\ast}
s_{\varsigma^{\prime}} 
[\Gamma_{\varsigma^{\prime}}]_{\sigma\sigma^{\prime}} 
[\hat{\Gamma}_{\varsigma}]_{\kappa^{\prime}\kappa}
[\Delta_{\varsigma^{\prime}}({\bf k})]_{l\mu,l^{\prime}\nu}, \\[0.5em]
[\bar{\hat{\Delta}}_{\varsigma}({\bf k})]_{b,b^{\prime}} & = & 
\sum_{\varsigma^{\prime}}
\sum_{l\mu,l^{\prime}\mu^{\prime}}
\sum_{\sigma,\sigma^{\prime}}
\sum_{\kappa,\kappa^{\prime}}
[\mathcal{U}(-{\bf k})]_{l\mu\sigma,b\kappa}
[\mathcal{U}({\bf k})]_{l^{\prime}\mu^{\prime}\sigma^{\prime},b^{\prime}\kappa^{\prime}}
s_{\varsigma} 
[\Gamma_{\varsigma^{\prime}}]_{\sigma\sigma^{\prime}} 
[\hat{\Gamma}_{\varsigma}]_{\kappa^{\prime}\kappa}
[\Delta_{\varsigma^{\prime}}({\bf k})]_{l\mu,l^{\prime}\nu}.
\ee
Multiplying the LGE \Eqref{eq:LGE_1} with the appropriate orbital-to-band matrix elements and performing the Fermi surface projection, we eventually arrive at ($ \hat{\Delta}_{\varsigma}^{b}({\bf k}) \equiv [\hat{\Delta}_{\varsigma}({\bf k})]_{b,b} $)
\be 
\label{eq:LGE_2}
\lambda \hat{\Delta}_{\varsigma}^{b}({\bf k}_{\mathrm{F}}) =
- \frac{1}{V_{\mathrm{BZ}}} \sum_{b^{\prime},\varsigma^{\prime}} \int_{\mathrm{FS}} \! dk^{\prime} \, 
\frac{1}{v_{\mathrm{F}}({\bf k}_{\mathrm{F}}^{\prime})} \,
[\hat{\Gamma}^{\varsigma,\varsigma^{\prime}}({\bf k}_{\mathrm{F}}, {\bf k}_{\mathrm{F}}^{\prime})]_{b^{\prime}b^{\prime}}^{b \,\, b} \,
\hat{\Delta}_{\varsigma^{\prime}}^{b^{\prime}}({\bf k}_{\mathrm{F}}^{\prime}),
\ee
with ${\bf k}_{\mathrm{F}}$ denoting the Fermi momentum, $v_{\mathrm{F}}({\bf k}_{\mathrm{F}}) = 
\left| \nabla_{\bf k} \epsilon_{b}({\bf k}) |_{{\bf k} = {\bf k_{\mathrm{F}}}}  \right|$ the Fermi velocity at Fermi momentum ${\bf k}_{\mathrm{F}}$, and $ [\hat{\Gamma}^{\varsigma,\varsigma^{\prime}}({\bf k}, {\bf k}^{\prime})]_{b_{3}b_{4}}^{b_{1} b_{2}} $ the pairing vertex in band and pseudo-spin singlet and triplet space. It is understood that the Fermi-surface projected version of the pairing vertex vanishes if the band indices $b$ do not correspond to a band contributing to the Fermi surface. The symbol $ \int_{\mathrm{FS}} \! dk $ stands for integration along the 1D Fermi surface segments and $ V_{\mathrm{BZ}} $ denotes the Brillouin zone volume (or rather area, for a 2D system). In writing the LGE in its Fermi-surface projected form \Eqref{eq:LGE_2}, we tacitly generalized it to an eigenvalue problem for the integration kernel defined by the right-hand side of \Eqref{eq:LGE_2}, where the corresponding eigenvalue is denoted by $\lambda$. We note, that we have absorbed a energy-cutoff and temperature dependent prefactor in the definition of $ \lambda $. Due the Fermi surface projection and the aforementioned redefinition of $ \lambda $, we strictly speaking lose quantitative control over the precise criterion that determines the onset of a Cooper instability within our approximation to the pairing problem. Had we generalized the LGE in its non-projected form \Eqref{eq:LGE_1} to an eigenvalue problem for the corresponding kernel, the criterion determining the onset of the Cooper instability is simply given by $ \lambda = 1 $. Finally, the band-space representation of the pairing vertex is obtained as
\be 
 [\hat{\Gamma}^{\varsigma,\varsigma^{\prime}}({\bf k}, {\bf k}^{\prime})]_{b_{3}b_{4}}^{b_{1} b_{2}} & = & 
\sum_{\varsigma^{\prime\prime},\varsigma^{\prime\prime\prime}}
\sum_{\kappa_{1},\dots,\kappa_{4}}
\sum_{\sigma_{1},\dots,\sigma_{4}}
\sum_{l_{1},\dots,l_{4}}
\sum_{\mu_{1},\dots,\mu_{4}} [\Gamma^{\varsigma^{\prime\prime},\varsigma^{\prime\prime\prime}}({\bf k},{\bf k}^{\prime})]^{l_{1}\mu_{1};l_2\mu_{2}}_{l_{3}\mu_{3};l_4\mu_{4}} \times
\nn \\
& &
\left( s_{\varsigma^{\prime\prime}}
[\mathcal{U}({\bf k})]_{l_{1}\mu_{1}\sigma_{1},b_{1}\kappa_{1}}^{\ast}
[\Gamma_{\varsigma^{\prime\prime}}]_{\sigma_{1}\sigma_{2}}
 [\hat{\Gamma}_{\varsigma}]_{\kappa_{2}\kappa_{1}}
[\mathcal{U}(-{\bf k})]_{l_{2}\mu_{2}\sigma_{2},b_{2}\kappa_{2}}^{\ast}
\right) \times \nn \\
& & 
\left(
[\mathcal{U}(-{\bf k}^{\prime})]_{l_{3}\mu_{3}\sigma_{3},b_{3}\kappa_{3}}
[\Gamma_{\varsigma^{\prime\prime\prime}}]_{\sigma_{3}\sigma_{4}}
[\hat{\Gamma}_{\varsigma^{\prime}}]_{\kappa_{4}\kappa_{3}}
[\mathcal{U}({\bf k}^{\prime})]_{l_{4}\mu_{4}\sigma_{4},b_{4}\kappa_{4}}
s_{\varsigma^{\prime}}
\right).
\ee
%

\section{Construction of the Pseudospin Basis}
\label{sec:pseudospin}

In Sec.~\ref{sec:gap} we have introduced the linearized gap equation in band space.
Due to the presence of SOC, spin is no longer a good quantum number to label the
degenerate single-particle eigenstates of the Bloch matrix $ h({\bf k}) $. The inclusion of SOC in the pairing problem has been discussed in various places in the literature. Here, we closely follow the logic presented in Ref.~\onlinecite{SMsigrist1991}. For a concrete example, see e.g. Ref.~\onlinecite{SMueda1985}. As we mentioned already above, the degenerate states at a given Bloch momentum $ {\bf k} $ can be labeled by the Kramers' index $ \kappa \in \{ +, - \} $. When diagonalizing the matrix $ h({\bf k}) $ in order to obtain the eigenvectors $ [\mathcal{U}({\bf k})]_{l\mu\sigma,b\kappa} $, we are, however, faced with an SU(2) `gauge' problem. As in our case the $ 20 \times 20 $ matrix $ h({\bf k}) $ can only be diagonalized numerically, the numerical diagonalization routine will choose \textit{some} orthonormal basis in each of the degenerate subspaces. There is no guarantee, however, that this arbitrary choice of bases will respect the symmetries of the system, nor that it provides a reasonably smooth dependence of the eigenvectors on momentum $ {\bf k} $. We note, that in the absence of SOC, this problem does typically not occur, as the Hamiltonian becomes block diagonal in spin space and diagonalization of one of the blocks might be sufficient. In this case, the SU(2)-gauge ambiguity can simply be removed by requiring the eigenstates of $ h({\bf k}) $ to have a well defined spin quantum number with respect to a given spin quantization axis. In other words, due to the complexity of $ h(\bf k) $ due to interorbital hybridization and SOC-induced loss of SU(2)-invariance, it is necessary to carefully construct a pseudospin degree of freedom in order to arrive at a well-defined formulation of the gap equation in band space. The construction of the pseudospin degree of freedom can be considered as finding an SU(2) gauge that satisfies i) as $ \lambda_{\mathrm{SOC}} \to 0 $ the eigenstates continuously evolve to eigenstates with well-defined spin quantum number and ii) the pseudospin has the transformation properties of a spin-1/2 degree of freedom. 

To implement this particular gauge, it is helpful to adopt a Wannier basis such that $ h({\bf k}) $ only has real eigenvectors. This can easily be achieved by taking an orbital basis with $ \mathrm{i} xz $ and $ \mathrm{i} yz $. We note, that such a change of the Wannier basis also leads to changes in certain interorbital matrix elements of the Hubbard-Hund interaction Hamiltonian that need to be taken into account. The next step in the construction consists of deriving the transformation properties of the eigenstates of $ h({\bf k}) $ under (non-symmorphic) space-group operations $ g $. The starting point for the derivation is the transformation rule for the Fock-space operator $ c_{l i \mu \sigma} $, which reads as
\be 
g \, c_{l i \mu \sigma} \, g^{-1} =
\sum_{l^{\prime}\mu^{\prime},\sigma^{\prime}} 
[\mathcal{M}_{g}^{\dagger}]_{l \mu \sigma, l^{\prime} \mu^{\prime} \sigma^{\prime}} \,
c_{l^{\prime} i^{\prime} \mu^{\prime}\sigma^{\prime}},
\ee
where $ i^{\prime} $ corresponds to the unit cell lattice vector $ {\bf r}_{i^{\prime}}  = R_{g} {\bf r}_{i} + {\bf t}_{g}$, with $ R_{g} $ and $ \mathcal{M}_{g} $ representation matrices for point-group transformations in real and sublattice $ \otimes $ orbital $\otimes $ spin space and ${\bf t}_{g}$ denoting a translation specific to the non-symmorphic spacegroup relevant for FeSCs~\cite{SMnourafkan2017}. Inserting the Bloch representation for the annihilation operator, we can derive the transformation rule
\be 
g \, c_{{\bf k} l \mu \sigma} \, g^{-1} =
\sum_{l^{\prime}\mu^{\prime},\sigma^{\prime}} 
\mathrm{e}^{-\mathrm{i} {\bf k}^{\prime} \cdot {\bf t}_{g}}
[\mathcal{M}_{g}^{\dagger}]_{l \mu \sigma, l^{\prime} \mu^{\prime} \sigma^{\prime}} \,
c_{{\bf k}^{\prime} l^{\prime}\mu^{\prime}\sigma^{\prime}},
\ee
with $ {\bf k}^{\prime} = R_{g} {\bf k} $. Defining $ \mathcal{M}_{g}({\bf k}) = \mathrm{e}^{+\mathrm{i} {\bf k} \cdot {\bf t}_{g}} \mathcal{M}_{g}$, on the level of the Bloch matrix, the transformation rule for a spacegroup transformation $ g $ implies
\be 
\mathcal{M}_{g}^{\dagger}({\bf k}^{\prime}) \, h({\bf k}^{\prime}) \, \mathcal{M}_{g}({\bf k}^{\prime}) = h({\bf k}).
\ee
Now inserting the representation in terms of the non-hybridizing band-space operators we obtain
\be 
g \, \Phi_{{\bf k} b \kappa} \, g^{-1} =
\sum_{b^{\prime},\kappa^{\prime}}
\sum_{l,\mu,\sigma}
\sum_{l^{\prime}\mu^{\prime}\sigma^{\prime}}
[\mathcal{U}({\bf k})]_{l\mu\sigma,b\kappa}^{\ast} 
[\mathcal{M}_{g}^{\dagger}({\bf k})]_{l\mu\sigma,l^{\prime}\mu^{\prime}\sigma^{\prime}} 
[\mathcal{U}(R_{g}{\bf k})]_{l^{\prime}\mu^{\prime}\sigma^{\prime},b^{\prime}\kappa^{\prime}} \,
\Phi_{R_{g}{\bf k} b^{\prime} \kappa^{\prime}}.
\ee
As we want the gauge to implement the spin-1/2 transformation for the pseudospin degree of freedom, we have to require
\be 
\sum_{l,\mu,\sigma}
\sum_{l^{\prime}\mu^{\prime}\sigma^{\prime}}
[\mathcal{U}({\bf k})]_{l\mu\sigma,b\kappa}^{\ast} 
[\mathcal{M}_{g}^{\dagger}({\bf k})]_{l\mu\sigma,l^{\prime}\mu^{\prime}\sigma^{\prime}} 
[\mathcal{U}(R_{g}{\bf k})]_{l^{\prime}\mu^{\prime}\sigma^{\prime},b^{\prime}\kappa^{\prime}} = [D_{g}^{\dagger}]_{\kappa\kappa^{\prime}} \delta_{b,b^{\prime}},
\ee
where $ D_{g} $ is the representation matrix for a spin rotation for a spin-1/2 degree of freedom corresponding to the transformation $ g $. For clarity, we rearrange this formula and arrive at
\be
\label{eq:trans_rule} 
[\mathcal{U}(R_{g}{\bf k})]_{l\mu\sigma,b\kappa} =
\sum_{\kappa^{\prime}}
\sum_{l^{\prime}\mu^{\prime}\sigma^{\prime}}
[\mathcal{M}_{g}({\bf k})]_{l\mu\sigma,l^{\prime}\mu^{\prime}\sigma^{\prime}}
[\mathcal{U}({\bf k})]_{l^{\prime}\mu^{\prime}\sigma^{\prime},b\kappa^{\prime}} 
[D_{g}^{\dagger}]_{\kappa^{\prime}\kappa}.
\ee
We can now turn \Eqref{eq:trans_rule} into a prescription of how to construct the eigenstates at ${\bf k}^{\prime} = R_{g} {\bf k}$ given those at $ {\bf k} $. Practically, we proceed as follows: We split the 2-Fe BZ into eight symmetry-related sectors of equal area. Each sector is defined by the the range of the angle $ \theta $, where $ \cos \theta = {\bf k} \cdot \hat{\mathrm{e}}_{x} /  | {\bf k} | $ with $ \hat{\mathrm{e}}_{x} = (1,0) $. The first sector is defined by taking all momenta in the first BZ with $ \theta \in [0, \pi/4) $. We then enumerate the different sectors in a counter-clockwise fashion. We now diagonalize $ h({\bf k})$, defined in the Wannier gauge introduced above, in order to obtain the corresponding eigenvectors at the Fermi surface. We then pick one of the two eigenvectors in the degenerate subspace as the $ \kappa = + $ state. To fix the U(1) gauge (even in the non-degenerate case, eigenvectors are only defined up to a phase), we arbitrarily require the $ l = A $, $ \mu = d_{x^{2}-y^{2}} $, $\sigma = \uparrow$ component of the eigenvectors to be real and positive. In the next step, we apply the combination of inversion ($\mathcal{P}$) and time-reversal ($\mathcal{T}$) to construct the $ \kappa = - $ state at ${\bf k}$. Similarly, we use $\mathcal{P}$ to define a $ \kappa = + $ state at $ - {\bf k} $, and $\mathcal{T}$ to obtain the corresponding $\kappa = -$ state. We proceed in this manner for the Fermi surface eigenstates in the first sector. We apply \Eqref{eq:trans_rule} to the 1st and 5th sector to obtain the states in sectors 2, 3, 4 and 6, 7, 8, respectively. In the notation of Ref.~\onlinecite{SMnourafkan2017}, we employ the symmetry operations $\{C_{2a} | 000 \}$, $\{ C_{4z}^{+} | 0\frac{1}{2}0 \}$, $ \{ C_{2x} | \frac{1}{2}00 \} $. In the construction of $ \mathcal{M}_{g} $ some care has to be taken in picking the correct rotation axes for orbitals (in our case defined with respect to the 1-Fe coordinate system) and momenta (defined with respect to the 2-Fe coordinate system).

By construction, \Eqref{eq:trans_rule} ensures that property ii) is satisfied and the pseudospin degree of freedom has the correct transformation behavior which allows for a classification of the band-space pairing fields in terms of pseudospin singlet and triplet. In order to make sure that property i) is satisfied by the eigenstates defined according to \Eqref{eq:trans_rule}, we make use of a small, regularizing Zeeman field $ h $, such that for $ \lambda_{\mathrm{SOC}} = 0 $ the numerical diagonalization routine can correctly identify the $\uparrow-$ and $\downarrow$-structure of eigenstates in spin space without additional modifications. In the limit $ \lambda_{\mathrm{SOC}} \to 0 $, the eigenstates thus satsify $ [\mathcal{U}({\bf k})]_{l\mu\sigma,b\kappa} = [\mathcal{U}({\bf k})]_{l\mu,b} \delta_{\sigma,\kappa} $ (which, strictly speaking, is valid only when also taking $ h \to 0 $; but since the Zeeman field is chosen to be on the order of $10^{-10}\,$eV, the spectral splitting and modifications of the eigenvectors are basically negligible). Using this expression in \Eqref{eq:trans_rule}, the decoupling of the spin degree of freedom from the transformation of the eigenstates becomes obvious. The spin part of the representation matrix $ \mathcal{M}_{g} $ exactly cancels the spin matrix $ D_{g}^{\dagger} $, leaving only the sublattice $ \otimes $ orbital part to transform non-trivially.

We note that while the regularizing effect of the Zeeman field facilitates the satisfaction of property i), it does in no way guarantee property ii). It is also worth emphasizing that for certain models of itinerant electrons, as for e.g. Sr$_{2}$RuO$_{4}$ in the presence of SOC as detailed in~\onlinecite{SMroemer2019}, the complications described above can be circumvented owing to a drastic simplification, namely the absence of interorbital hybridization in the hopping part of $ h({\bf k}) $. In this case it is possible to define the pseudospin degree of freedom in a $ {\bf k} $ independent way, which is easily implemented by a unitary transformation which leads to a block-diagonal form for the Bloch matrix $ h({\bf k}) $. In the more realistic case of finite interorbital hybridization, the use of \Eqref{eq:trans_rule} (or alternative gauge-fixing procedures, see e.g. Ref.~\onlinecite{SMnomoto2016}) becomes mandatory (actually, it depends on the precise structure of the SOC Hamiltonian and which orbitals hybridize due to interorbital hopping, wether a ${\bf k}$ independent definition of pseudospin is possible or not, see e.g. Ref.~\onlinecite{SMzhang2018} for a Sr$_{2}$RuO$_{4}$ model with finite interorbital hybridization). As a side remark, we would like to mention that in the Sr$_{2}$RuO$_{4}$ model used in Ref.~\onlinecite{SMroemer2019}, the SOC term involves a projection of the angular momentum operator to the $t_{2g}$ manifold. One can show that the corresponding matrices satisfy an angular momentum $ \ell = 1$ algebra when the sign of one of the components of the angular momentum operator is flipped. The sign flip in turn can be absorbed in a redefinition of the spin frame. Taking this additional structure into account leads to a slightly different transformation rule than the one written down in \Eqref{eq:trans_rule}.

\section{Fluctuation-Mediated Pairing Vertex in the Random-Phase Approximation}
\label{sec:vertex}

Here, for the sake of providing a self-contained supplementary to the main text, we briefly outline the construction of the fluctuation mediated pairing vertex in the random-phase approximation (RPA). We start out by providing the definition of the microscopic interaction Hamiltonian $ H_{\mathrm{int}} $ of Hubbard-Hund type. It reads as
\be
\label{eq:interaction}
H_{\mathrm{int}} & = &  
U \sum_{l,i,\mu} n_{l i \mu \uparrow} n_{l i \mu \downarrow} + 
\left(U^{\prime} \!-\! \frac{J}{2}\right) \sum_{l,i,\mu < \nu, \sigma,\sigma^{\prime}} n_{li \mu \sigma} n_{li \nu \sigma^{\prime}}
- 2 J \!\! \sum_{l,i, \mu < \nu}{\bf S}_{li\mu}\cdot{\bf S}_{li\nu}  + 
\frac{J^{\prime}}{2}\!\!\!\! \sum_{l,i, \mu \neq \nu,\sigma}\left(c_{li\mu\sigma}^{\dagger}c_{li\mu\bar{\sigma}}^{\dagger}c_{li\nu\bar{\sigma}}c_{li\nu\sigma}+\mathrm{h.c.}\right). \nn
\ee
The Hamiltonian \Eqref{eq:interaction} is parameterized by an intraorbital Hubbard $U$, an interorbital coupling $U^{\prime}$, Hund's coupling $J$ and pair hopping $J^{\prime}$, satisfying $U^{\prime} = U - 2J$, $J = J^{\prime}$ due to orbital rotational invariance of the Coulomb matrix elements with respect to the Wannier basis functions. The operators for local charge and spin are $n_{li\mu} = n_{li\mu\uparrow} + n_{li\mu\downarrow}$ with $n_{li\mu\sigma} = c_{li\mu\sigma}^{\dagger} c_{li\mu\sigma}$ and ${\bf S}_{li\mu} = 1/2\sum_{\sigma\sigma^{\prime}} c_{li\mu\sigma}^{\dagger} {\boldsymbol \sigma}_{\sigma\sigma^{\prime}}c_{li\mu\sigma^{\prime}}$, respectively. At the level of the RPA, the bare interaction vertex defined by the interaction Hamiltonian above provides the 2-particle irreducible (2PI) vertex in the particle-hole channel. Concretely, we can re-write the interaction Hamiltonian as (normal-ordering is implied)
\be
H_{\mathrm{int}} =
-\frac{1}{2}\sum_{i} \sum_{l_{1},\dots ,l_{4}}\sum_{\mu_{1},\dots ,\mu_{4}} \sum_{\sigma_{1},\dots ,\sigma_{4}}
[U]^{l_{1}\mu_{1}\sigma_{1};l_{2}\mu_{2}\sigma_{2}}_{l_{3}\mu_{3}\sigma_{3};l_{4}\mu_{4}\sigma_{4}}
c_{l_{1}i\mu_{1}\sigma_{1}}^{\dagger}
c_{l_{2}i\mu_{2}\sigma_{2}}
c_{l_{3}i\mu_{3}\sigma_{3}}^{\dagger}
c_{l_{4}i\mu_{4}\sigma_{4}},
\ee
where  $ [U]^{l_{1}\mu_{1}\sigma_{1};l_{2}\mu_{2}\sigma_{2}}_{l_{3}\mu_{3}\sigma_{3};l_{4}\mu_{4}\sigma_{4}} $ denotes the bare particle-hole vertex. Within the RPA approach to pairing, it describes how single-particle excitations scatter of collective excitations.

As the interaction Hamiltonian \Eqref{eq:interaction} is spin-rotation invariant, the bare interaction vertex can be decomposed into charge ($ U_{\mathrm{c}} $) and spin vertices ($ U_{\mathrm{s}} $) in the particle-hole channel as
\be
[U]^{l_{1}\mu_{1}\sigma_{1};l_{2}\mu_{2}\sigma_{2}}_{l_{3}\mu_{3}\sigma_{3};l_{4}\mu_{4}\sigma_{4}}
= - \frac{1}{2} \delta_{l_{1},l_{2}}\delta_{l_{2},l_{3}}\delta_{l_{3},l_{4}}
\left( 
[U_{\mathrm{c}}]^{\mu_{1}\mu_{2}}_{\mu_{3}\mu_{4}} \,
\delta_{\sigma_{1}\sigma_{2}}\delta_{\sigma_{3}\sigma_{4}}  - 
[U_{\mathrm{s}}]^{\mu_{1}\mu_{2}}_{\mu_{3}\mu_{4}} \,
{\boldsymbol \sigma}_{\sigma_{1}\sigma_{2}} \cdot {\boldsymbol \sigma}_{\sigma_{3}\sigma_{4}}
\right),
\ee
which in turn are defined by
\be
[U_{\mathrm{s}}]^{\mu\mu}_{\mu\mu} = U,
\quad
[U_{\mathrm{s}}]^{\nu\mu}_{\mu\nu} = U^{\prime},
\quad
[U_{\mathrm{s}}]^{\nu\nu}_{\mu\mu} = J,
\quad
[U_{\mathrm{s}}]^{\mu\nu}_{\mu\nu} = J^{\prime},
\quad\text{with}\,\mu \neq \nu,
\ee
and
\be
[U_{\mathrm{c}}]^{\mu\mu}_{\mu\mu} = U,
\quad
[U_{\mathrm{c}}]^{\nu\mu}_{\mu\nu} = 2J - U^{\prime},
\quad
[U_{\mathrm{c}}]^{\nu\nu}_{\mu\mu} = 2U^{\prime}-J,
\quad
[U_{\mathrm{c}}]^{\mu\nu}_{\mu\nu} = J^{\prime},
\quad\text{with}\,\mu \neq \nu,
\ee
and zero otherwise.

Building on the formalism developed in Refs.~\onlinecite{SMscherer2016} and \onlinecite{SMscherer2018}, we can formulate the RPA propagator for collective particle-hole fluctuations as
\be 
[\mathcal{D}(\mathrm{i}\omega_{n},{\bf q})]_{X_{3};X_{4}}^{X_{1};X_{2}} = 
[U]_{X_{3};X_{4}}^{X_{1};X_{2}} + 
[ U \, [\mathbbm{1} - \chi_{0}(\mathrm{i}\omega_{n},{\bf q}) U]^{-1} \, 
\chi_{0}(\mathrm{i}\omega_{n},{\bf q}) U ]_{X_{3};X_{4}}^{X_{1};X_{2}},
\ee
where we introduced the combined index $X \equiv (l,\mu,\sigma)$ by collecting sublattice, orbital and spin indices. In the equation above $ \chi_{0}(\mathrm{i}\omega_{n},{\bf q}) $ denotes the irreducible bubble in the particle-hole channel, defined as
\be
\label{eq:bare_bubble} 
[\chi_{0}(\mathrm{i}\omega_n,{\bf q})]^{X_1;X_2}_{X_3;X_4} =
\frac{1}{\mathcal{N}}\int_{0}^{\beta} \! d\tau \, \mathrm{e}^{\mathrm{i}\omega_n \tau}\sum_{{\bf k},{\bf k}^{\prime}}
\langle \mathcal{T}_{\tau} 
c_{{\bf k} - {\bf q}X_1}^{\dagger}(\tau) 
c_{{\bf k}X_2}(\tau)
c_{{\bf k}^{\prime} + {\bf q}^{\prime}X_3}^{\dagger}(0) 
c_{{\bf k}^{\prime}X_4}(0)
\rangle_{c,0}.
\ee
Vertices and bubbles are to be understood as matrices in sublattice and orbital space. Their product is defined as
\be
[A \, B]^{X_{1};X_{2}}_{X_{3};X_{4}} = 
\sum_{Y_{1},Y_{2}}
[A]^{X_{1};X_{2}}_{Y_{1};Y_{2}} \, [B]^{Y_{2};Y_{3}}_{X_{3};X_{4}}.
\ee
In \Eqref{eq:bare_bubble}, the symbol $\mathcal{T}_{\tau}$ denotes the time-ordering operator with respect to the imaginary-time variable $\tau \in [0,\beta)$, with $\beta$ the inverse temperature. The particle-hole propagator $ \mathcal{D}(\mathrm{i}\omega_{n},{\bf q}) $ can be expressed in terms of the generalized RPA susceptibility $ \chi(\mathrm{i}\omega_n,{\bf q}) $ which is the solution to the equation
\be 
\label{eq:RPAequation}
[\chi(\mathrm{i}\omega_n,{\bf q}) ]^{X_1;X_2}_{X_3; X_4}=
[\chi_{0}(\mathrm{i}\omega_n,{\bf q})]^{X_1;X_2}_{X_3;X_4} +
[\chi_{0}(\mathrm{i}\omega_n,{\bf q}) \,
U \,
\chi(\mathrm{i}\omega_n,{\bf q})]^{X_1;X_2}_{X_3;X_4}.
\ee
We can then write the RPA propagator for collective fluctuations as
\be 
[\mathcal{D}(\mathrm{i}\omega_{n},{\bf q})]_{X_{3};X_{4}}^{X_{1};X_{2}} = 
[U]_{X_{3};X_{4}}^{X_{1};X_{2}} + 
[ U \, \chi(\mathrm{i}\omega_{n},{\bf q}) \, U ]_{X_{3};X_{4}}^{X_{1};X_{2}}.
\ee
Assuming the electrons are coupled to the collective fluctuations, the effective, fluctuation-mediated interaction (in other words the 2PI vertex in the Cooper channel) in the static approximation can be obtained as
\be
\label{eq:vertex_definition}
[\Gamma^{\varsigma,\varsigma^{\prime}}({\bf k},{\bf k}^{\prime})]^{l_{1}\mu_{1};l_2\mu_{2}}_{l_{3}\mu_{3};l_4\mu_{4}} =
\frac{s_{\varsigma^{\prime}}}{2} 
\sum_{\sigma_{1},\dots,\sigma_{4}}
[\Gamma_{\varsigma}]_{\sigma_{2}\sigma_{1}}
\left[
[\mathcal{D}(0,{\bf k} + {\bf k}^{\prime})]^{l_{1}\mu_{1}\sigma_{1};l_{3}\mu_{3}\sigma_{3}}_{l_{2}\mu_{2}\sigma_{2};l_{4}\mu_{4}\sigma_{4}}
-
[\mathcal{D}(0,{\bf k} - {\bf k}^{\prime})]^{l_{1}\mu_{1}\sigma_{1};l_{4}\mu_{4}\sigma_{4}}_{l_{2}\mu_{2}\sigma_{2};l_{3}\mu_{3}\sigma_{3}}
\right]
[\Gamma_{\varsigma^{\prime}}]_{\sigma_{4}\sigma_{3}}.
\ee
In \Eqref{eq:vertex_definition}, we have explicitly written out the antisymmetrization of the 2PI vertex with respect to momenta and sublattice, orbital and spin indices due to the symmetries of fermionic vertices under exchange of quantum numbers.

\newpage

\section{LaFeAsO: Spin Susceptibility and Pairing Solution in Sublattice $ \otimes $ Orbital Space}
\label{sec:orbital}

Here, we collect some additional plots for the LaFeAsO model for various dopings, showing the static charge and spin susceptibility, as well as plots for the leading pairing solution as obtained from the Fermi surface projected LGE, transformed to sublattice $ \otimes $ orbital space by virtue of Eqs.~\ref{eq:transformation_1} and ~\ref{eq:transformation_2}.


\vfill

\begin{figure}[h!]
\centering
\begin{minipage}{0.80\columnwidth}
\centering
\includegraphics[width=1\columnwidth]{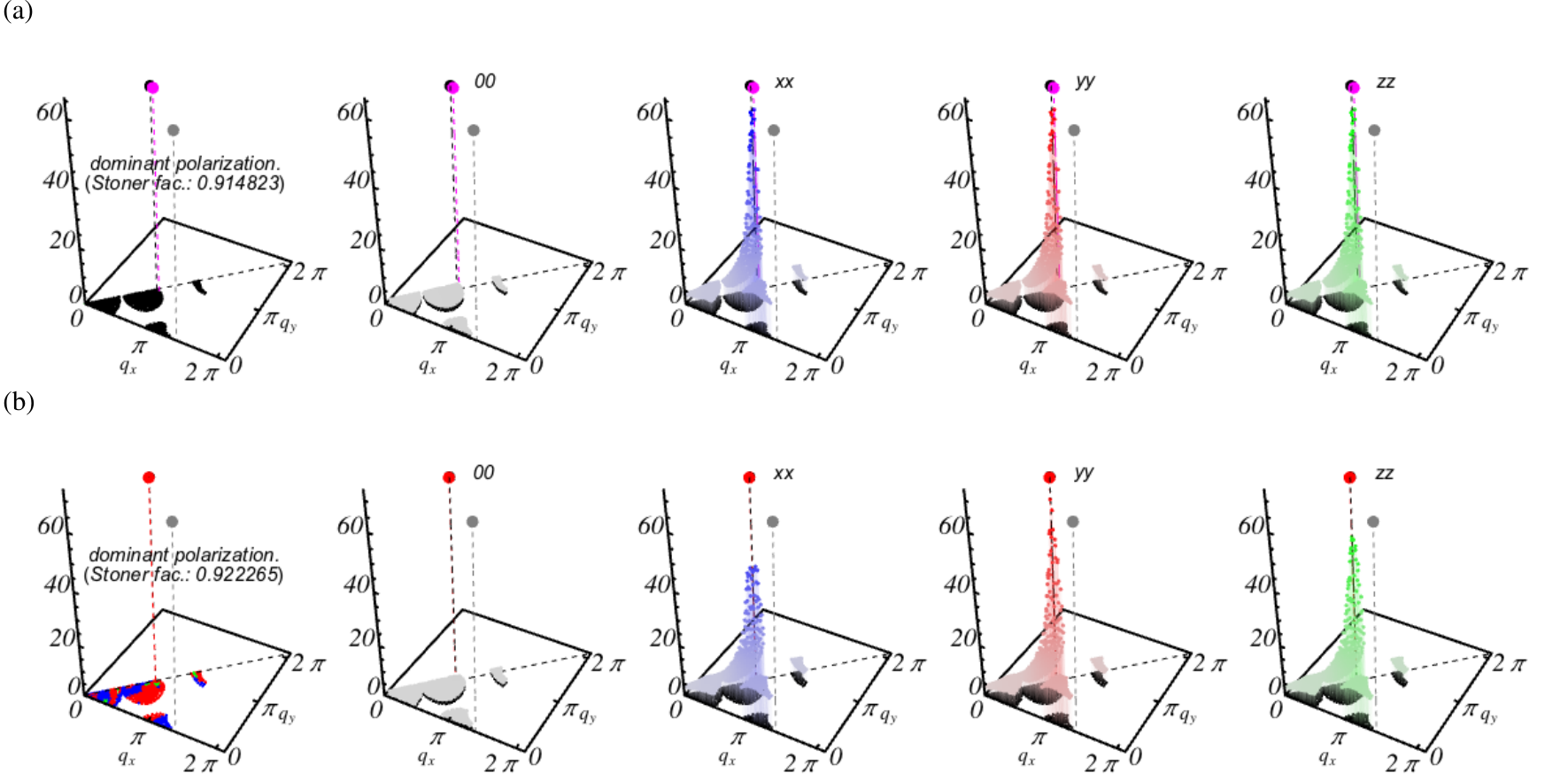}
\end{minipage}
\caption{Static charge ($00$) spin susceptibility (polarizations $xx$, $yy$, $zz$) in a section of momentum space (with 2-Fe BZ coordinates) for (a) $ \lambda_{\mathrm{SOC}} = 0\,$meV, and (b) $ \lambda_{\mathrm{SOC}} = 75\,$meV, with $\mu_{0} = 0\,$eV, $ T = 0.01\, $eV, $ U = 0.80 \, $eV. We note that the susceptibilities are shown only for momenta that actually enter the Fermi surface projected LGE. The first plot in each row shows the dominant spin polarization at a given momentum. The black dot together with the black dotted line shows marks the momentum $ {\bf Q}_{2} $, while the grey dot with the grey dashed line marks $ {\bf Q} = (\pi,\pi) $ (in 1-Fe notation). Magenta dot and line in (a) and red dot and line in (b) mark the momentum with largest spin susceptibility component (here $xx$).}
\label{fig:susceptibility_undoped}
\end{figure}

\vfill

\begin{figure}[h!]
\centering
\begin{minipage}{0.90\columnwidth}
\centering
\includegraphics[width=1\columnwidth]{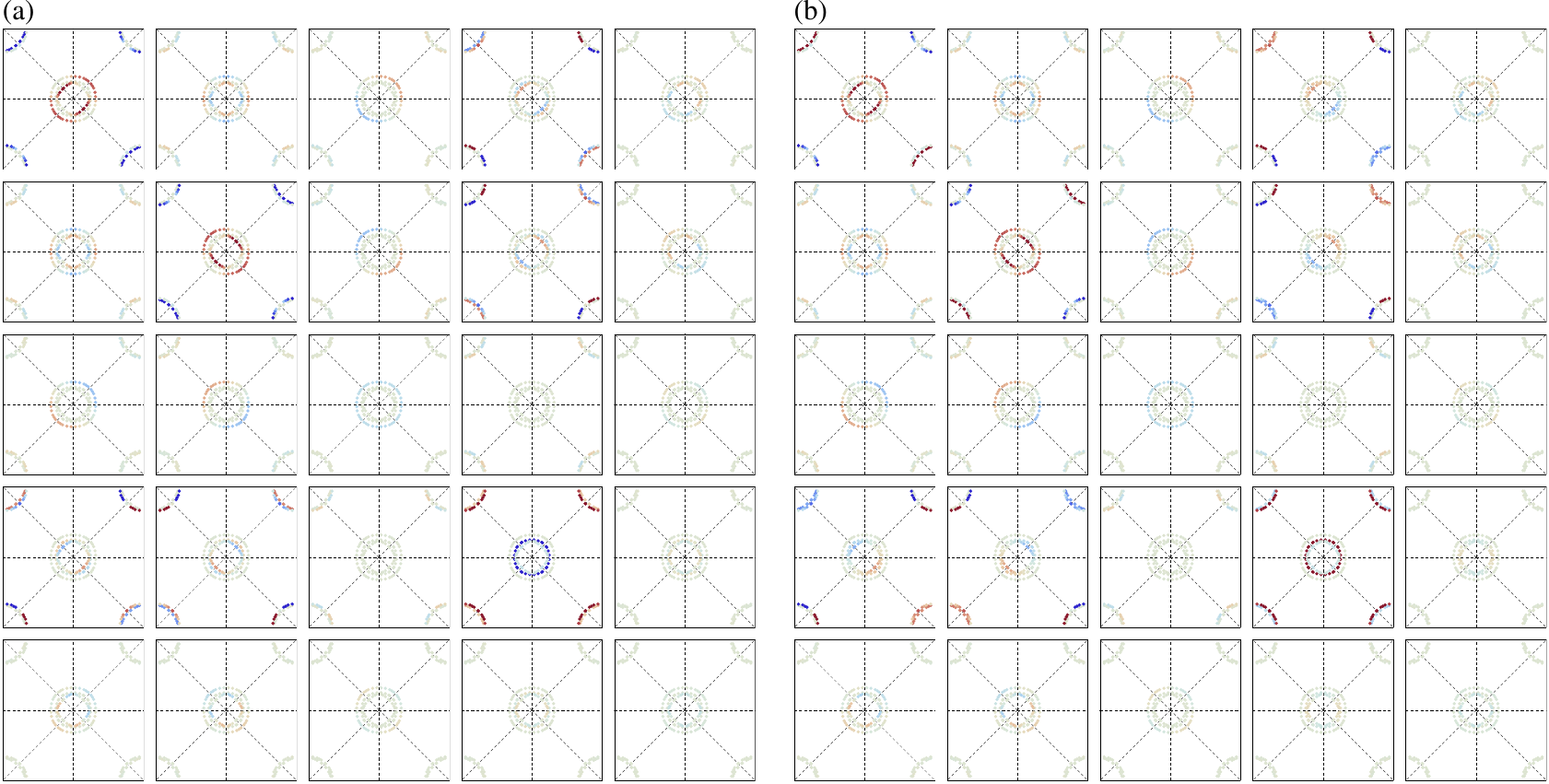}
\end{minipage}
\caption{Visualization of the spin-singlet part of the $ s_{+-} $ pairing solution in sublattice $ \otimes $ orbital space in the 2-Fe BZ for (a) intrasublattice ($ l = A, l^{\prime} = A $) and (b) intersublattice ($ l = A, l^{\prime} = B $) components, where the orbital components are ordered as $ d_{xz} $, $ d_{yz} $, $d_{x^{2}-y^{2}}$, $d_{xy}$, $d_{3z^2-r^2}$. The LGE was solved for $ \mu_{0} = 0 \, $eV, $ \lambda = 0\, $eV, $ T = 0.01 \, $eV, $ U = 0.80 \,$eV, $ J = U/4$.}
\label{fig:spin_singlet_1_undoped}
\end{figure}

\vfill

\begin{figure}[h!]
\centering
\begin{minipage}{0.90\columnwidth}
\centering
\includegraphics[width=1\columnwidth]{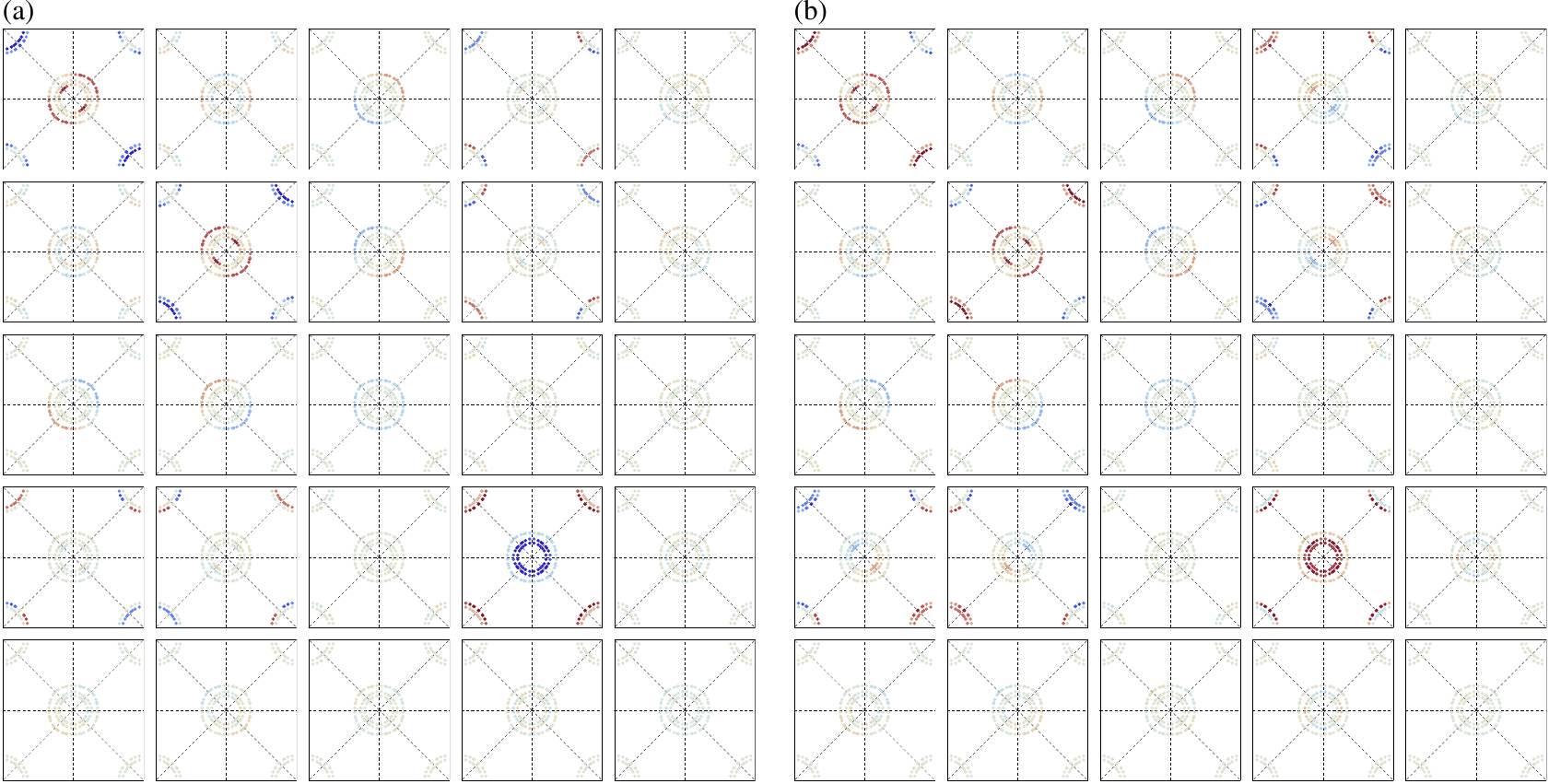}
\end{minipage}
\caption{Visualization of the spin-singlet part of the $ s_{+-} $ pairing solution in sublattice $ \otimes $ orbital space in the 2-Fe BZ for (a) intrasublattice ($ l = A, l^{\prime} = A $) and (b) intersublattice ($ l = A, l^{\prime} = B $) components, where the orbital components are ordered as $ d_{xz} $, $ d_{yz} $, $d_{x^{2}-y^{2}}$, $d_{xy}$, $d_{3z^2-r^2}$. The LGE was solved for $ \mu_{0} = 0 \, $eV, $ \lambda = 75\, $meV, $ T = 0.01 \, $eV, $ U = 0.80 \,$eV, $ J = U/4$.}
\label{fig:spin_singlet_2_undoped}
\end{figure}

\vfill

\begin{figure}[h!]
\centering
\begin{minipage}{0.90\columnwidth}
\centering
\includegraphics[width=1\columnwidth]{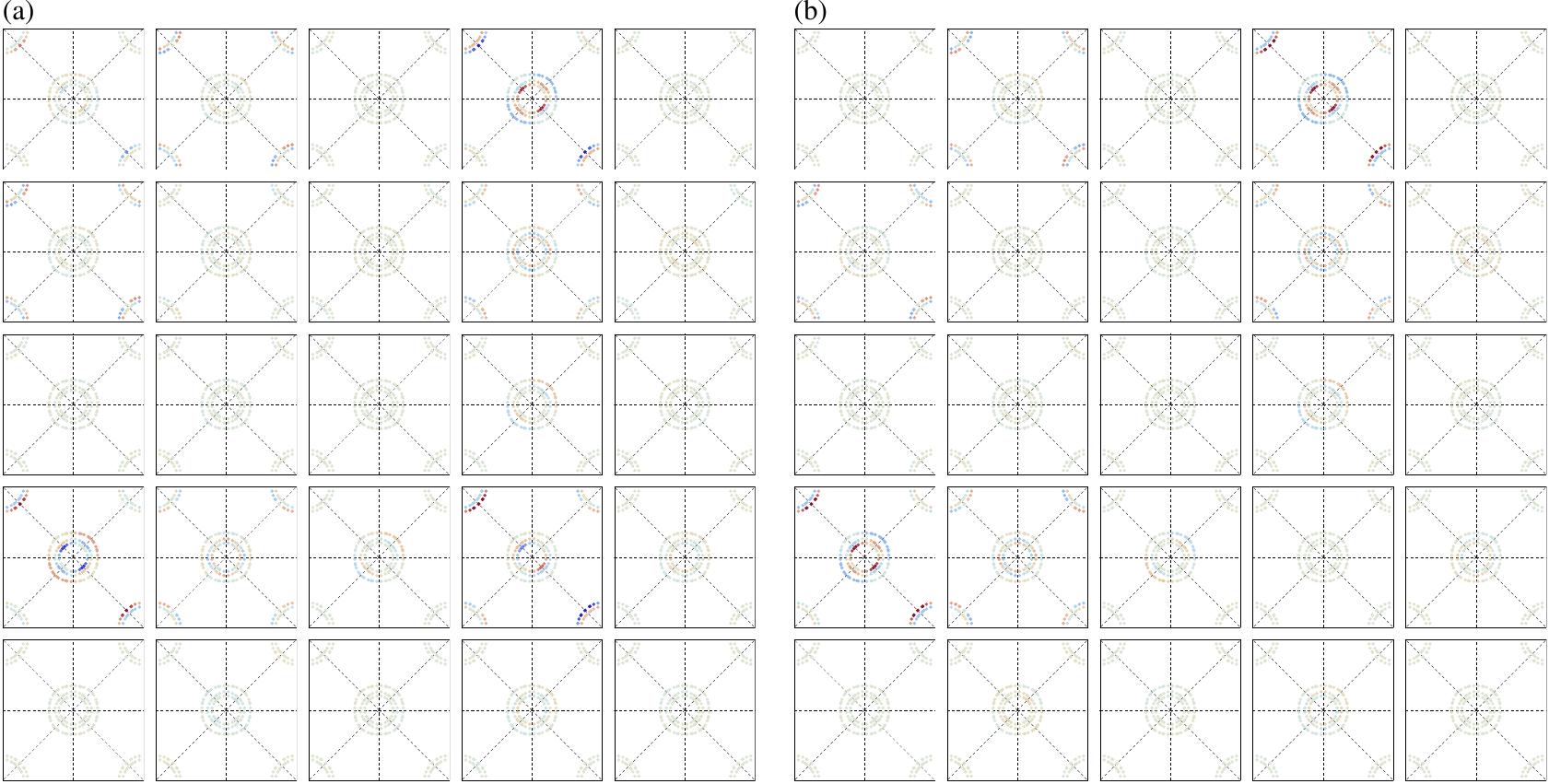}
\end{minipage}
\caption{Visualization of the spin-triplet-$x$ part of the $ s_{+-} $ pairing solution in sublattice $ \otimes $ orbital space in the 2-Fe BZ for (a) intrasublattice ($ l = A, l^{\prime} = A $) and (b) intersublattice ($ l = A, l^{\prime} = B $) components, where the orbital components are ordered as $ d_{xz} $, $ d_{yz} $, $d_{x^{2}-y^{2}}$, $d_{xy}$, $d_{3z^2-r^2}$. The LGE was solved for $ \mu_{0} = 0 \, $eV, $ \lambda = 75\, $meV, $ T = 0.01 \, $eV, $ U = 0.80 \,$eV, $ J = U/4$.}
\label{fig:spin_triplet_2_undoped}
\end{figure}
%


\vfill

\begin{figure}[h!]
\centering
\begin{minipage}{0.80\columnwidth}
\centering
\includegraphics[width=1\columnwidth]{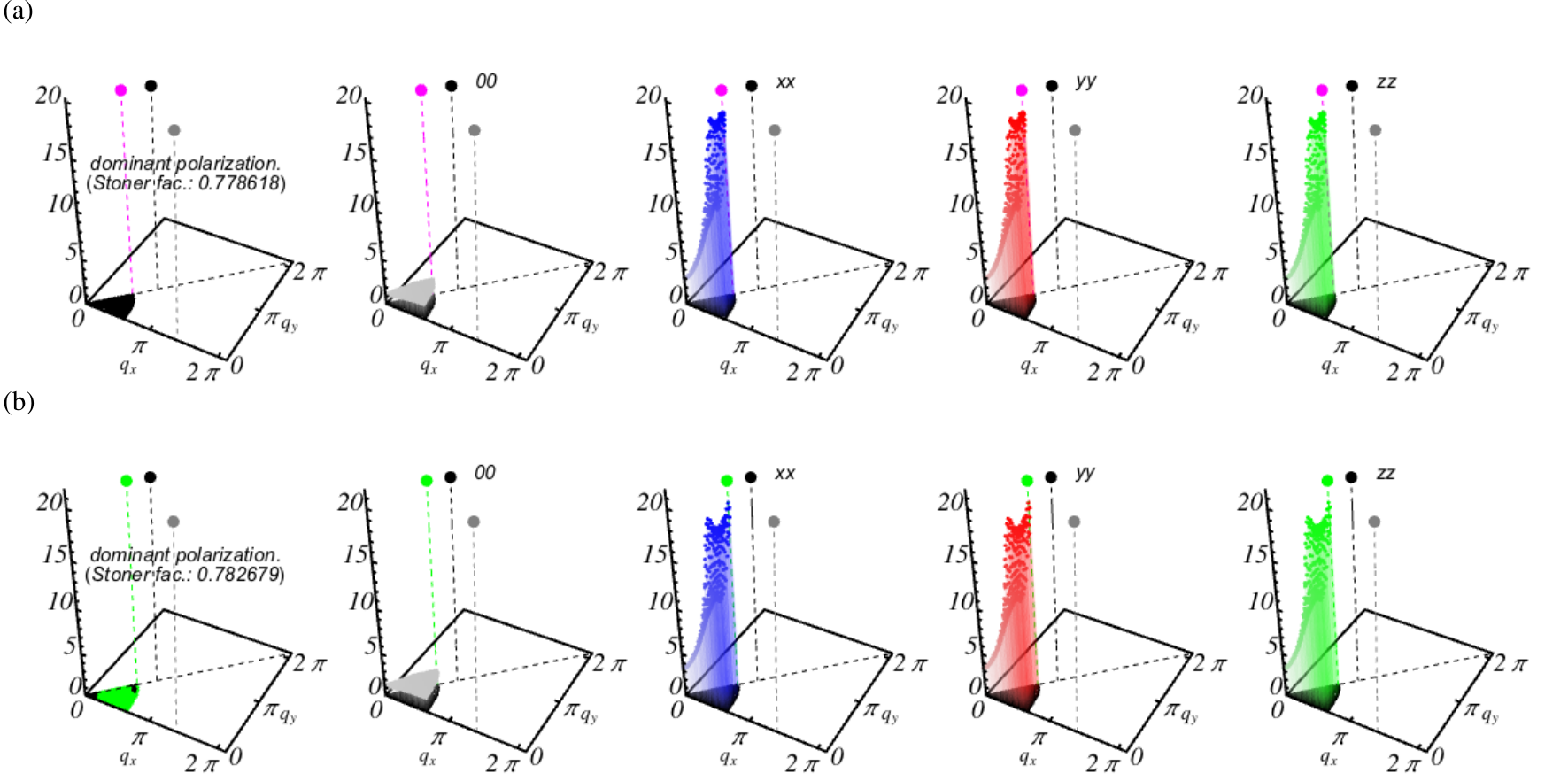}
\end{minipage}
\caption{Static charge ($00$) spin susceptibility (polarizations $xx$, $yy$, $zz$) in a section of momentum space (with 2-Fe BZ coordinates) for (a) $ \lambda_{\mathrm{SOC}} = 0\,$meV, and (b) $ \lambda_{\mathrm{SOC}} = 75\,$meV, with $\mu_{0} = -150\,$meV, $ T = 0.01\, $eV, $ U = 0.80 \, $eV. We note that the susceptibilities are shown only for momenta that actually enter the Fermi surface projected LGE. The first plot in each row shows the dominant spin polarization at a given momentum. The black dot together with the black dotted line shows marks the momentum $ {\bf Q}_{2} $, while the grey dot with the grey dashed line marks $ {\bf Q} = (\pi,\pi) $ (in 1-Fe notation). Magenta dot and line in (a) and green dot and line in (b) mark the momentum with largest spin susceptibility (here $yy$).}
\label{fig:susceptibility_heavily_hdoped}
\end{figure}

\vfill

\begin{figure}[h!]
\centering
\begin{minipage}{0.90\columnwidth}
\centering
\includegraphics[width=1\columnwidth]{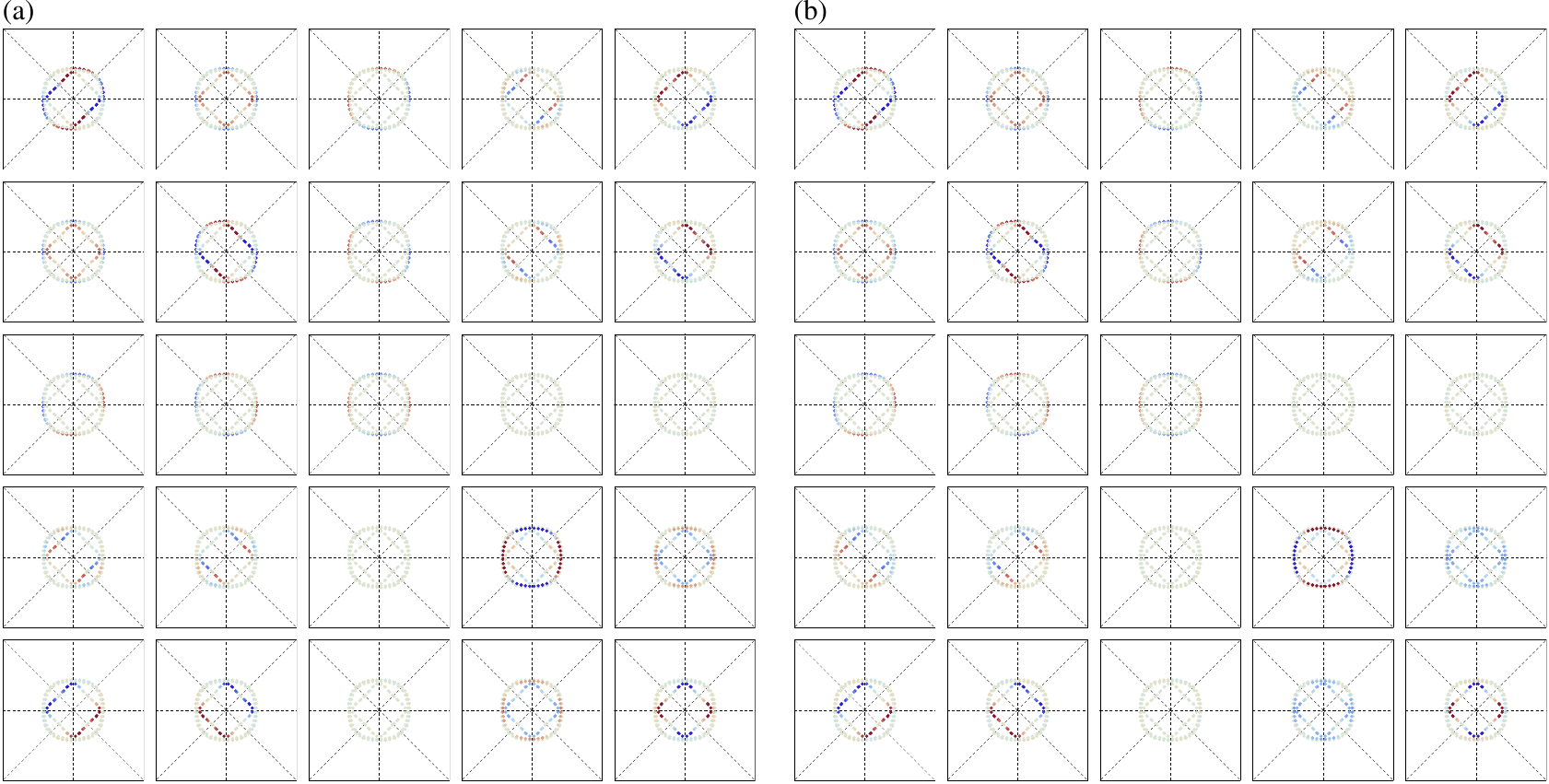}
\end{minipage}
\caption{Visualization of the spin-singlet part of the $ d $-wave pairing solution in sublattice $ \otimes $ orbital space in the 2-Fe BZ for (a) intrasublattice ($ l = A, l^{\prime} = A $) and (b) intersublattice ($ l = A, l^{\prime} = B $) components, where the orbital components are ordered as $ d_{xz} $, $ d_{yz} $, $d_{x^{2}-y^{2}}$, $d_{xy}$, $d_{3z^2-r^2}$. The LGE was solved for $ \mu_{0} = -150 \, $meV, $ \lambda = 0\, $eV, $ T = 0.01 \, $eV, $ U = 0.80 \,$eV, $ J = U/4$.}
\label{fig:spin_singlet_1_heavily_hdoped}
\end{figure}

\vfill

\begin{figure}[h!]
\centering
\begin{minipage}{0.90\columnwidth}
\centering
\includegraphics[width=1\columnwidth]{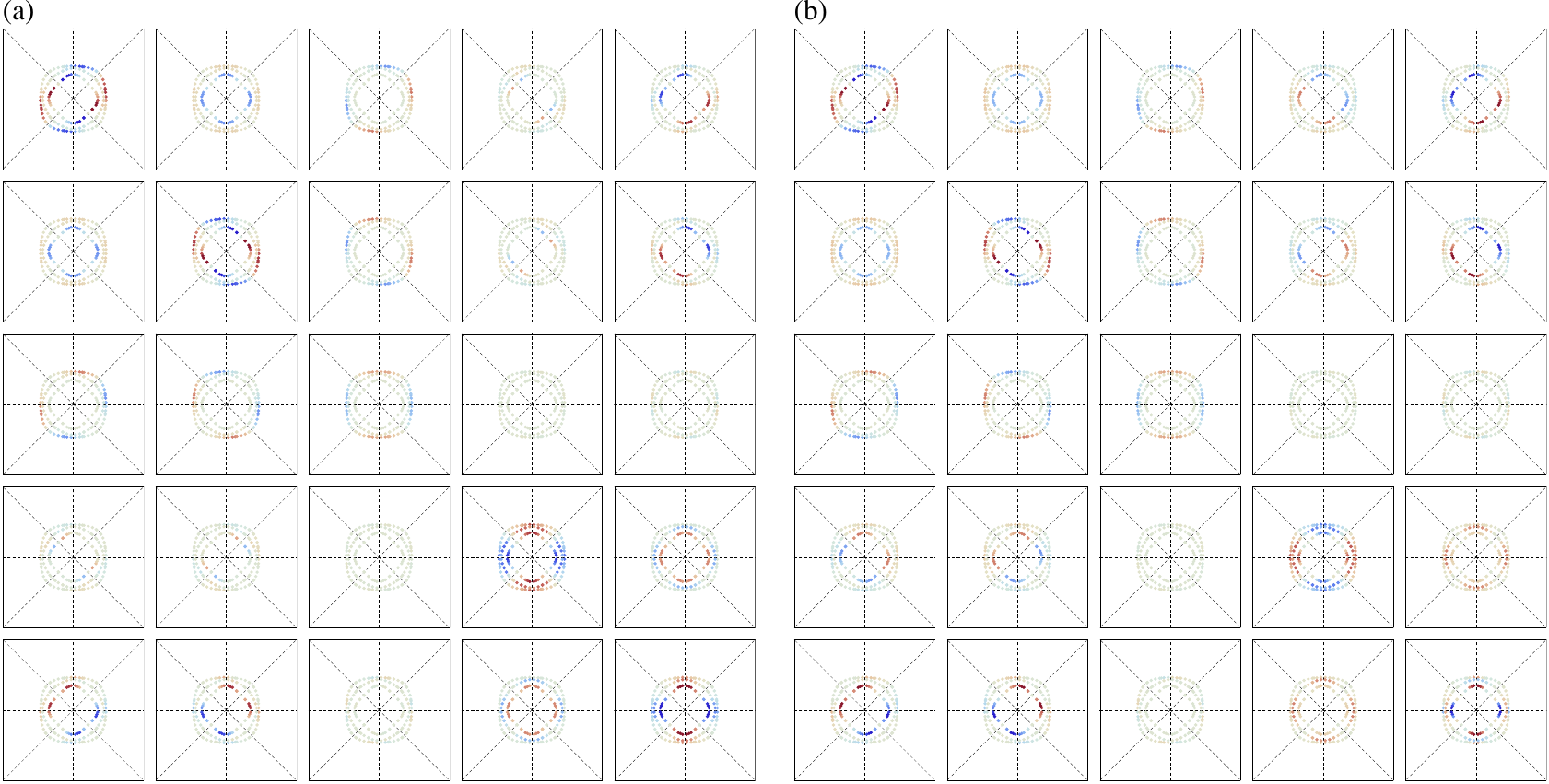}
\end{minipage}
\caption{Visualization of the spin-singlet part of the $ d $-wave pairing solution in sublattice $ \otimes $ orbital space in the 2-Fe BZ for (a) intrasublattice ($ l = A, l^{\prime} = A $) and (b) intersublattice ($ l = A, l^{\prime} = B $) components, where the orbital components are ordered as $ d_{xz} $, $ d_{yz} $, $d_{x^{2}-y^{2}}$, $d_{xy}$, $d_{3z^2-r^2}$. The LGE was solved for $ \mu_{0} = -150 \, $meV, $ \lambda = 75\, $meV, $ T = 0.01 \, $eV, $ U = 0.80 \,$eV, $ J = U/4$.}
\label{fig:spin_singlet_2_heavily_hdoped}
\end{figure}

\vfill

\begin{figure}[h!]
\centering
\begin{minipage}{0.90\columnwidth}
\centering
\includegraphics[width=1\columnwidth]{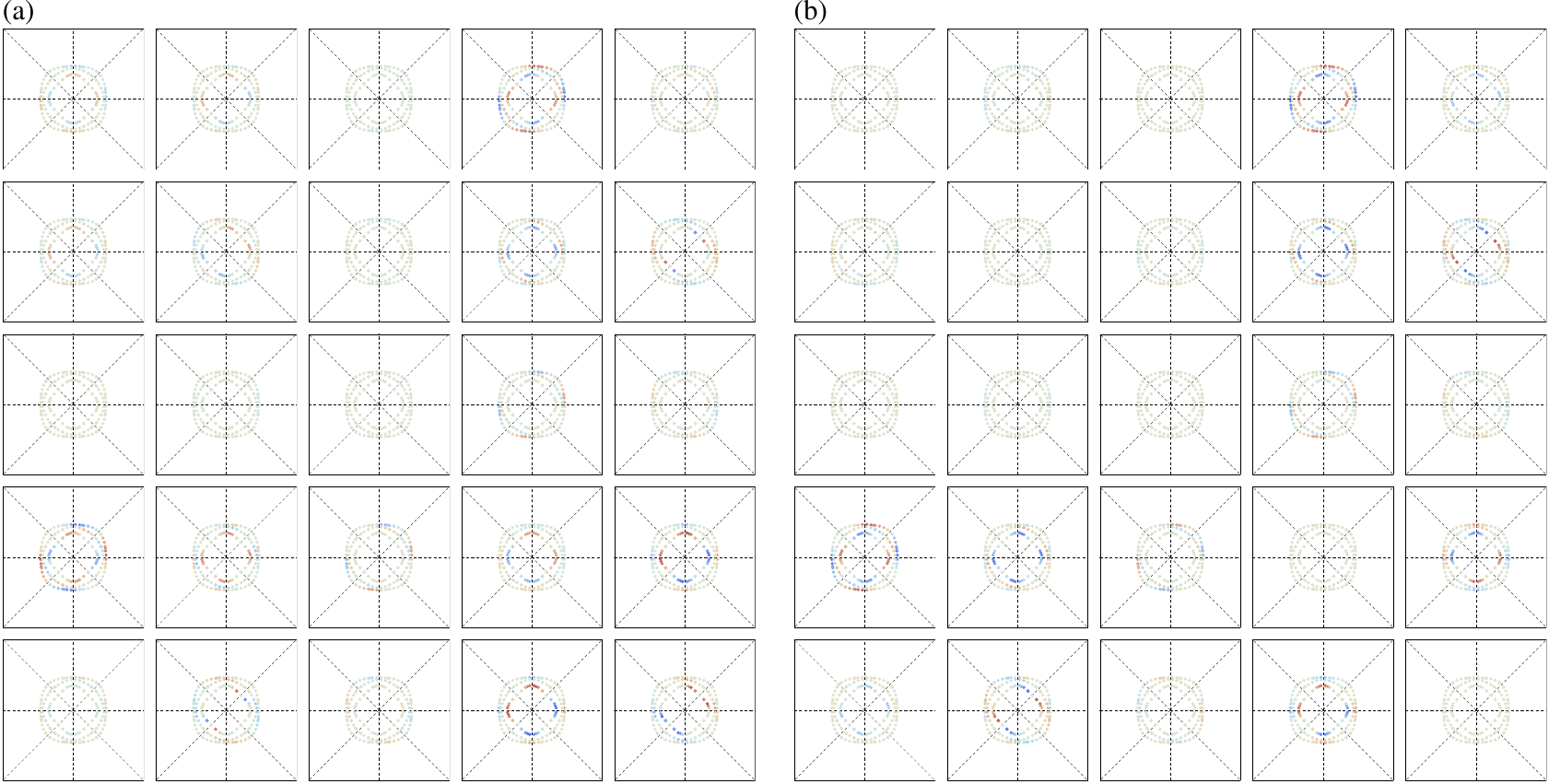}
\end{minipage}
\caption{Visualization of the spin-triplet-$x$ part of the $ d $-wave pairing solution in sublattice $ \otimes $ orbital space in the 2-Fe BZ for (a) intrasublattice ($ l = A, l^{\prime} = A $) and (b) intersublattice ($ l = A, l^{\prime} = B $) components, where the orbital components are ordered as $ d_{xz} $, $ d_{yz} $, $d_{x^{2}-y^{2}}$, $d_{xy}$, $d_{3z^2-r^2}$. The LGE was solved for $ \mu_{0} = -150 \, $meV, $ \lambda = 75\, $meV, $ T = 0.01 \, $eV, $ U = 0.80 \,$eV, $ J = U/4$.}
\label{fig:spin_triplet_2_heavily_hdoped}
\end{figure}
%


\vfill

\begin{figure}[h!]
\centering
\begin{minipage}{0.80\columnwidth}
\centering
\includegraphics[width=1\columnwidth]{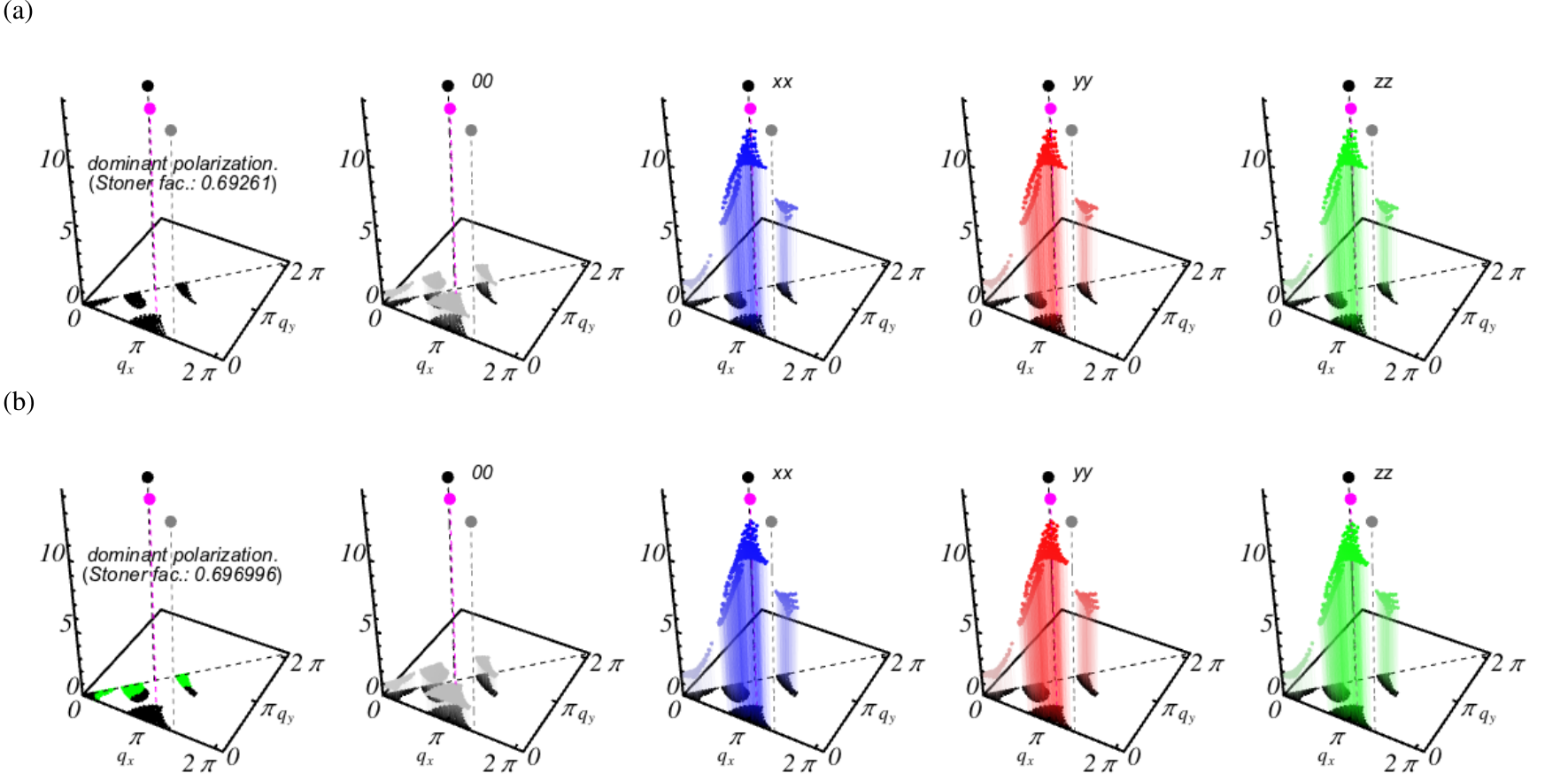}
\end{minipage}
\caption{Static charge ($00$) spin susceptibility (polarizations $xx$, $yy$, $zz$) in a section of momentum space (with 2-Fe BZ coordinates) for (a) $ \lambda_{\mathrm{SOC}} = 0\,$meV, and (b) $ \lambda_{\mathrm{SOC}} = 75\,$meV, with $\mu_{0} = 20\,$meV, $ T = 0.01\, $eV, $ U = 0.80 \, $eV. We note that the susceptibilities are shown only for momenta that actually enter the Fermi surface projected LGE. The first plot in each row shows the dominant spin polarization at a given momentum. The black dot together with the black dotted line shows marks the momentum $ {\bf Q}_{2} $, while the grey dot with the grey dashed line marks $ {\bf Q} = (\pi,\pi) $ (in 1-Fe notation). Magenta dot and line in (a) and magenta dot and line in (b) mark the momentum with largest spin susceptibility (here in-plane with equal $xx$ and $yy$ components).}
\label{fig:susceptibility_heavily_edoped}
\end{figure}

\vfill

\begin{figure}[h!]
\centering
\begin{minipage}{0.90\columnwidth}
\centering
\includegraphics[width=1\columnwidth]{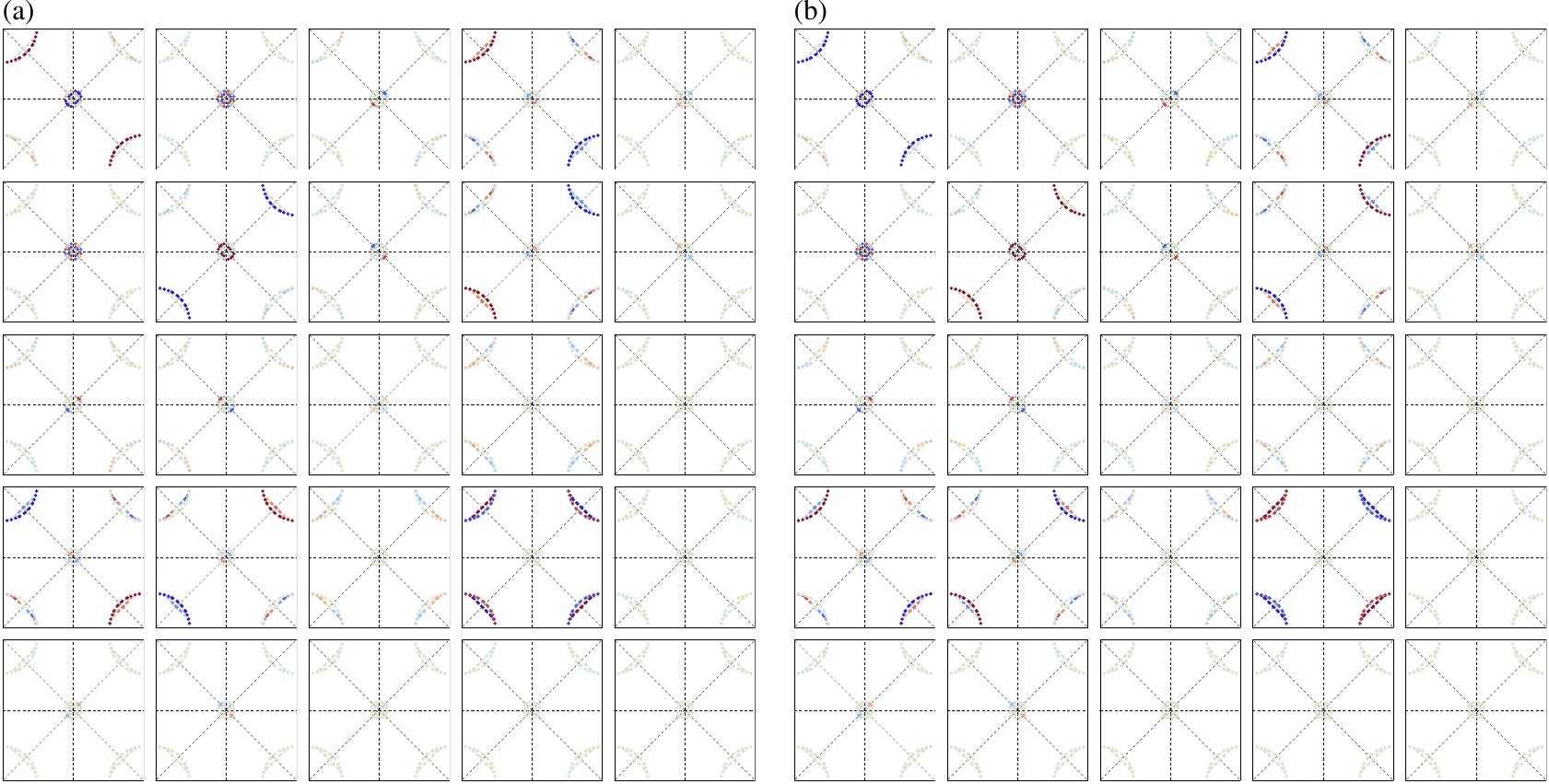}
\end{minipage}
\caption{Visualization of the spin-singlet part of the $ d $-wave pairing solution in sublattice $ \otimes $ orbital space in the 2-Fe BZ for (a) intrasublattice ($ l = A, l^{\prime} = A $) and (b) intersublattice ($ l = A, l^{\prime} = B $) components, where the orbital components are ordered as $ d_{xz} $, $ d_{yz} $, $d_{x^{2}-y^{2}}$, $d_{xy}$, $d_{3z^2-r^2}$. The LGE was solved for $ \mu_{0} = 20 \, $meV, $ \lambda = 0\, $eV, $ T = 0.01 \, $eV, $ U = 0.80 \,$eV, $ J = U/4$.}
\label{fig:spin_singlet_1_heavily_edoped}
\end{figure}

\vfill

\begin{figure}[h!]
\centering
\begin{minipage}{0.90\columnwidth}
\centering
\includegraphics[width=1\columnwidth]{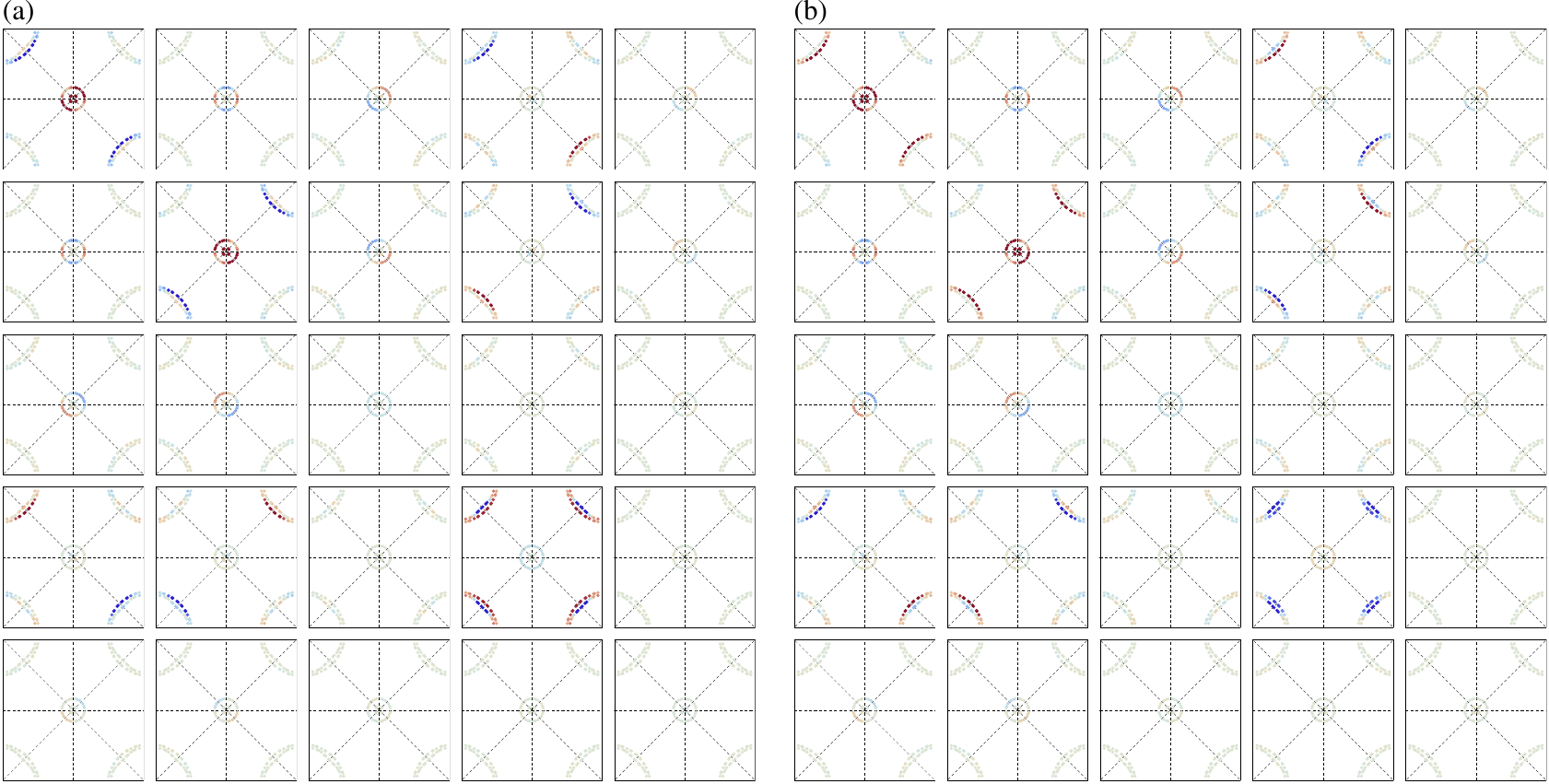}
\end{minipage}
\caption{Visualization of the spin-singlet part of the $ s $-wave pairing solution in sublattice $ \otimes $ orbital space in the 2-Fe BZ for (a) intrasublattice ($ l = A, l^{\prime} = A $) and (b) intersublattice ($ l = A, l^{\prime} = B $) components, where the orbital components are ordered as $ d_{xz} $, $ d_{yz} $, $d_{x^{2}-y^{2}}$, $d_{xy}$, $d_{3z^2-r^2}$. The LGE was solved for $ \mu_{0} = 20 \, $meV, $ \lambda = 75\, $meV, $ T = 0.01 \, $eV, $ U = 0.80 \,$eV, $ J = U/4$.}
\label{fig:spin_singlet_2_heavily_edoped}
\end{figure}

\vfill

\begin{figure}[h!]
\centering
\begin{minipage}{0.90\columnwidth}
\centering
\includegraphics[width=1\columnwidth]{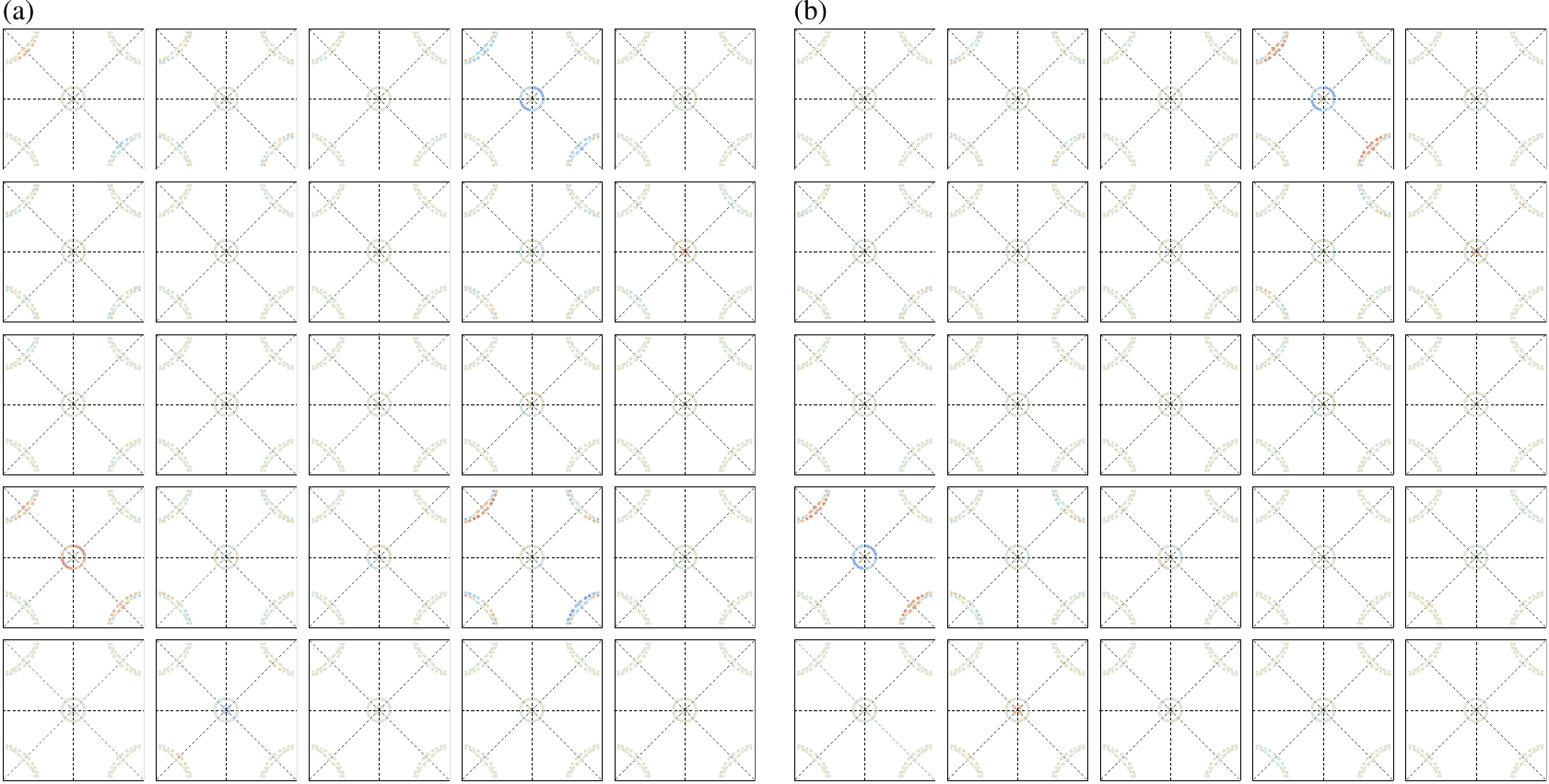}
\end{minipage}
\caption{Visualization of the spin-triplet-$x$ part of the $ s $-wave pairing solution in sublattice $ \otimes $ orbital space in the 2-Fe BZ for (a) intrasublattice ($ l = A, l^{\prime} = A $) and (b) intersublattice ($ l = A, l^{\prime} = B $) components, where the orbital components are ordered as $ d_{xz} $, $ d_{yz} $, $d_{x^{2}-y^{2}}$, $d_{xy}$, $d_{3z^2-r^2}$. The LGE was solved for $ \mu_{0} = 20 \, $meV, $ \lambda = 75\, $meV, $ T = 0.01 \, $eV, $ U = 0.80 \,$eV, $ J = U/4$.}
\label{fig:spin_triplet_2_heavily_edoped}
\end{figure}

\vfill

\clearpage

\section{FeSe: Spin Susceptibility and Pairing Solution in Sublattice $ \otimes $ Orbital Space}
\label{sec:orbital_FeSe}

Here, we collect some additional plots for the FeSe model, showing the static charge and spin susceptibility, as well as plots for the leading pairing solution as obtained from the Fermi surface projected LGE, transformed to sublattice $ \otimes $ orbital space by virtue of Eqs.~\ref{eq:transformation_1} and ~\ref{eq:transformation_2}.

\vfill

\begin{figure}[h!]
\centering
\begin{minipage}{0.80\columnwidth}
\centering
\includegraphics[width=1\columnwidth]{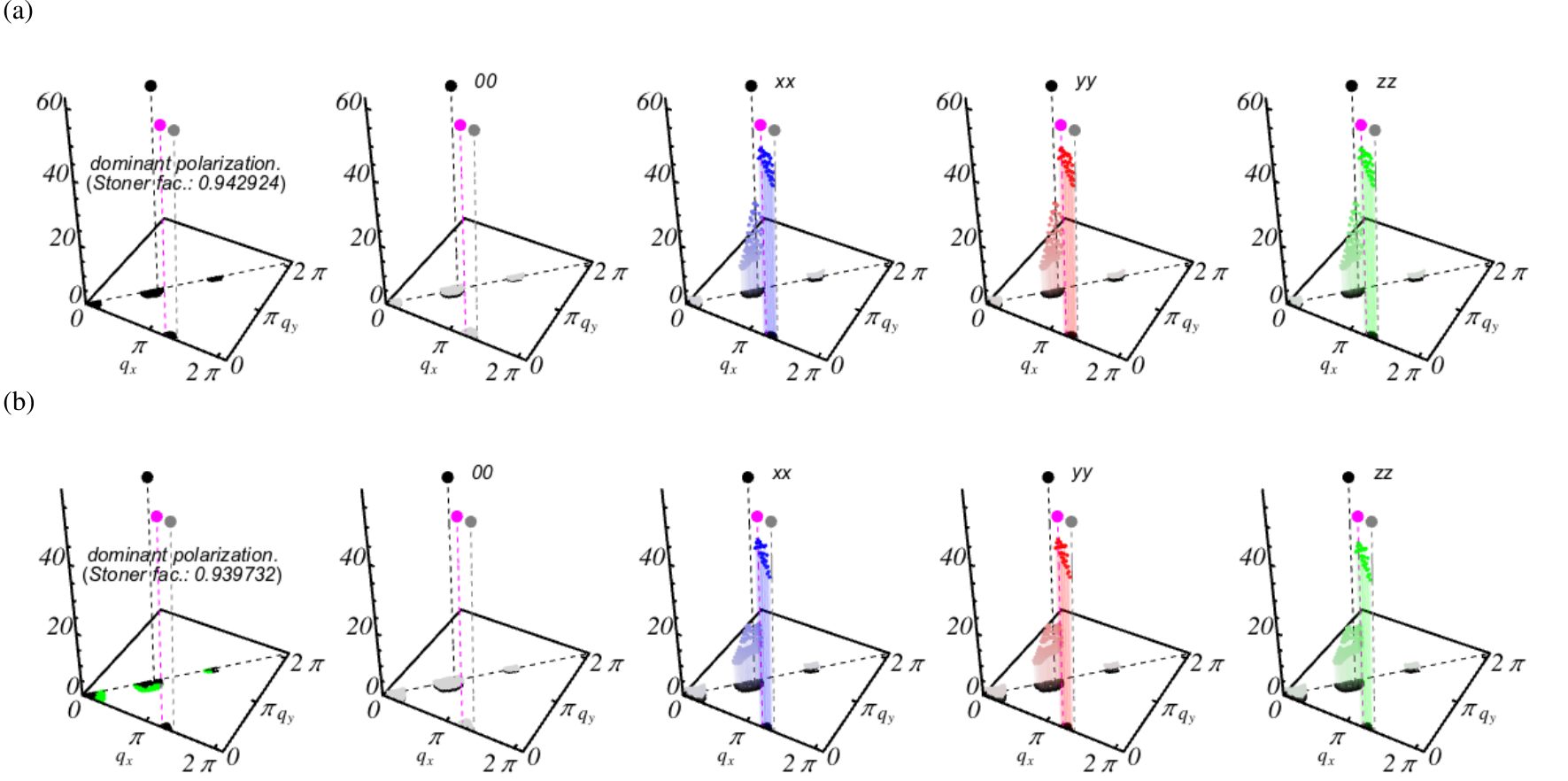}
\end{minipage}
\caption{Static charge ($00$) spin susceptibility (polarizations $xx$, $yy$, $zz$) in a section of momentum space (with 2-Fe BZ coordinates) for (a) $ \lambda_{\mathrm{SOC}} = 0\,$meV, and (b) $ \lambda_{\mathrm{SOC}} = 75\,$meV, with $\mu_{0} = 0\,$eV, $ T = 0.01\, $eV, $ U = 1.30 \, $eV. We note that the susceptibilities are shown only for momenta that actually enter the Fermi surface projected LGE. The first plot in each row shows the dominant spin polarization at a given momentum. The black dot together with the black dotted line shows marks the momentum $ {\bf Q}_{2} $, while the grey dot with the grey dashed line marks $ {\bf Q} = (\pi,\pi) $ (in 1-Fe notation). Magenta dot and line in (a) and magenta dot and line in (b) mark the momentum with largest spin susceptibility component (here in plane with equal $xx$ and $yy$ components).}
\label{fig:susceptibility_FeSe}
\end{figure}

\vfill

\begin{figure}[h!]
\centering
\begin{minipage}{0.90\columnwidth}
\centering
\includegraphics[width=1\columnwidth]{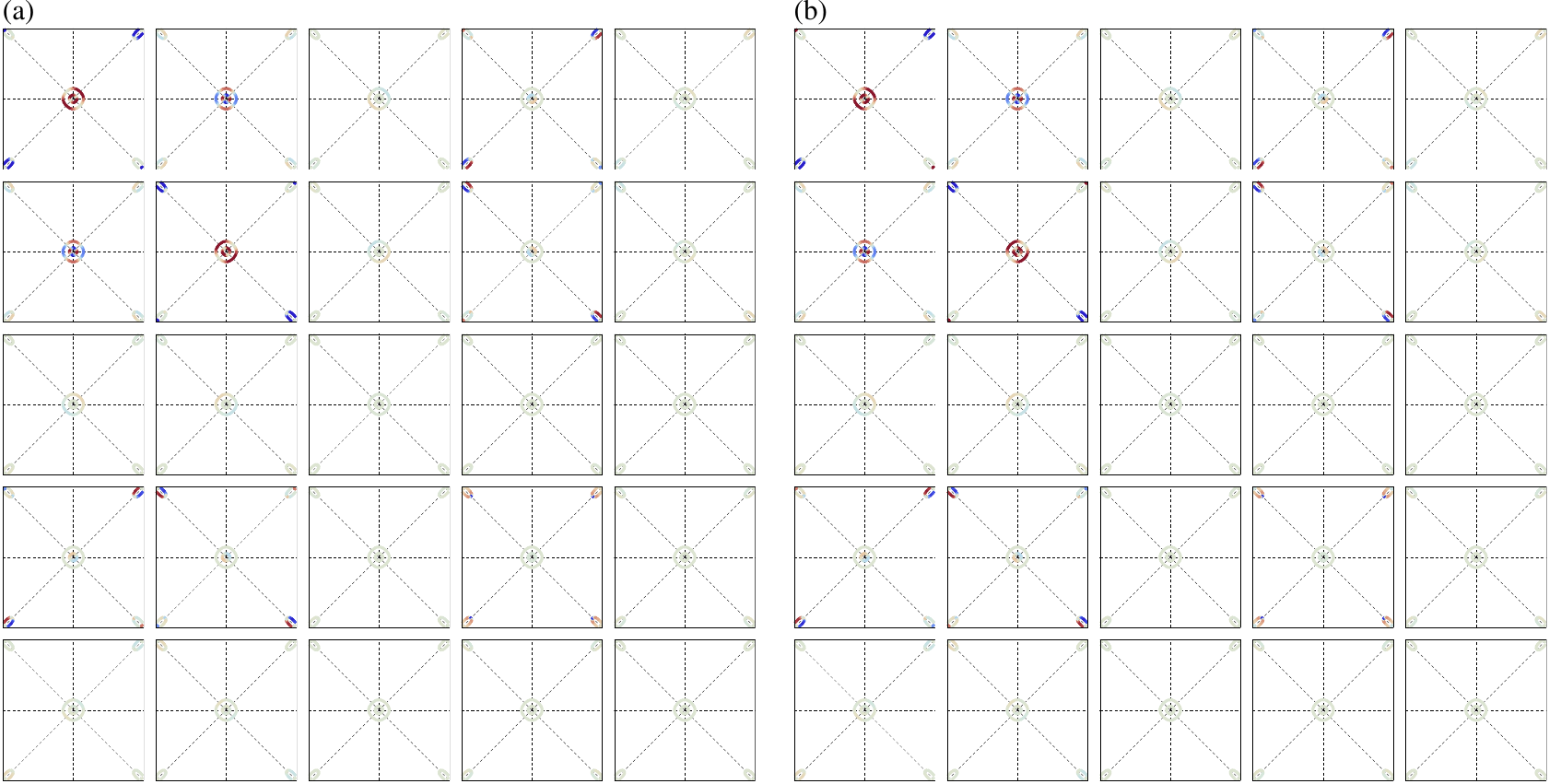}
\end{minipage}
\caption{Visualization of the spin-singlet part of the $ s_{+-} $ pairing solution in sublattice $ \otimes $ orbital space in the 2-Fe BZ for (a) intrasublattice ($ l = A, l^{\prime} = A $) and (b) intersublattice ($ l = A, l^{\prime} = B $) components, where the orbital components are ordered as $ d_{xz} $, $ d_{yz} $, $d_{x^{2}-y^{2}}$, $d_{xy}$, $d_{3z^2-r^2}$. The LGE was solved for $ \mu_{0} = 0 \, $eV, $ \lambda = 0\, $eV, $ T = 0.01 \, $eV, $ U = 1.30 \,$eV, $ J = U/4$.}
\label{fig:spin_singlet_1_FeSe}
\end{figure}

\vfill

\begin{figure}[h!]
\centering
\begin{minipage}{0.90\columnwidth}
\centering
\includegraphics[width=1\columnwidth]{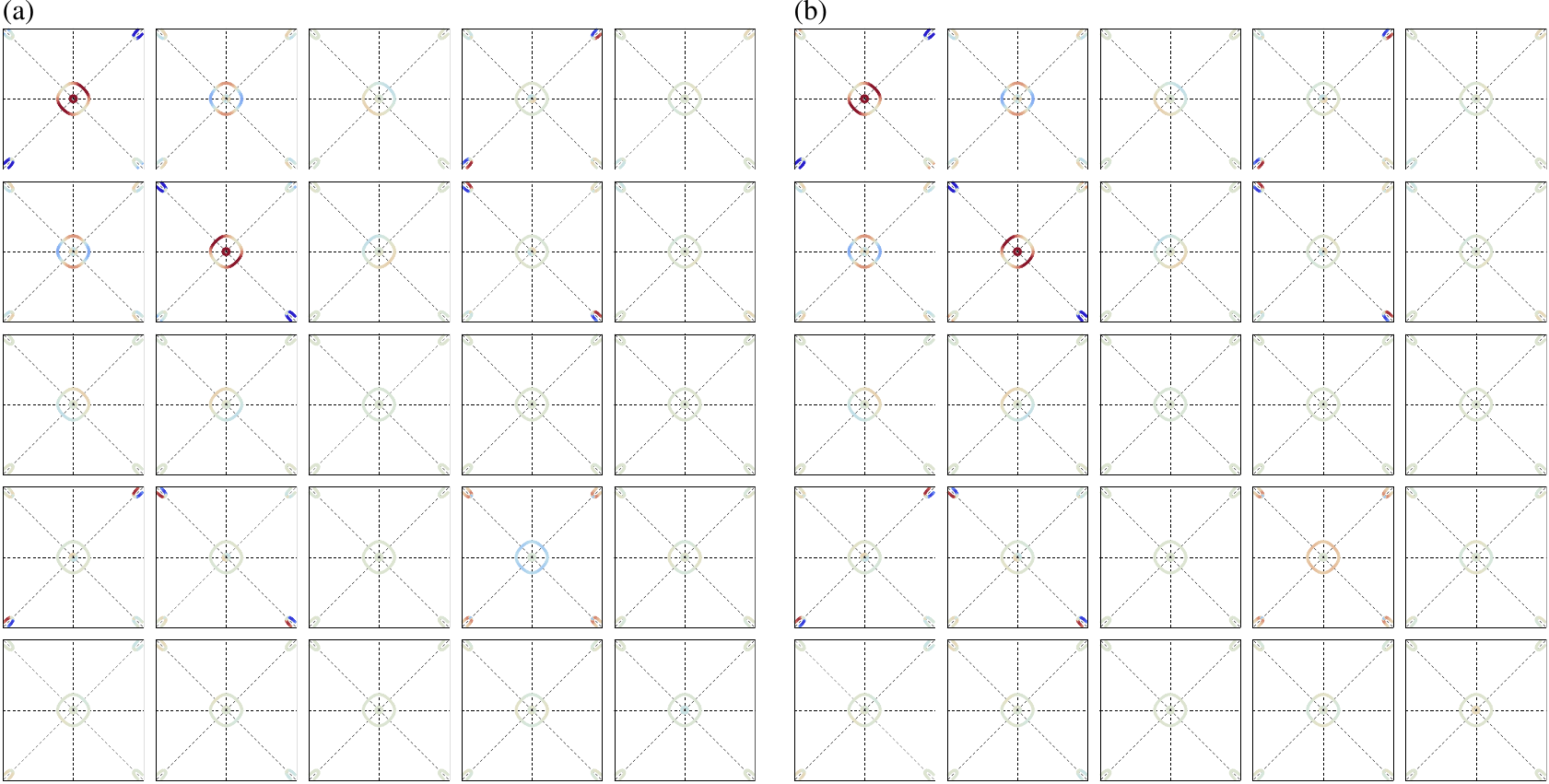}
\end{minipage}
\caption{Visualization of the spin-singlet part of the $ s_{+-} $ pairing solution in sublattice $ \otimes $ orbital space in the 2-Fe BZ for (a) intrasublattice ($ l = A, l^{\prime} = A $) and (b) intersublattice ($ l = A, l^{\prime} = B $) components, where the orbital components are ordered as $ d_{xz} $, $ d_{yz} $, $d_{x^{2}-y^{2}}$, $d_{xy}$, $d_{3z^2-r^2}$. The LGE was solved for $ \mu_{0} = 0 \, $eV, $ \lambda = 75\, $meV, $ T = 0.01 \, $eV, $ U = 1.30 \,$eV, $ J = U/4$.}
\label{fig:spin_singlet_2_FeSe}
\end{figure}

\vfill

\begin{figure}[h!]
\centering
\begin{minipage}{0.90\columnwidth}
\centering
\includegraphics[width=1\columnwidth]{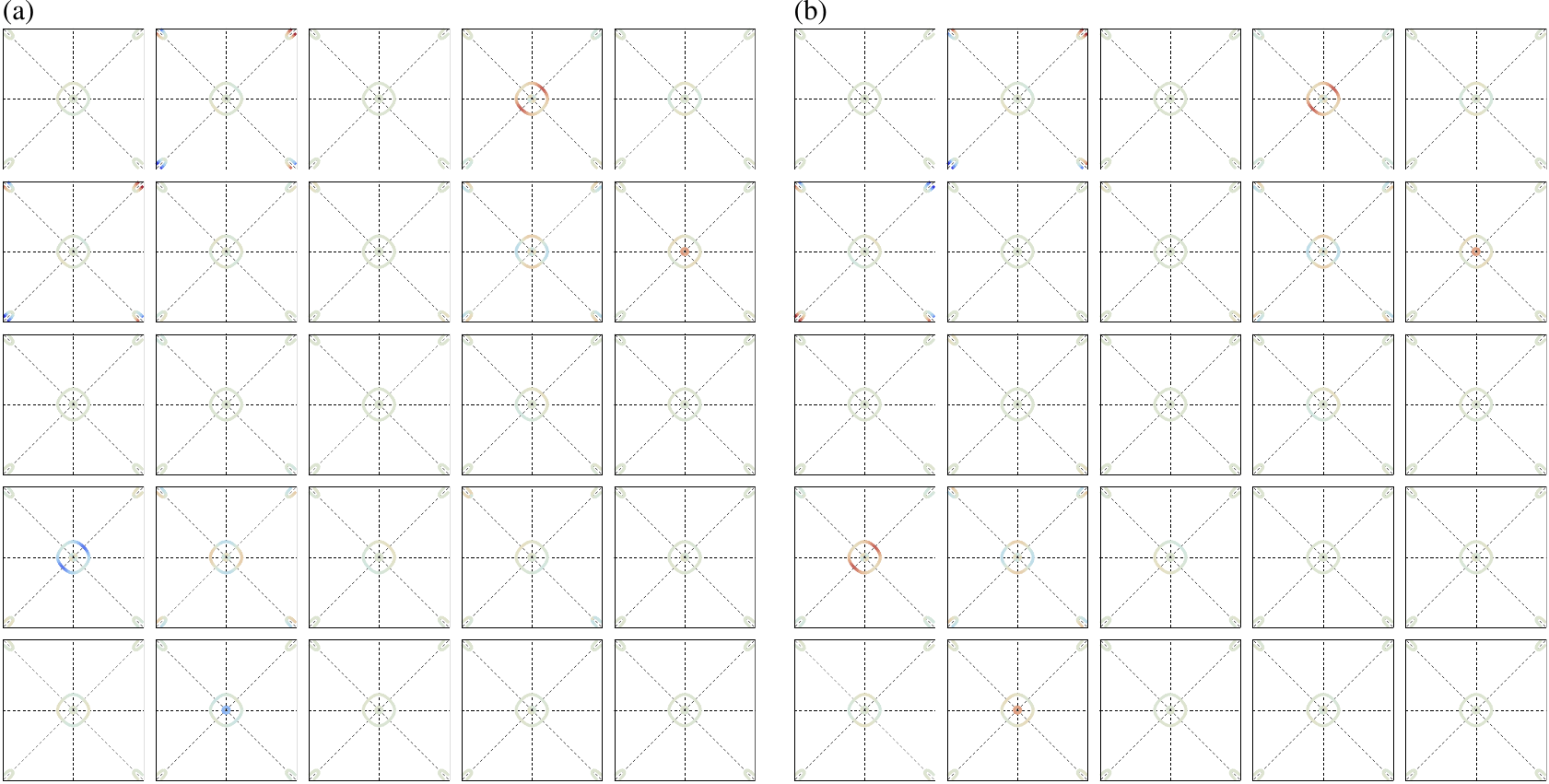}
\end{minipage}
\caption{Visualization of the spin-triplet-$x$ part of the $ s_{+-} $ pairing solution in sublattice $ \otimes $ orbital space in the 2-Fe BZ for (a) intrasublattice ($ l = A, l^{\prime} = A $) and (b) intersublattice ($ l = A, l^{\prime} = B $) components, where the orbital components are ordered as $ d_{xz} $, $ d_{yz} $, $d_{x^{2}-y^{2}}$, $d_{xy}$, $d_{3z^2-r^2}$. The LGE was solved for $ \mu_{0} = 0 \, $eV, $ \lambda = 75\, $meV, $ T = 0.01 \, $eV, $ U = 1.30 \,$eV, $ J = U/4$.}
\label{fig:spin_triplet_2_FeSe}
\end{figure}

\vfill

\clearpage



\end{widetext}


\begin{thebibliography}{n}

\bibitem{scalapinoreview} 
D. J. Scalapino, 
Rev. Mod. Phys. {\bf 84}, 1383 (2012).

\bibitem{Taillefer2019} 
L. Taillefer, 
Annu. Rev. Condens. Matter Phys. {\bf 1}, 51 (2010).

\bibitem{TBG} 
Y. Cao, V. Fatemi, S. Fang, K. Watanabe, T. Taniguchi, E. Kaxiras, and P. Jarillo-Herrero, 
Nature {\bf 556}, 43 (2018).

\bibitem{kohn} 
W. Kohn and J. H. Luttinger, 
Phys. Rev. Lett. {\bf 15}, 524 (1965).

\bibitem{berk66} 
N. F. Berk and J. R. Schrieffer, 
Phys. Rev. Lett. {\bf 17}, 433 (1966).

\bibitem{scalapino86} 
D. J. Scalapino, E. Loh Jr., and J. E. Hirsch, 
Phys. Rev. B {\bf 34}, 8190 (1986).

\bibitem{beal-monod86} 
M. T. Beal-Monod, C. Bourbonnais, and V. J. Emery, 
Phys. Rev. B {\bf 34}, 7716 (1986).

\bibitem{miyake86} 
K. Miyake, S. Schmitt-Rink, and C.M. Varma, 
Phys. Rev. B {\bf 34}, 6554 (1986).

\bibitem{maiti} 
S. Maiti and A. V. Chubukov, 
AIP Conf. Proceedings {\bf 1550}, 3 (2013).

\bibitem{hirschfeld2011} 
P. J. Hirschfeld, M. M. Korshunov, and I. I. Mazin, 
Rep. Prog. Phys.  {\bf 74}, 124508 (2011).

\bibitem{chubukov2012} 
A. V. Chubukov, 
Annu. Rev. Condens. Matter Phys. {\bf 3}, 57 (2012).

\bibitem{hirschfeld2015} 
P. J. Hirschfeld, D. Altenfeld, I. Eremin, and I. I. Mazin, 
Phys. Rev. B  {\bf 92}, 184513 (2015).

\bibitem{martiny} 
J. H. J. Martiny, A. Kreisel, P. J. Hirschfeld, and B. M. Andersen, 
Phys. Rev. B  {\bf 95}, 184507 (2017).

\bibitem{sprau} 
P. O. Sprau, A. Kostin, A. Kreisel, A. E. B\"{o}hmer, P. C. Taufour, V. Canfield, S. Mukherjee, P. J. Hirschfeld, B. M.
Andersen, and J. Davis, 
Science  {\bf 357}, 75 (2017).

\bibitem{HHWen} 
Z. Du, D. Altenfeld, Q. Gu, H. Yang, I. Eremin, P. J.
Hirschfeld, I. I. Mazin, H. Lin, X. Zhu, and H. H. Wen,
Nat. Phys. {\bf 14}, 134 (2017).

\bibitem{hirschfeld2016} 
P. Hirschfeld, 
C. R. Phys. {\bf 17}, 197 (2016).

\bibitem{borisenkolifeas1} 
S. V. Borisenko, V. B. Zabolotnyy, D. V. Evtushinsky, T. K. Kim, I. V. Morozov, A. N. Yaresko, A. A. Kordyuk, G. Behr, A. Vasiliev, R. Follath, and B. B\"{u}chner, 
Phys. Rev. Lett.  {\bf 105}, 067002 (2010).

\bibitem{borisenkolifeas2} 
S. V. Borisenko, V. B. Zabolotnyy, A. A. Kordyuk, D. V. Evtushinsky, T. K. Kim, I. V. Morozov, R. Follath,
and B. B\"{u}chner, 
Symmetry  {\bf 4}, 251 (2012).

\bibitem{Umezawa} 
K. Umezawa, Y. Li, H. Miao, K. Nakayama, Z.-H. Liu, P. Richard, T. Sato, J. B. He, D.-M. Wang, G. F. Chen, H. Ding, T. Takahashi, and S.-C. Wang, 
Phys. Rev. Lett. {\bf 108}, 037002 (2012).

\bibitem{Allan} 
M. P. Allan, A. W. Rost, A. P. Mackenzie, Y. Xie, J. C. Davis, K. Kihou, C. H. Lee, A. Iyo, H. Eisaki, and T.-M. Chuang, 
Science {\bf 336}, 563 (2012).

\bibitem{Evtushinsky} 
D. V. Evtushinsky, V. B. Zabolotnyy, T. K. Kim, A. A. Kordyuk, A. N. Yaresko, J. Maletz, S. Aswartham, S.Wurmehl, A. V. Boris, D. L. Sun, C. T. Lin, B. Shen, H. H. Wen, A.Varykhalov, R. Follath, B. Buchner, and S. V. Borisenko, 
Phys. Rev. B {\bf 89}, 064514 (2014).

\bibitem{Terashima} T. Terashima,
M. Kimata, N. Kurita, H. Satsukawa, A. Harada, K. Hazama, M. Imai, A. Sato, K. Kihou, C.-H. Lee, H. Kito, H. Eisaki, A. Iyo, T. Saito, H. Fukazawa, Y. Kohori, H. Harima, and S. Uji, 
J. Phys. Soc. Jpn. {\bf 79}, 053702 (2010).

\bibitem{Dong} J. K. Dong, S. Y. Zhou, T. Y. Guan, H. Zhang, Y. F. Dai, X.
Qiu, X. F.Wang, Y. He, X. H. Chen, and S. Y. Li, Phys. Rev.
Lett. {\bf 104}, 087005 (2010).

\bibitem{Reid} J.-Ph. Reid, M. A. Tanatar, A.
Juneau-Fecteau, R. T. Gordon, S. R. de Cotret, N. Doiron-
Leyraud, T. Saito, H. Fukazawa, Y. Kohori, K. Kihou, C. H.
Lee, A. Iyo, H. Eisaki, R. Prozorov, and L. Taillefer, Phys.
Rev. Lett. {\bf 109}, 087001 (2012).

\bibitem{Okazaki} K. Okazaki, Y. Ota, Y. Kotani, W. Malaeb, Y. Ishida, T.
Shimojima, T. Kiss, S. Watanabe, C.-T. Chen, K. Kihou,
C. H. Lee, A. Iyo, H. Eisaki, T. Saito, H. Fukazawa, Y.
Kohori, K. Hashimoto, T. Shibauchi, Y. Matsuda, H. Ikeda,
H. Miyahara, R. Arita, A. Chainani, and S. Shin, Science
{\bf 337}, 1314 (2012).

\bibitem{Bohm} T. B\"{o}hm, A. F. Kemper, B. Moritz, F. Kretzschmar, B.
Muschler, H.-M. Eiter, R. Hackl, T. P. Devereaux, D. J.
Scalapino, and H.-H. Wen, Phys. Rev. X {\bf 4}, 041046 (2014).

\bibitem{Hardy} F. Hardy,  R. Eder, M. Jackson, D. Aoki, C. Paulsen, T. Wolf, P. Burger, A. B\"{o}hmer, P. Schweiss, P. Adelmann, R. A. FisheR, and C. Meingast, J. Phys. Soc. Jpn. {\bf 83}, 014711 (2014).

\bibitem{Coldeareview} A. I. Coldea and M. D. Watson, Annu. Rev. Condens. Matter Phys. {\bf 9}, 125 (2018)

\bibitem{Bohmerreview} A. E. B\"{o}hmer and A. Kreisel, J. of Phys.: Cond. Mat. {\bf 30}, 023001 (2017).

\bibitem{Kostin}  A. Kostin, P. O. Sprau, A. Kreisel, Y. X. Chong, A. E. B\"{o}hmer, P. C. Canfield, P. J. Hirschfeld, B. M. Andersen, J. C. S\'{e}amus Davis, Nature Materials {\bf 17}, 869 (2018).

\bibitem{onari} S. Onari, Y. Yamakawa, and H. Kontani, Phys. Rev. Lett. {\bf 116}, 227001 (2016).

\bibitem{kreisel2017} A. Kreisel, B. M. Andersen, P. O. Sprau, A. Kostin, J. C. S\'{e}amus Davis, and P. J. Hirschfeld
Phys. Rev. B  {\bf 95}, 174504 (2017).

\bibitem{she} J.-H. She, M. J. Lawler, and E.-A. Kim, Phys. Rev. Lett. {\bf 121}, 237002 (2018).

\bibitem{kang2018} J. Kang, R. M. Fernandes, and A. Chubukov, Phys. Rev. Lett. {\bf 120}, 267001 (2018).

\bibitem{benfatto} L. Benfatto, B. Valenzuela, and L. Fanfarillo, npj Quantum Mater. {\bf 3}, 56 (2018).

\bibitem{rhodes} L. C. Rhodes, M. D. Watson, A. A. Haghighirad, D. V. Evtushinsky, M. Eschrig, and T. K. Kim, Phys. Rev. B {\bf 98}, 180503 (2018).

\bibitem{kreisel2018} A. Kreisel, B. M. Andersen, and P. J. Hirschfeld, Phys. Rev. B {\bf 98}, 214518 (2018).

\bibitem{Jiang} K. Jiang, J. Hu, H. Ding, and Z. Wang, Phys. Rev. B {\bf 93}, 115138 (2016).

\bibitem{scherer2017} D.~D.~Scherer, A.~C.~Jacko, C.~Friedrich, E. \c{S}a\c{s}{\i}o\u{g}lu, S. Bl\"{u}gel, R.~Valent\'{i}, B.~M.~Andersen,
Phys. Rev. B {\bf 95}, 094504 (2017).

\bibitem{borisenko} 
S. V. Borisenko, D. V. Evtushinsky, Z.-H. Liu, I. Morozov, R. Kappenberger, S. Wurmehl, B. B\"{u}chner, A. N. Yaresko, T. K. Kim, M. Hoesch, T. Wolf, and N. D. Zhigadlo, 
Nat. Phys. {\bf 12}, 311 (2016).

\bibitem{day} R. Day, G. Levy, M. Michiardi, B. Zwartsenberg, M. Zonno, F. Ji, E. Razzoli, F. Boschini, S. Chi, R. Liang, P. Das, I. Vobornik, J. Fujii, W. Hardy, D. Bonn, I. Elfimov, and A. Damascelli, Phys. Rev. Lett. {\bf 121}, 076404 (2018).

\bibitem{saito} T. Saito, Y. Yamakawa, S. Onari, and H. Kontani, Phys. Rev. B {\bf 92}, 134522 (2015).

\bibitem{liu_2017} L. Liu, K. Okazaki, T. Yoshida, H. Suzuki, M. Horio, L. C. C. Ambolode, II, J. Xu, S. Ideta, M. Hashimoto, D. H. Lu, Z.-X. Shen, Y. Ota, S. Shin, M. Nakajima, S. Ishida, K. Kihou, C. H. Lee, A. Iyo, H. Eisaki, T. Mikami, T. Kakeshita, Y. Yamakawa, H. Kontani, S. Uchida, and A. Fujimori, Phys. Rev. B {\bf 95}, 104504 (2017).

\bibitem{khodas} M. Khodas and A. V. Chubukov, Phys. Rev. Lett. {\bf 108}, 247003 (2012).

\bibitem{cvetkovic} V. Cvetkovic and O. Vafek, Phys. Rev. B  {\bf 88}, 134510 (2013).

\bibitem{vafek} O. Vafek and A. V. Chubukov, Phys. Phys. Lett.  {\bf 118}, 087003 (2017).

\bibitem{eugenio} P. M. Eugenio and O. Vafek, Phys. Rev. B  {\bf 98}, 014503 (2018).

\bibitem{Boker} J. B\"{o}ker, P. A. Volkov, P. J. Hirschfeld, and I. Eremin, ArXiv:1903.05935.

\bibitem{scherer2018}
D. D. Scherer and B. M. Andersen,
Phys. Rev. Lett. {\bf 121}, 037205 (2018).

\bibitem{christenson2008} A. D. Christianson,  E. A. Goremychkin, R. Osborn, S. Rosenkranz, M. D. Lumsden, C. D. Malliakas, I. S. Todorov, H. Claus, D. Y. Chung, M. G. Kanatzidis, R. I. Bewley, and T. Guidi, Nature (London) {\bf 456}, 930 (2008).

\bibitem{lumsden} M. D. Lumsden and A. D. Christianson, J. Phys.: Condens. Matter {\bf 22}, 203203 (2010).

\bibitem{dai} 
Pengcheng Dai, 
Rev. Mod. Phys. {\bf 87}, 855 (2015).

\bibitem{Inosov2016} D. S. Inosov, 
Comptes Rendus Physique  \textbf{17}, 60 (2016).

\bibitem{scherer2019}
D. D. Scherer and B. M. Andersen, arXiv:1906.08566.

\bibitem{astrid19}
A. T. R\o mer, D. D. Scherer, I. M. Eremin, P. J. Hirschfeld, and B. M. Andersen, 
arXiv:1905.04782.

\bibitem{hongding1} 
Peng Zhang, Koichiro Yaji, Takahiro Hashimoto, Yuichi Ota, Takeshi Kondo, Kozo Okazaki, Zhijun Wang, Jinsheng Wen, G. D. Gu, Hong Ding, Shik Shin, 
Science  {\bf 360}, 182 (2018).

\bibitem{hongding2} 
Dongfei Wang, Lingyuan Kong, Peng Fan, Hui Chen, Yujie Sun, Shixuan Du, J. Schneeloch, R.D. Zhong, G.D. Gu, Liang Fu, Hong Ding, Hongjun Gao, 
Science  {\bf 362}, 333 (2018).

\bibitem{jiang2019} Kun Jiang, Xi Dai, and Ziqiang Wang, Phys. Rev. X {\bf 9}, 011033 (2019).

\bibitem{pan} J.-X. Yin, Zheng Wu, J-H. Wang, Z-Y. Ye, Jing Gong, X-Y. Hou, Lei Shan, Ang Li, X-J. Liang, X-X. Wu, Jian Li, C-S. Ting, Z-Q. Wang, J-P. Hu, P-H. Hor, H. Ding, S. H. Pan, Nat. Physics {\bf 11}, 543 (2015).

\bibitem{ikeda2010}
H.~Ikeda, R.~Arita, and J.~Kune$\check{\mathrm{s}}$, 
Phys. Rev. B {\bf 81}, 054502 (2010).

\bibitem{maier2009}
T.~A.~Maier, S.~Graser, D.~J.~Scalapino, and P.~Hirschfeld, 
Phys. Rev. B \textbf{79}, 134520 (2009).

\bibitem{Watson_Hub} M. D. Watson, S. Backes, A. A. Haghighirad, M. Hoesch, T. K.
Kim, A. I. Coldea, and R. Valent\'{i}, Phys. Rev. B {\bf 95}, 081106 (2017).

\bibitem{Evtushinsky_Hub} D. V. Evtushinsky, M. Aichhorn, Y. Sassa, Z.-H. Liu, J. Maletz,
T. Wolf, A. N. Yaresko, S. Biermann, S. V. Borisenko, and B.
B\"{u}chner, arXiv:1612.02313.

\bibitem{luca} L. de' Medici, Hund's metal explained, in The Physics of Correlated
Insulators, Metals, and Superconductors, Modeling and
Simulation Vol. 7, edited by E. Pavarini, E. Koch, R. Scalettar,
and R. Martin (Verlag des Forschungszentrum J\"{u}lich, J\"{u}lich,
2017).

\bibitem{kreisel_2015} A. Kreisel, S. Mukherjee, P. J. Hirschfeld, and B. M. Andersen, Phys.
Rev. B {\bf 92}, 224515 (2015).

\bibitem{rahn} M. C. Rahn, R. A. Ewings, S. J. Sedlmaier, S. J. Clarke, and
A. T. Boothroyd, Phys. Rev. B {\bf 91}, 180501
(2015).

\bibitem{Wang_NS1} Q. Wang, Y. Shen, B. Pan, Y. Hao, M. Ma, F. Zhou, P. Steffens,
K. Schmalzl, T. R. Forrest, M. Abdel-Hafiez, X. Chen,
D. A. Chareev, A. N. Vasiliev, P. Bourges, Y. Sidis, H. Cao,
and J. Zhao, Nat. Mater. {\bf 15}, 159
(2016).

\bibitem{Wang_NS2} Q. Wang, Y. Shen, B. Pan, X. Zhang, K. Ikeuchi, K. Iida, A. D.
Christianson, H. C. Walker, D. T. Adroja, M. Abdel-Hafiez, X.
Chen, D. A. Chareev, A. N. Vasiliev, and J. Zhao, Nat. Commun. {\bf 7}, 12182 (2016).

\bibitem{Tong} T. Chen, Y. Chen, A. Kreisel, X. Lu, A. Schneidewind, Y. Qiu, J. Park, T. G. Perring, J. R. Stewart, H. Cao, R. Zhang, Y. Li, Y. Rong, Y. Wei, B. M. Andersen, P. J. Hirschfeld, C. Broholm, and P. Dai, Nat. Mater. {\bf 18}, 709 (2019).

\bibitem{MingYi} M. Yi, Y. Zhang, H. Pfau, T. Chen, Zi. Ye, M. Hashimoto, R. Yu, Q. Si, D.-H. Lee, P. Dai, Z. -X. Shen, D. Lu, R. J. Birgeneau, arXiv:1903.04557.



\end{thebibliography}

\begin{thebibliography}{00}

%
\bibitem{SMsigrist1991}
M.~Sigrist and K.~Ueda,
Rev. Mod. Phys. {\bf 63}, 239 (1991).

%
\bibitem{SMueda1985}
K.~Ueda and T.~M.~Rice,
Phys. Rev. B {\bf 31}, 7114 (1985).

%
\bibitem{SMnourafkan2017}
R.~Nourafkan and A.-M.~S.~Tremblay
Phys. Rev. B {\bf 96}, 125140 (2017).

%
\bibitem{SMroemer2019}
A.~T.~R\o mer, D.~D.~Scherer, I.~M.~Eremin, P.~J. Hirschfeld, B.~M. Andersen,
arXiv:1905.04782.

%
\bibitem{SMzhang2018}
Li-Da Zhang, Wen Huang, Fan Yang, and Hong Yao
Phys. Rev. B {\bf 97}, 060510(R) (2018).

%
\bibitem{SMnomoto2016}
T. Nomoto, K. Hattori, and H. Ikeda
Phys. Rev. B {\bf 94}, 174513 (2016).

%
\bibitem{SMscherer2016}
D.~D.~Scherer, I.~Eremin, and B.~M.~Andersen
Phys. Rev. B \textbf{94}, 180405(R) (2016).

\bibitem{SMscherer2018}
D.~D.~Scherer and B.~M.~Andersen,
Phys. Rev. Lett. \textbf{121}, 037205 (2018).

\end{thebibliography}
\end{document}